\documentclass[twocolumn,10pt,a4paper]{article}

\usepackage{palatino}
\usepackage{amsmath}
\usepackage{amssymb}
\usepackage{graphicx}
\usepackage{graphics}
\usepackage{subfigure}

\begin{document}
\title{\sf \bf Hierarchical Characterization of Complex Networks}

\author{
    Luciano da Fontoura Costa and Filipi Nascimento Silva \footnote{Cybernetic Vision Research
    Group, GII-IFSC, Universidade de S\~ ao Paulo, S\~{a}o Carlos, SP,
    Caixa Postal 369, 13560-970, Brasil, luciano@if.sc.usp.br.  }
    }

\maketitle

\begin{abstract}
While the majority of approaches to the characterization of complex
networks has relied on measurements considering only the immediate
neighborhood of each network node, valuable information about the
network topological properties can be obtained by considering
further neighborhoods.  The current work discusses on how the
concepts of hierarchical node degree and hierarchical clustering
coefficient (introduced in cond-mat/0408076), complemented by new
hierarchical measurements, can be used in order to obtain a powerful
set of topological features of complex networks.  The interpretation
of such measurements is discussed, including an analytical study of
the hierarchical node degree for random networks, and the potential
of the suggested measurements for the characterization of complex
networks is illustrated with respect to simulations of random,
scale-free and regular network models as well as real data
(airports, proteins and word associations).  The enhanced
characterization of the connectivity provided by the set of
hierarchical measurements also allows the use of agglomerative
clustering methods in order to obtain taxonomies of relationships
between nodes in a network, a possibility which is also illustrated
in the current article.
\end{abstract}

\section{Introduction}

Graph theory and statistical mechanics are well-established areas in
mathematics and physics, respectively.  Since its beginnings in the
XVIII century, with the solution of the bridges problem by L. Euler,
graph theory has progressed all the way to the forefront of
theoretical and applied investigations in mathematics and computer
science.  Much of the importance of this broad area stems from the
\emph{generality} of graphs as representational models.  As a matter
of fact, most discrete structures including matrices, trees, queues,
among many others, are but particular cases of graphs.  The potential
of graphs is further extended by models where features are assigned to
nodes, different types of nodes and/or edges are allowed to co-exist,
synchronization schemes are incorporated, and so on (see, for
instance,~\cite{Newman:surv}).  At the same time, statistical
mechanics, also drawing on a rich past of accomplishments, provides
concepts and tools for bridging the gap between dynamics in the micro
and macro realms.  Of particular interest have been the investigations
on phase transitions and complex systems, which represent a major area
of development today.

While graph theory provides effective means for characterizing,
modeling and simulating the \emph{structure} of natural phenomena,
statistical mechanics contains the methods for analyzing the
\emph{dynamics} of natural phenomena along several scales. The novel
area of \emph{complex networks}~\cite{AB:surv,Newman:surv} can be
understood as a fortunate intersection between those two major areas,
therefore allowing a natural and powerful means for integrating
structure and dynamics.  With origins extending back to the pioneering
developments of Flory~\cite{Flory}, Rapoport~\cite{Rapoport} and
Erd\"os and R\'enyi~\cite{Erdos_Renyi}, the area of complex networks
was boosted more recently by the advances by Watts and
Strogatz~\cite{Watts_Strogatz,Small_Worlds} and Barab\'asi and
collaborators~\cite{Barabasi}.

Complex network investigations frequently involve the measurement of
topological features of the analyzed structures, such as the
\emph{node degree} (namely the number of edges attached to a node) and
the clustering coefficient (quantifying the connectivity among the
immediate neighbors of a node).  Although degenerated, in the sense
that they do not allow a one-to-one identification of the possible
network architectures, such a pair of measurements does provide a rich
characterization of the connectivity of the networks.  As a matter of
fact, particularly interesting network models, such as the
small-world~\cite{Newman:surv,AB:surv,Small_Worlds,Watts_Strogatz} and
scale-free (Barab\'asi-Albert)~\cite{AB:surv,Newman:surv,Barabasi},
are characterized in terms of specific types of node degree
distributions (logarithmic and power-law, respectively).

Although such distributions emphasize important properties of the
analyzed networks, further valuable topological information can be
gathered not only by considering the clustering coefficient, but also
by analyzing such features along the \emph{hierarchical levels} of the
networks~\cite{PRL:Costa,Generalized}. While some attention has been
focused on the relevant issue of hierarchy in complex networks
(e.g.~\cite{Ravasz_Barab:2002, ravasz_etal:2003, Caldarelli:2003,
Barab_Oltvai:2003, Trusina_etal:2003, Boss_etal:2003,
Barthelemy_etal:2003, Zhou:2003, Vazquez:2003, Steyvers:2003,
Goldshtein,Cohen:2003,newman:2001,newman:physreve}), and hierarchical
extensions of the node degree and clustering coefficient were only
more recently formalized in~\cite{PRL:Costa,Generalized} by using
concepts derived from mathematical
morphology~\cite{Lotufo,Vincent:1989,Vincent:1992} including
\emph{dilations} and \emph{distance transforms} in graphs. Despite
their recent introduction, such concepts have already yielded valuable
results when applied to essentiality of protein-protein interaction
networks~\cite{heart}, bone structure characterization~\cite{bones},
and community finding~\cite{hub_based,Bollt}.

The purpose of the current article is to review and further extend
the concepts of hierarchical measurements, which is done by the
consideration of the concepts of \emph{radial reference system} and
\emph{hierarchical common degree}, as well as the introduction of
the measurements of \emph{hierchical edge degree}, \emph{inter-ring
degree}, \emph{intra-ring degree}, \emph{convergence ratio}, and
emph{edge clustering coefficient}. The extensions of these
measurement (excluding the clustering coefficient) to weighted and
directed networks are also described in this work.  We start by
presenting the basic concepts and discussing hierarchies in complex
networks in terms of \emph{virtual nodes} and proceed by describing,
interpreting and discussing the hierarchical measurements.  An
analytical characterization of the general shape of the hierarchical
node degree in random networks is also presented, and the potential
of the reported concepts and methods is illustrated with respect to
the characterization of simulated random, scale-free and regular
network models.  Such a potential is further illustrated with
respect to real networks, including word associations, airports, and
protein-protein interactions.  Because the hierarchical measurements
provide a rich characterization of the connectivity around each
network node, it becomes possible to use clustering
methods~\cite{DudaHart:book,LCosta:book} in order to organize the
nodes in a network into a taxonomical scheme reflecting the
similarities between their connectivity.  This possibility is also
illustrated in the present article.

\section{Notation and Basic Concepts}

Let the graph or network $\Gamma$ of interest contain $N$ nodes and
$e$ edges, and the connections between any two nodes $i$ and $j$ be
represented as $(i,j)$.  Although non-oriented graphs are assumed
henceforth, all reported concepts and methods can be immediately
extended to digraphs and weighted networks.  We henceforth assume the
complete absence of loops (i.e. self-connections). A non-oriented
graph can be completely specified in terms of its \emph{adjacency
matrix} $K$, with each connection $(i,j)$ implying $K(i,j)=K(j,i)=1$.
The absence of a connection between nodes $i$ and $j$ is represented
as $K(i,j)=K(j,i)=0$. Now, the \emph{node degree} $k(i)$ of a node
$i$ of $\Gamma$ can be defined as

\begin{equation}
k(i) = \sum_{j=1}^{N}K(i,j) =  \sum_{j=1}^{N}K(j,i).
\end{equation}

Observe that the degree of node $i$ corresponds to the number of edges
attached to that node, representing a direct measurement of the
\emph{connectivity} of that specific node.  Indeed the overall
connectivity of a specific network can be quantified in terms of its
average node degree $\left< k \right>$.  While a random network is
characterized by a typical average node degree with relatively low
standard deviation, a scale-free model will present a power-law log-log
distribution of node degrees, favoring the existence of hubs
(i.e. nodes with high node degree).

The \emph{clustering coefficient} of a network node $i$ can be defined
as quantifying the connectivity among the immediate neighbors of $i$,
which are henceforth represented by the set $R_1(i)$. More
specifically, in case that node has $n_1(i)$ immediate neighbors
(i.e., the cardinality of $R(i)$), implying a maximum number
$e_T(i)=n_1(n_1-1)/2$ of connections between such nodes, and $e(i)$
connections are observed among such neighbors, the clustering
coefficient of $i$ can be calculated as

\begin{equation}
cc(i)= \frac{e(i)}{e_T(i)} = 2 \frac{e(i)}{n_1(i)(n_1(i)-1)} \label{eq:cc}.
\end{equation}

Observe that $0 \leq cc(i) \leq 1$, with the minimum and maximum
values being achieved for complete absence of connections (for
$cc(i)=0$) and complete connectivity among the neighbors of $i$ (for
$cc(i)=1$).

\begin{figure}
 \begin{center}
   \includegraphics[scale=0.3,angle=-90]{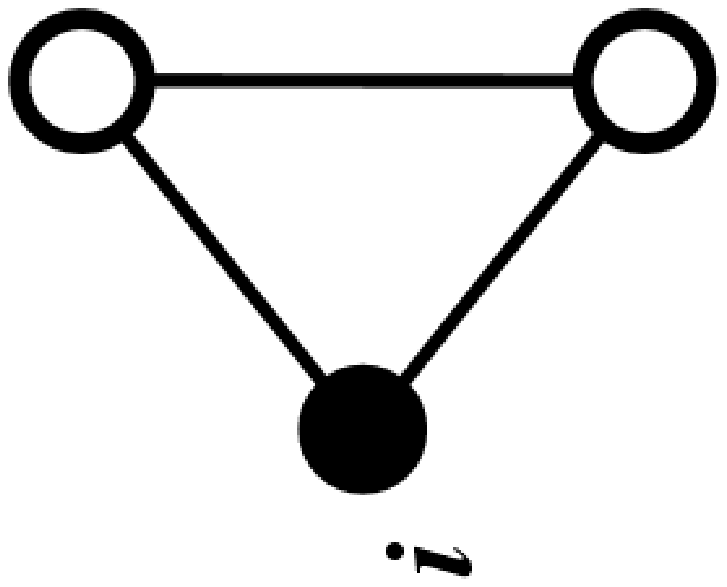}
   \hspace{2cm}
   \includegraphics[scale=0.3,angle=-90]{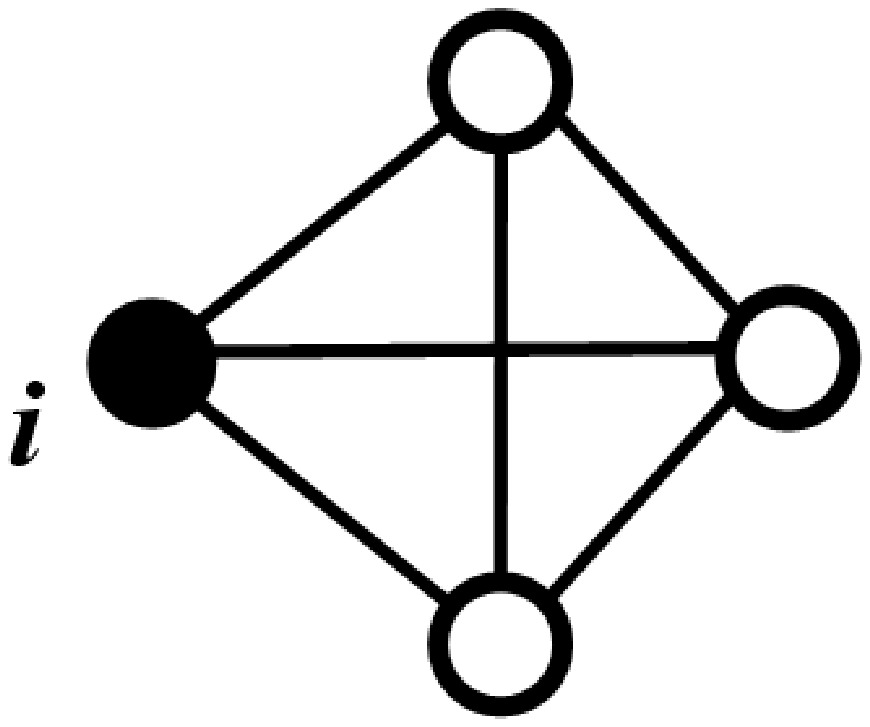} \\
   \includegraphics[scale=0.3,angle=-90]{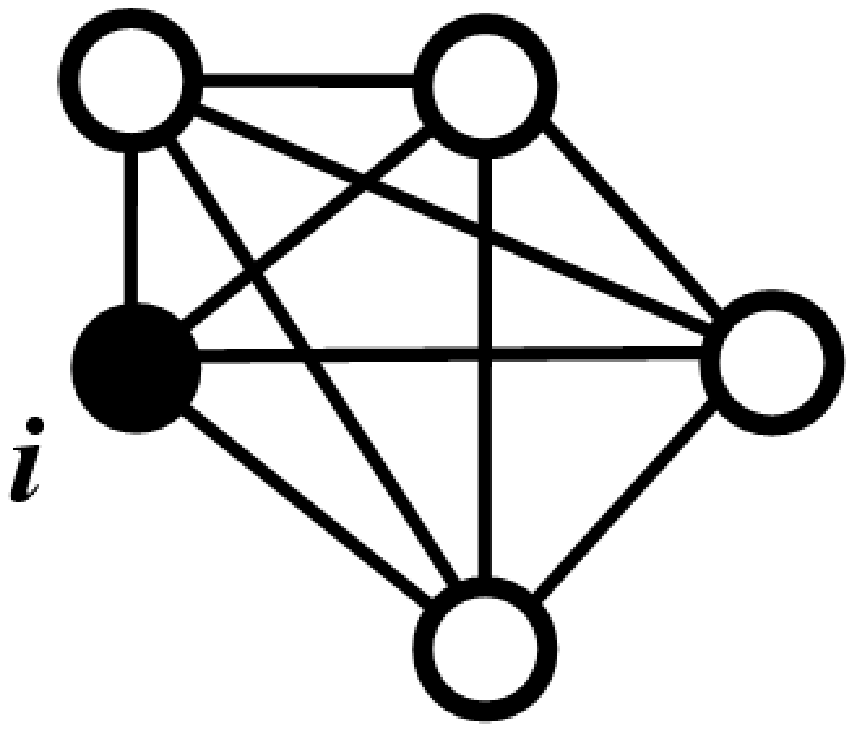} \\
   \caption{Three situations yielding the same clustering coefficient
   (equal to 1) for the reference node $i$.~\label{fig:ex1}}
  \end{center}
\end{figure}

Although the clustering coefficient provides a powerful indication
about the connectivity among the neighbors of the reference node,
several different situations (see Figure~\ref{fig:ex1}) may yield the
same clustering coefficient value (1 for these examples), which is a
consequence of the fact that this measurement is relative to the total
number of connections among the elements of $S(i)$. Such situations
can be distinguished by considering the respective value of $n_1(i)$.

\section{Virtual Edges and Hierarchies}

\begin{figure}
 \begin{center}
   \includegraphics[scale=0.7,angle=-90]{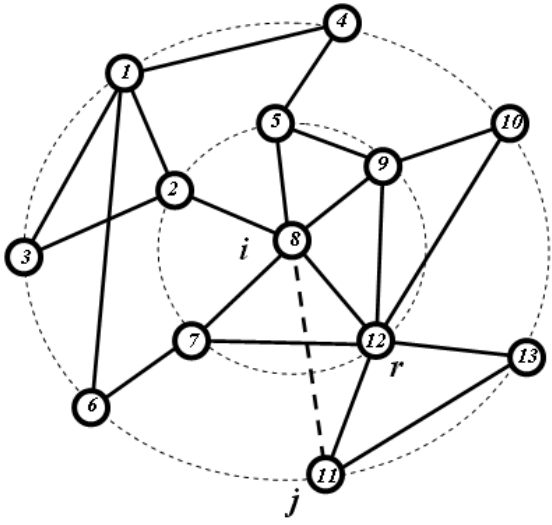} \\
   \caption{A small network and a reference node $i$. The
   \emph{virtual edge} between nodes $i$ and $j$, one of the many of
   such a kind in this network, is represented by the dashed
   line.~\label{fig:ex}}
  \end{center}
\end{figure}

Consider the situation depicted in Figure~\ref{fig:ex}, where a
reference node $i=1$ is connected to several other network nodes.  The
set of immediate neighbors of $i$, hence $R_1(i)$, is identified by
the innermost ellipsis.  Observe that although no connection is
observed between nodes $i$ and $j$, information from the former node
can propagate to the latter through the \emph{relay} node $r$, which
is indicated by the \emph{virtual edge}~\cite{PRL:Costa} shown as a
dashed line.

In the case of weighted networks, the \emph{virtual edges} may take
into account the cumulative effect of the respective weights.  For
instance, in case we had in Figure~\ref{fig:virtual} $W(i,r)=3$ and
$W(r,j)=4$, the weight of the virtual edge extending from $i$ to $j$
would be $W(i,j)=(3)(4)=12$.

The concept of virtual edge can be immediately extended by considering
further distances $d$ from the reference node.  Such an extension can
be naturally defined in terms of the weight matrix $W$ representing
the complex network of interest (observe that $W=K$ for weightless
networks).  Let $\vec{v}(i)$ be a column vector with $N$ elements
equal to zero, except that at the $i-th$ position (recall that $i$ is
the label of the reference node), which is assigned unit value.  Let
the vector $\vec{v}_1(i)$ be defined as

\begin{equation}
  \vec{v}_1(i) = W \vec{v}(i),
\end{equation}

and let the generalized Kronecker delta $\vec{a}=\delta(\vec{b})$ be
the operator acting on a vector $\vec{a}$ in order to produce a vector
$\vec{b}$ such that each element $b(j)$ of $\vec{b}$ is one if and
only $a(j)$ is different from zero, and zero otherwise.  By applying
such operator on $\vec{v}_1(i)$ we obtain

\begin{equation}
  \vec{p}_1(i) = \delta(\vec{v}_1(i)).
\end{equation}

The set of \emph{immediate neighbors} of $i$, i.e. $R_1(i)$, can now
be obtained as corresponding to the indices of the elements of
$\vec{p}_1(i)$ which are equal to 1.  For example, we have for the
situation depicted in Figure~\ref{fig:virtual} that $R_1(i=8)=\{ 2, 5,
7, 9, 12\}$.

The above matrix framework can be extended to any neighborhood of $i$
by introducing the vector $\vec{v}_d(i)$ defined as

\begin{equation}  \label{eq:vd}
  \vec{v}_d(i) = W^d \vec{v}(i).
\end{equation}

The weights of the \emph{virtual edges} between $i$ and the remainder
network nodes at distance $d$ are given by the successive entries of
$\vec{v}_d$, i.e. $W_d(i,j) = v_d(j)$.  Observe that the
\emph{distance} $d$ between two nodes $i$ and $j$ is henceforth
understood as corresponding to the number of edges along the shortest
path between those two nodes.

The set of neighbors of $i$ placed at distances varying from 0 to $d$
from the reference node $i$, henceforth represented as $B_d(i)$ and
referred to as the \emph{ball} of radius $d$ centered at $i$, can be
verified to correspond to the non-zero entries in the vector
$\vec{p}_d(i)$ defined as follows

\begin{equation}  \label{eq:pd}
  \vec{p}_d(i) = \delta \left( \sum_{j=1}^{d} \vec{p}_j(i)+ \vec{v}(i) \right).
\end{equation}

For instance, the ball of radius 2 centered at $i=8$ in
Figure~\ref{fig:ex} corresponds to the whole network in that
figure.  Now, the set of network nodes which are exactly at distance
$d$ from the reference node $i$ can be obtained as the unit entries in
the vector

\begin{equation} \label{eq:rd}
  \vec{r}_d(i) = \vec{p}_d(i)-\vec{p}_{d-1}(i).
\end{equation}

The set obtained from the above vector has also been
called~\cite{Generalized} the \emph{ring} of radius $d$ centered at
$i$, being henceforth represented as $R_d(i)$.  Observe that the ring
of radius 2 centered at $i=8$ in Figure~\ref{fig:ex} is
$R_2(8) =\{ 1, 3, 4, 6, 10, 11, 13\}$.

The subnetwork defined by the nodes at a specific ring $R_d(i)$,
together with the edges between them, is henceforth represented as
$\gamma_d(i)$.  We are now ready to define the \emph{hierarchical
level} $d$ of a complex network as corresponding to the nodes in
$\gamma_d(i)$ and the edges extending from such nodes and the nodes in
$\gamma_{d+1}(i)$.  The two hierarchical levels of nodes existing in
the network shown in Figure~\ref{fig:ex} are identified by the
inner and outermost ellipsis, respectively.  Observe that the
hierarchies $d$ provide a \emph{radial reference frame} or coordinate
system which can be used to partially identify nodes and edges with
respect to the reference node $i$.  The concept o hierarchy in a
complex network is also related to the concept of
\emph{roles}~\cite{roles} and the \emph{distance
transform}~\cite{Vincent:1989,Vincent:1992} of the nodes in the
original network $\Gamma$ with respect to the reference
node~\cite{Generalized}.

Observe that statistics of the number of hierarchical levels $d$ while
considering several nodes in a complex network provide a valuable
characterization of its topology.  Generally speaking, $d$ tends do
increase with the density of connections up to a peak, decreasing
afterwards.  At the same time, as will become clear along the
remainder of this article, the more connected the network is, the less
hierarchical levels it tends to have.  It should be also observed that
algorithmic implementation of hierarchy identification, such as those
reported in~\cite{PRL:Costa} and~\cite{Generalized} (see
also~\cite{Cormen}), are typically more computationally efficient than
the use of the matrix arithmetic presented in this Section.

\section{Hierarchical Measurements}

The concept of hierarchical level introduced above allows a natural
and powerful extension of traditional measurements such as the node
degree and clustering coefficient.  This section defines such
features as well as ancillary measurements which can be used in
order to obtain a more complete characterization of complex
networks. The considered measures can be generalized for weighted
networks taking some modifications as described along the measures.
When considering oriented graphs, a new network can be obtained
retrieving only the In or Out connections of each node.

The \emph{hierarchical node degree} of a reference node $i$ at
distance $d$ is henceforth defined as corresponding to the number of
edges extending between the nodes in $R_d(i)$ and $R_{d+1}(i)$. This
measurement is henceforth represented as $k_d(i)$.  As an example,
in Figure~\ref{fig:ex} we have that $k_0(8)=5$ (corresponding to the
traditional node degree) and $k_1(8)=8$. Observe that the
hierarchical node degree is \emph{not} averaged among the number of
nodes in $R_d(i)$.  Actually, this measurement can be understood as
the traditional node degree where the reference node is understood
as corresponding to the ball $B_d(i)$ (i.e. the nodes in this ball
are merged into a subsumed node). This measure can be extended to
weighted networks by taking the sum of the weight values for every
connection between these nodes and the nodes of the next level.

Let the number of edges in the subnetwork $\gamma_d(i)$ be expressed
as $e_d(i)$, and the number of elements of the ring $R_d(i)$ be
represented as $n_d(i)$.  The \emph{hierarchical clustering
coefficient} of node $i$ at distance $d$, hence $cc_d(i)$, can be
obtained in terms of the immediate generalization of
Equation~\ref{eq:cc}

\begin{equation}
  cc_d(i)= 2 \frac{e_d(i)}{n_d(i)(n_d(i)-1)} \label{eq:hcc}.
\end{equation}

For node $i=8$ in the simple network shown in Figure~\ref{fig:ex}
we have that $cc_1(8)=0.3$ and $cc_2(8) \approx 0.19$.

Other interesting hierarchical measurements which can be obtained with
respect to the reference node $i$ and which can be used to diminish
the degeneracy of the node degree and clustering coefficient include
the following:

{\bf Convergence ratio ($C_d(i)$):} Corresponds to the ratio between
the hierarchical node degree of node $i$ at distance $d$ and the
number of nodes in the ring at next level distance, i.e.

\begin{equation}
 C_d(i)=\frac{k_{d}(i)}{n_{d+1}(i)}.
\end{equation}

This measurement quantifies the average number of edges received by
each node in the hierarchical level $d+1$. We have necessarily that
$C_0(i)=1$ for whatever node selected as the reference $i$.  In the
case illustrated in Figure~\ref{fig:ex}, we have $C_0(8)=1$ and
$C_1(8)=8/7$, indicating a low level of edge convergence into the
nodes in $R_d(i)$.

{\bf Intra-ring degree ($A_d(i)$):} This measurement is obtained by
taking the average among the degrees of the nodes in the subnetwork
$\gamma_d(i)$.  Observe that only those edges between the nodes in
such a subnetwork are considered, therefore overlooking the
connections established by such nodes with the nodes in the
hierarchical levels at $d-1$ and $d+1$. For instance, we have for
the situation in Figure~\ref{fig:ex} that $A_1(8)=6/5$ and
$A_2(8)=8/7$. For weighted networks the value of intra-ring is the
average of weights of all nodes in such subnetwork.

{\bf Inter-ring degree ($E_d(i)$):} The average of the number of
connections between each node in ring $R_d(i)$ and those in
$R_{d+1}(i)$.  For instance, for Figure~\ref{fig:ex} we have
$E_0(8)=5$, $E_1(8)=8/5$ and $E_2(8)=0$.  Observe that
$E_d(i)=k_d(i)/n_d(i)$.

{\bf Hierarchical common degree ($H_d(i)$):} The average node degree
among the nodes in $R_d(i)$, considering all edges in the original
network.  For Figure~\ref{fig:ex} we have $H_1(8)=18/5$ and
$H_2(8)=16/7$.  The hierarchical common degree expresses the average
node degree at each hierarchical level, indicating how the network
node degrees are distributed along the network hierarchies.

It is also interesting to eventually consider versions of the above
described measurements considering the \emph{ball} $B_d(i)$, and not
the ring $R_d(i)$. Table~\ref{tab:meas} summarizes the hierarchical
measurements reviewed/introduced in the current article, all of which
are defined with respect to one of the network nodes, identified by
$i$, taken as a reference and at a distance $d$ from that node.
Observe that most measurements are averaged among the number of nodes
in $R_d(i)$, except the first three features in Table~\ref{tab:meas}.

\begin{table}
  \begin{center}
  \vspace{1cm}
  \begin{tabular}{||l|r||}  \hline
    $e_d(i)$    &  hier. number of edges among the nodes \\
                &  in the ring $R_d(i)$ \\  \hline
    $n_d(i)$    &  hier. number of nodes in the ring $R_d(i)$ \\   \hline
    $k_d(i)$    &  hierarchical degree of node \\
                &  $i$ at distance $d$ \\   \hline
    $cc_d(i)$   &  hier. clustering coefficient of node \\
                &  $i$ at distance $d$ \\   \hline
    $C_d(i)$    &  convergence rate at \\
                &  hierarchical level $d$  \\  \hline
    $A_d(i)$    &  intra-ring node degree of node \\
                & $i$ at distance $d$ \\  \hline
    $E_d(i)$    &  inter-ring node degree of node \\
                & $i$ at distance $d$  \\ \hline
    $H_d(i)$    &  hierarchical common degree of node \\
                & $i$ at distance $d$  \\ \hline
  \end{tabular}
  \caption{The hierarchical measurements considered in the current
  article.~\label{tab:meas}}
  \end{center}
\end{table}

\section{Edge Degree and Edge Clustering Coefficient}

One important thing about the traditional node degree and clustering
coefficient is that these concepts have been defined with respect to a
network \emph{node} and its immediate neighbors.  It would be
interesting to extend such concepts with respect to network
\emph{edges}.  The generalization of the node degree and clustering
coefficient to any subset of nodes in a complex network reported
in~\cite{Generalized} provides an immediate means to obtain the above
extensions.

Such a generalization can be immediately obtained by considering more
general vectors $\vec{v}(i)$ in the equations in the previous two
sections.  More specifically, instead of assigning the value one only
to the vector element whose index corresponds to the label of the
reference node, we assign ones to the elements whose indices
correspond to the labels of \emph{all} nodes in the subnetwork of
interest.  For instance, in case we define the subnetwork $\gamma$ as
including the nodes $\{1, 11 \}$ and respective edges in the network
in Figure~\ref{fig:ex}, we have $\vec{v}(\gamma) = (1, 0, 0, 0,
0, 0, 0, 0, 0, 1, 0, 0)^T$. Let us obtain the ring centered at
$\gamma$ at distance $2$.  By applying Equation~\ref{eq:vd} we have

\begin{equation}
 \vec{v}_1(\gamma)=(0, 1, 1, 1, 0, 1, 0, 0, 0, 0, 0, 11, 11)^T  \nonumber
\end{equation}

 and

\begin{equation}
 \vec{v}_2(\gamma)=(4, 1, 1, 0, 1, 0, 12, 12, 11, 11, 22, 11, 11)^T   \nonumber
\end{equation}

and, through Equation~\ref{eq:pd}, we obtain

\begin{equation}
  \vec{p}_1(\gamma)=(1, 1, 1, 1, 0, 1, 0, 0, 0, 0, 1, 1, 1)^T   \nonumber
\end{equation}

  and

\begin{equation}
  \vec{p}_2(\gamma)=(1, 1, 1, 1, 1, 1, 1, 1, 1, 1, 1, 1, 1)^T  \nonumber.
\end{equation}

The vector specifying the ring centered at $\gamma$ at distance $d=2$
is now obtained by using Equation~\ref{eq:rd} as
$\vec{r}_2(\gamma)=\vec{p}_2-\vec{p}_1 = (0, 0, 0, 0, 1, 0, 1, 1, 1,
1, 0, 0, 0)^T$, from which we finally obtain $R_2(\gamma) = \{ 5, 7,
8, 9, 10\}$.

The extension of the hierarchical node degree and hierarchical
clustering coefficient to an edge (instead of a node) can now be
easily obtained by first identifying the two nodes $i$ and $j$
defining the edge of interest and making the nodes in $\gamma$ to
correspond to those two nodes.  The hierarchical node degree and
hierarchical clustering coefficient can be obtained by using immediate
extensions of their respective definitions.

\section{Analytical Results for Random Networks}

This section presents a mean-field analytical investigation of the
typical values and behavior of the main measurements
reviewed/introduced in the previous sections of this work.

\begin{figure*}
 \begin{center}
   \includegraphics[scale=1.2,angle=-90]{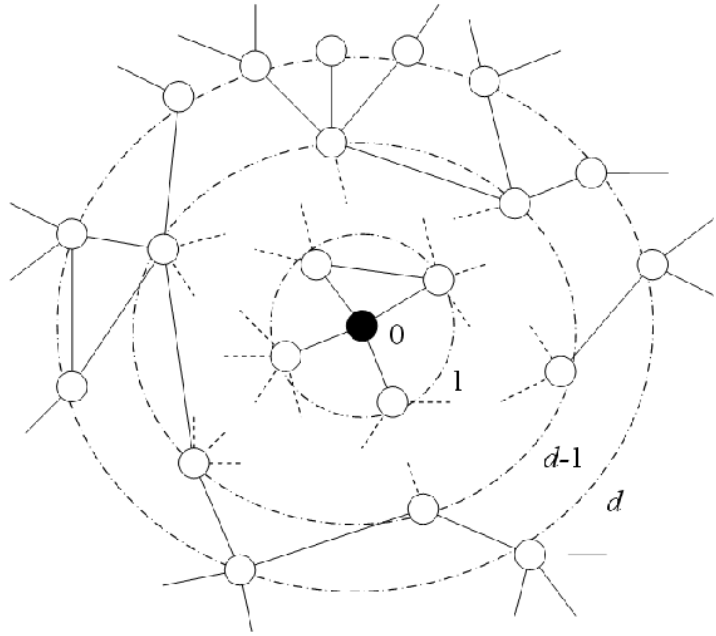} \\
   \caption{A generic situation in a complex network involving a
   reference node $i$ (in black) and the respectively defined
   hierarchical levels.~\label{fig:virtual}}
  \end{center}
\end{figure*}

Consider the generic situation depicted in Figure~\ref{fig:virtual},
including a reference node $i$ and the several respectively defined
hierarchical levels, extending from 0 (corresponding to the reference
node) to $d$, and further.  Recall that the subnetwork $\gamma_d(i)$
is the subgraph obtained by considering the $n_d(i)$ nodes at level
$d$ (i.e. the ring $R_d(i)$) and the $e_d(i)$ edges among those nodes.
It can be shown that the following mean-field recursive approximation
holds for a random network with overall mean degree $\left< k \right>$

\begin{eqnarray}
  \left\{  \begin{array}{l}
     n_d(i) \approx \eta(k_{d-1},N-N_{d-1}) \\
     N_d(i) \approx N_d(i) + n_d(i) \\
     k_d(i) \approx \left( \frac{N - C_d(i)}{N} \right) \left( \sum_{j \in R_d(i)} k_j \right) n_d(i)
  \end{array}
  \right.
\end{eqnarray}

where $N_d(i)$ is the cumulative number of nodes from the hierarchical
level 0 up to level $d$ (inclusive), i.e. $N_d = \sum_{j=0}^{d}
n_d(i)$, and the function $\eta(a,b)$ gives the average number of
manners $b$ objects can be taken, with repetition, to fill $a$ slots.
Now, the average and variance of the hierarchical node degree of node
$i$ at distance $d$ can be respectively approximated as

\begin{eqnarray}
  E \left[ k_d(i) \right] \approx \left( \frac{N - N_d(i)}{N} \right) \left<
  k \right> n_d(i) \\
  Var \left\{ k_d(i) \right\} \approx \left( \frac{N - N_d(i)}{N} \right)^2 \left<
  k \right> n_d(i)^2 \label{eq:analyt}
\end{eqnarray}

\begin{figure}
   \resizebox{8cm}{2cm}{\includegraphics[]{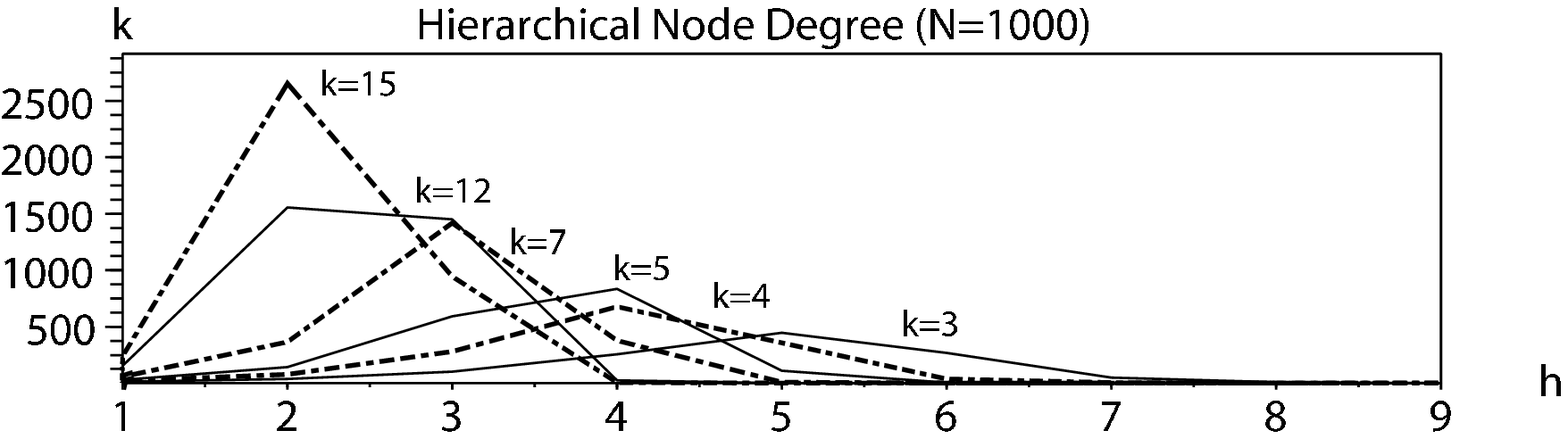}}
   \resizebox{8cm}{2cm}{\includegraphics[]{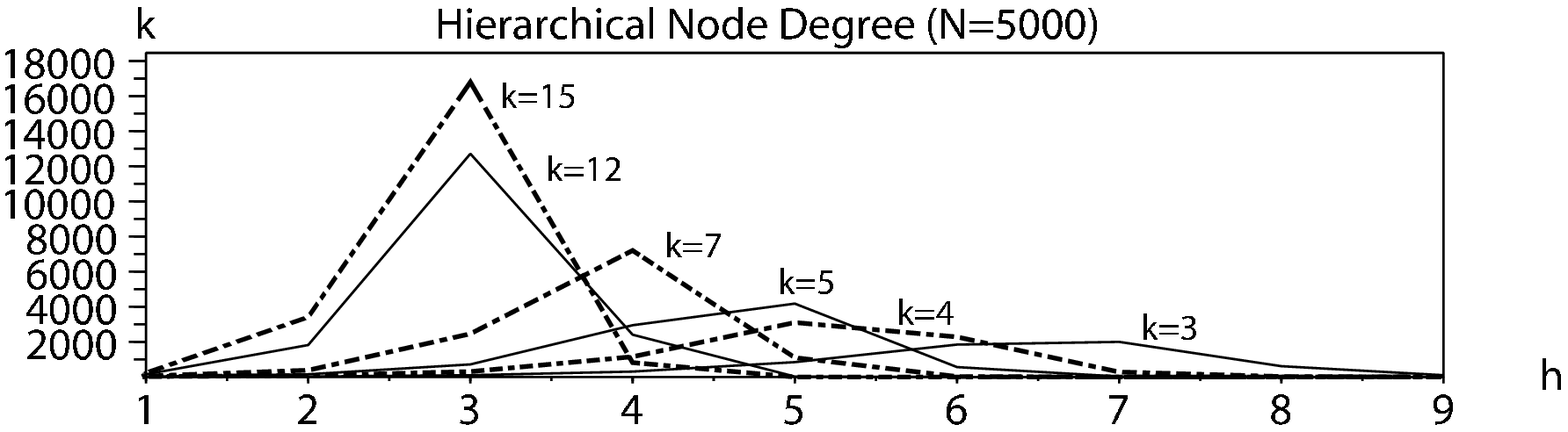}}
   \resizebox{8cm}{2cm}{\includegraphics[]{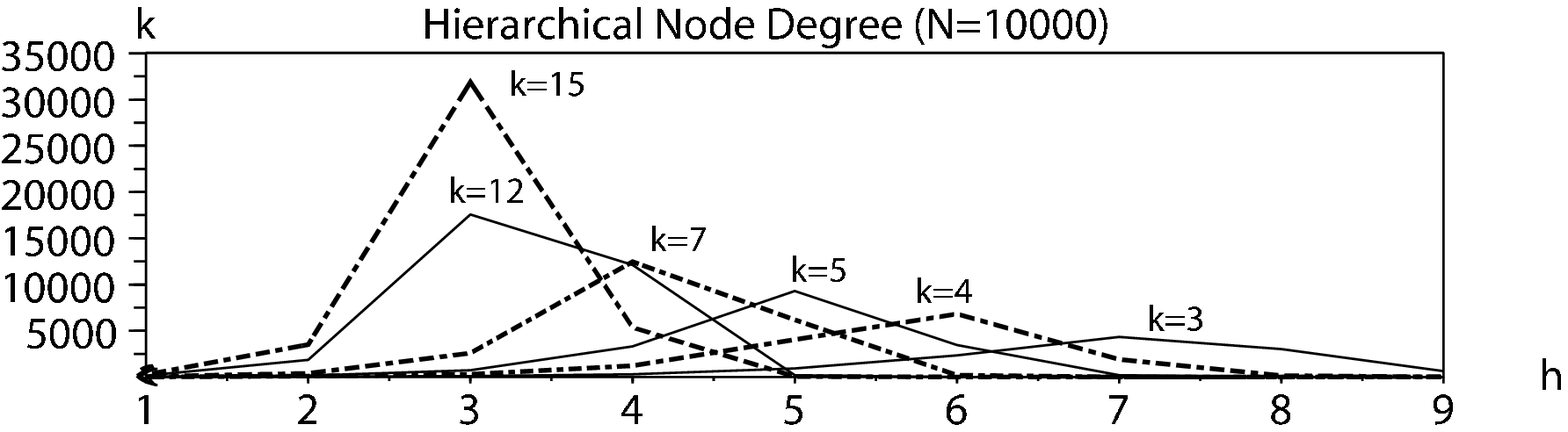}}
   \caption{The hierarchical node degree for several configurations of $\left< k \right>$ and
   $N$.  Observe that such curves are always characterized by a peak,
   which is a consequence if the finite size of the considered
   networks.  Observe also that increased connectivity, implied by
   larger values of $\left< k \right>$ tends to reduce the number of
   hierarchical levels in the network.~\label{fig:profils}}
\end{figure}

Figures~\ref{fig:profils}(a-i) show the hierarchical node degree for
several combinations of $\left< k \right>$ and $N$.  It is clear from
this figure that the hierarchical node degree curves are approximately
symmetric with respect to the abscissa $P$ of the respective peak
value, which is a consequence of the finite size of the considered
networks.  Actually, the following three situations can be identified
during the dynamic evolution of the hierarchical node degree for a
specific network node: (i) the hierarchical node degree increases as
more nodes imply links to more nodes; (ii) a peak is achieved with
abscissa $P$; and (iii) the node degree decreases because of the
finite size of the network, which implies the `saturation' of the
hierarchical expansion.  Observe also that higher connectivity,
implied by large values of $\left< k \right>$, tends to reduce the
value of $P$ and, consequently, the hierarchical levels of the
networks.  Such an effect is usually accompanied by an increase of the
heights of the respective curves, in order to conserve the average
node degree.  As a matter of fact, it can be shown that also important
is the fact that the standard deviation tends to increase with the
values of the hierarchical node degree.

\begin{figure*}
 \begin{center}
   \includegraphics[scale=1.5,angle=-90]{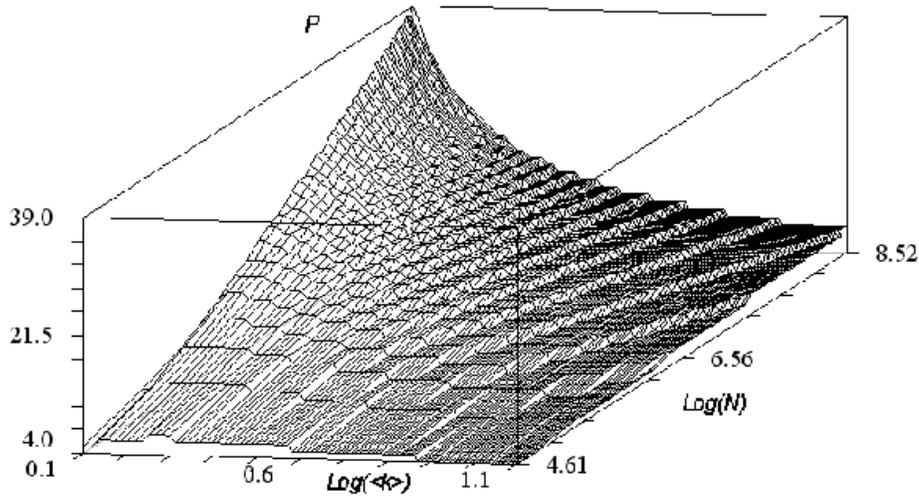} \\
   \caption{The values of the abscissa of the peak hierarchical node
   degree for several values of $Log(\left< k \right>)$ and $Log(N)$.
   ~\label{fig:peak}}
  \end{center}
\end{figure*}

Figure~\ref{fig:peak} shows the values of $P$, obtained by
simulations using the Equation~\ref{eq:analyt}, for several values
of $\left< k \right>$ and $N$.  Observe that, for a fixed average
degree $\left< k \right>$, we have that $P \approx c Log(N)$, for
some real constant $c$.  It is clear from Figure~\ref{fig:peak} that
the hierarchical levels are much more speedily reduced with the
increase of $\left< k \right>$ than with the reduction of $N$, an
effect which can also be appreciated from Figure~\ref{fig:profils}.

\section{Characterization of Complex Networks Models}

In order to further illustrate the potential of the hierarchical
measurements discussed so far in this work, they have been used to
characterize, through simulations, random, scale-free
(i.e. Barab\'asi-Albert -- BA) and regular network models.

The random networks are generated by selecting edges with uniform
probability $p$.  The BA networks are produced as described
in~\cite{AB:surv}, i.e. starting with $m0$ randomly interconnected
nodes and adding new nodes with $m$ edges which are attached to the
existing nodes with probability proportional to their respective node
degrees. The considered regular networks are characterized by each
node being connected exactly to 8 other nodes.  Two types of networks
have been studied in this article: one with border effects, where the
nodes at its border have a lesser degree; and another without border
effects, i.e. considering toroidal connections. In both cases, the
nodes are organized into an $L \times L$ array, and each internal node
(i.e. non-border node), specified by its position $(x,y)$ in such an
array, is connected to its 8-neighbors $(x-1,y)$, $(x+1,y)$,
$(x-1,y-1)$, $(x+1,y-1)$, $(x-1,y+1)$, $(x+1,y+1)$, $(x,y-1)$,
$(x,y+1)$.  The random model assumes $\left< k \right>=15$, $\left< k
\right>=5$ and $\left< k \right>=3$, and the BA model considers
$\left< k \right>=16$, $\left< k \right>=6$ and $\left< k \right>=4$.
These two models assume $N~=10000$. In the case of the regular
networks, $N=10000$ (i.e. $L=100$) and $\left< k \right> = 8$. Observe
that the average node degree of the regular network differs from those
for the other two models as an unavoidable consequence of the
intrinsic topology of that network.

The remainder of this section presents the hierarchical measurements
obtained for each of the complex networks types described above. For
the sake of comprehensiveness, three instances of each model were
considered respectively to decreasing average node degree, namely
$k=15$, $5$, and $3$ for Radom Graph Results; $k=16$, $6$, and $4$ for
Barab\'asi-Albert model.

\begin{figure}
   \resizebox{8cm}{2cm}{\includegraphics[]{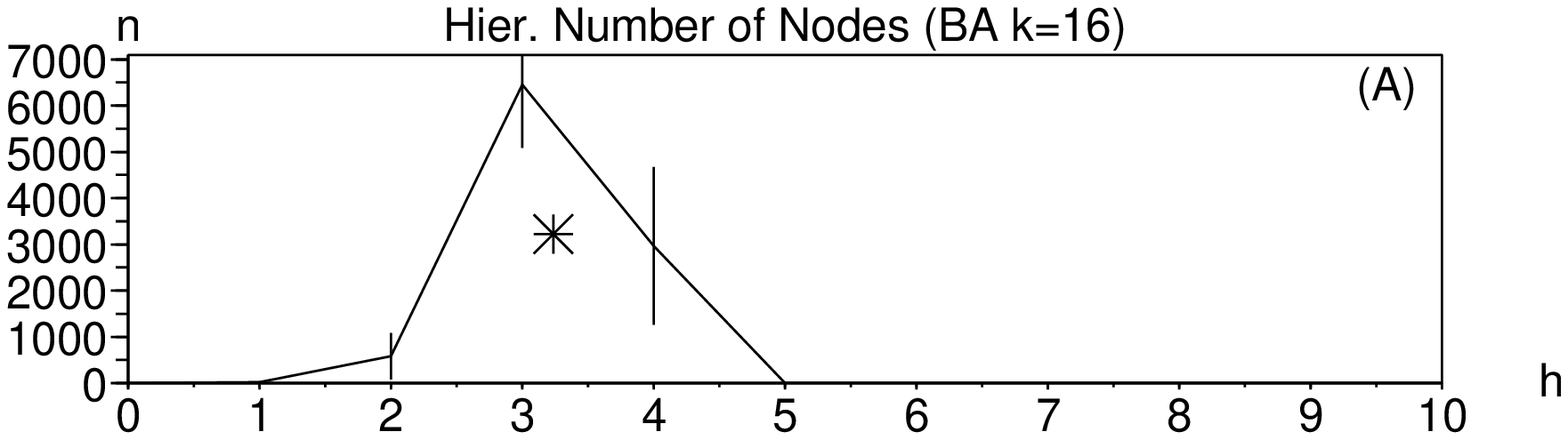}}
   \resizebox{8cm}{2cm}{\includegraphics[]{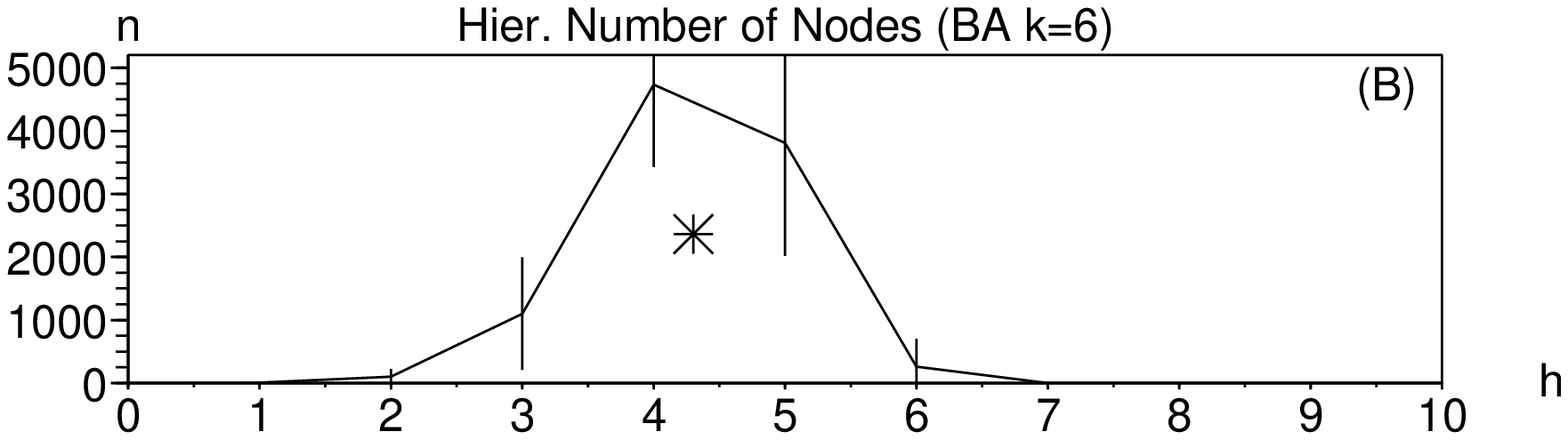}}
   \resizebox{8cm}{2cm}{\includegraphics[]{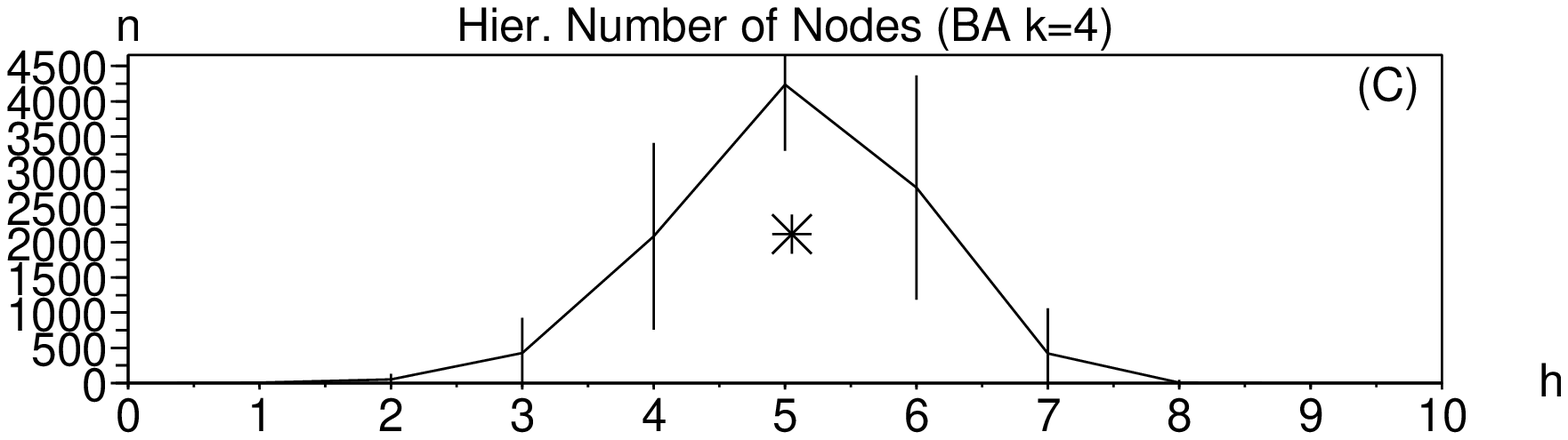}}
   \resizebox{8cm}{2cm}{\includegraphics[]{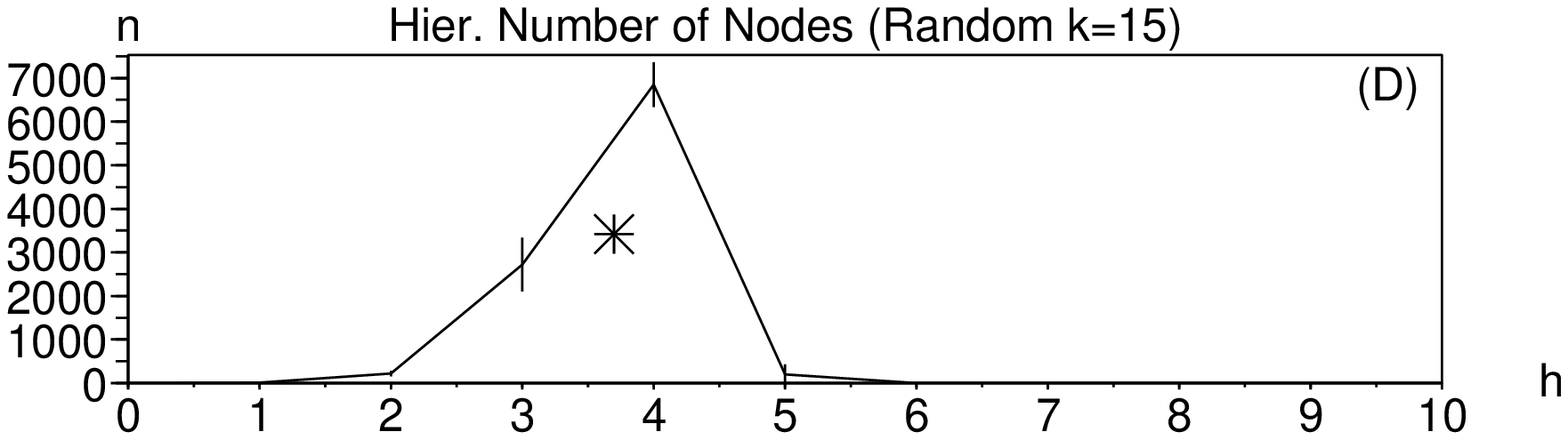}}
   \resizebox{8cm}{2cm}{\includegraphics[]{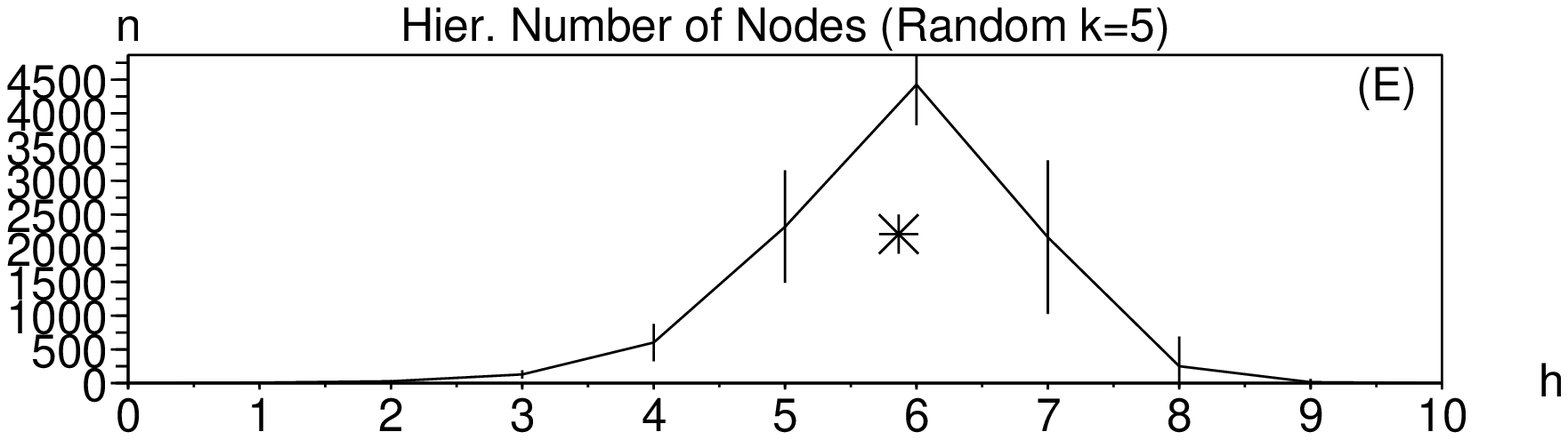}}
   \resizebox{8cm}{2cm}{\includegraphics[]{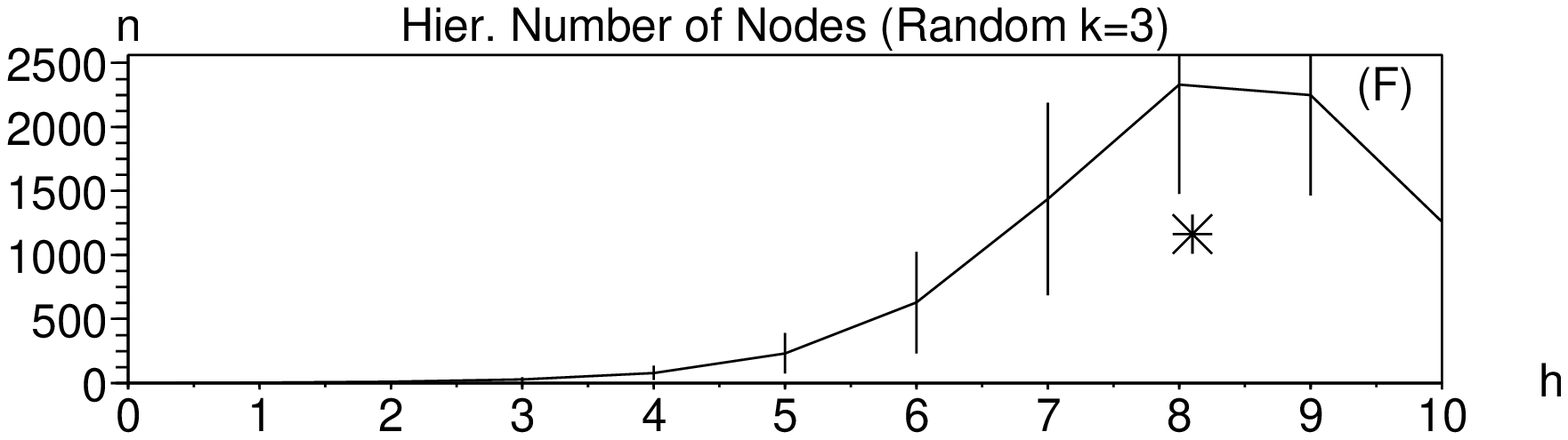}}
   \resizebox{8cm}{2cm}{\includegraphics[]{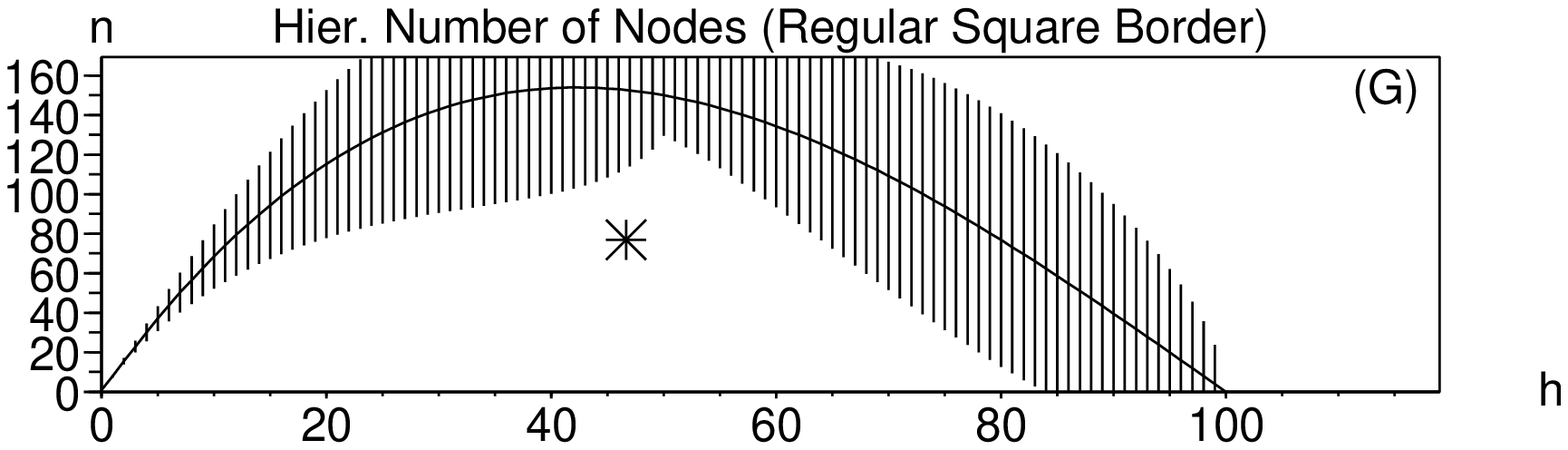}}
   \resizebox{8cm}{2cm}{\includegraphics[]{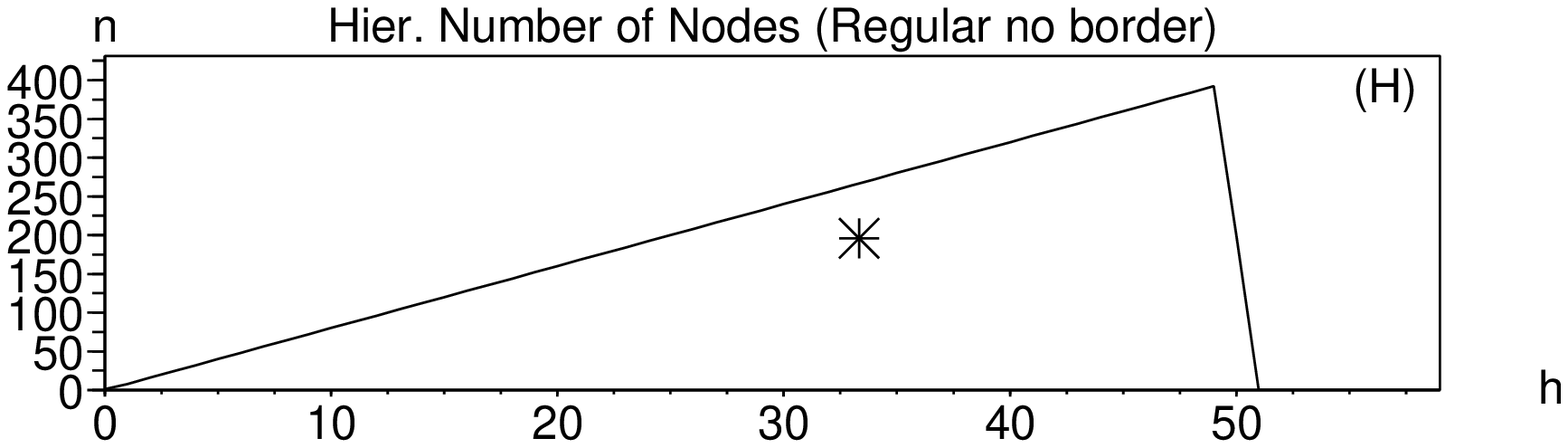}}

   \caption{Hierarchical number of nodes (average $\pm$ standard
   deviation) for all considered networks, which are identified above
   each graph. Observe that most curves are characterized by a
   peak. The average value of the shortes path between any two nodes
   is marked by an asterisk.  ~\label{fig:sim1}}

\end{figure}

Figure~\ref{fig:sim1} shows the hierarchical number of nodes (average
$\pm$ standard deviation) obtained for the considered network models,
including three average degree values in the case of the BA and random
cases, while taking all nodes into account. The asterisks indicate the
position of the average shortest path between any pair of nodes, which
are included in order to provide a reference for the hierarchical
analysis. All curves are characterized by a peak, except for the
regular graph with no border effects. The values of the hierarchical
number of nodes obtained for the random models are more susceptible to
the change of mean degree (i.e. Figs.~\ref{fig:sim1}a-c) than those
values obtained for the Barab\'asi-Albert networks. For a decrease
from $k=16$ to $k=6$, the peak of the Barab\'asi-Albert model shows a
change of only one hierarchical level, while in the Random model,
decreasing from $k=15$ to $k=5$, such a displacement involves four
levels.  For a reduction from $k=5$ to $k=3$ ($k=6$ to $k=4$ for BA),
the change was one level for Barab\'asi-Albert, and 3 levels for the
random models. This is a direct consequence of the fact that
scale-free structures are less susceptible to the removal of random
edges (same as reducing the mean degree) than the random models. Hubs
in BA model establish shortcuts between nodes, reducing the weight of
other edges distances in the average minimal distance. The regular
networks without border effects yielded hierarchical number of nodes
which are linearly increasing, reflecting the basic structure of such
models. However, the regular networks with borders were characterized
by a wide peak and high variance of measurements.  Interestingly, the
peaks obtained for the hierarchical number of nodes occur near the
average shortest path marked by the asterisks. Note that the last
level with a non-zero value corresponds to the graph
diameter~\cite{Bollobas:book}.

\begin{figure}
   \resizebox{8cm}{2cm}{\includegraphics[]{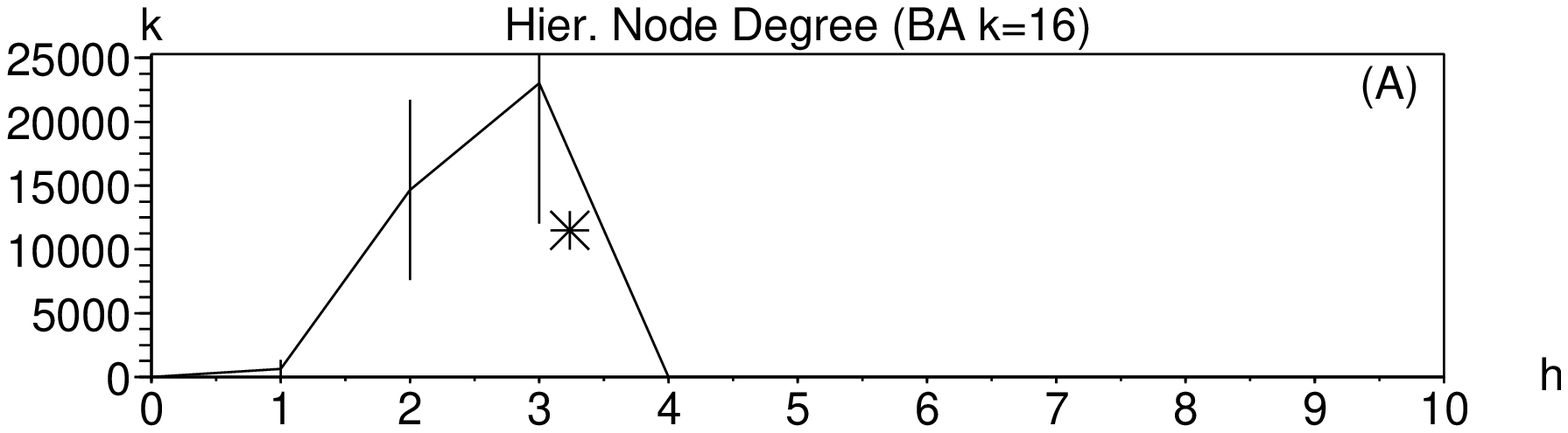}}
   \resizebox{8cm}{2cm}{\includegraphics[]{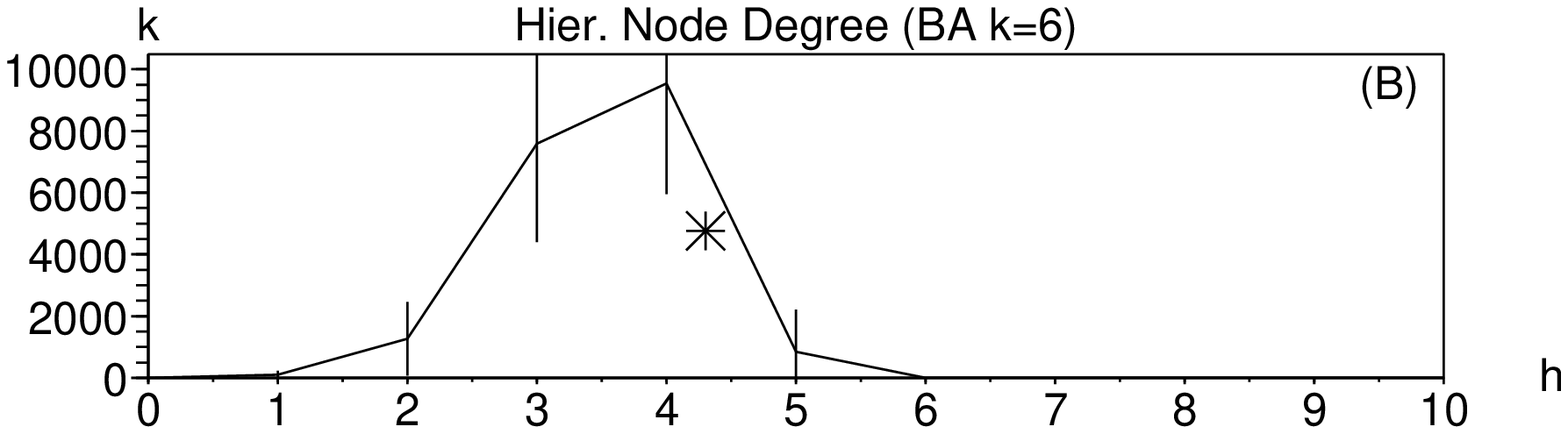}}
   \resizebox{8cm}{2cm}{\includegraphics[]{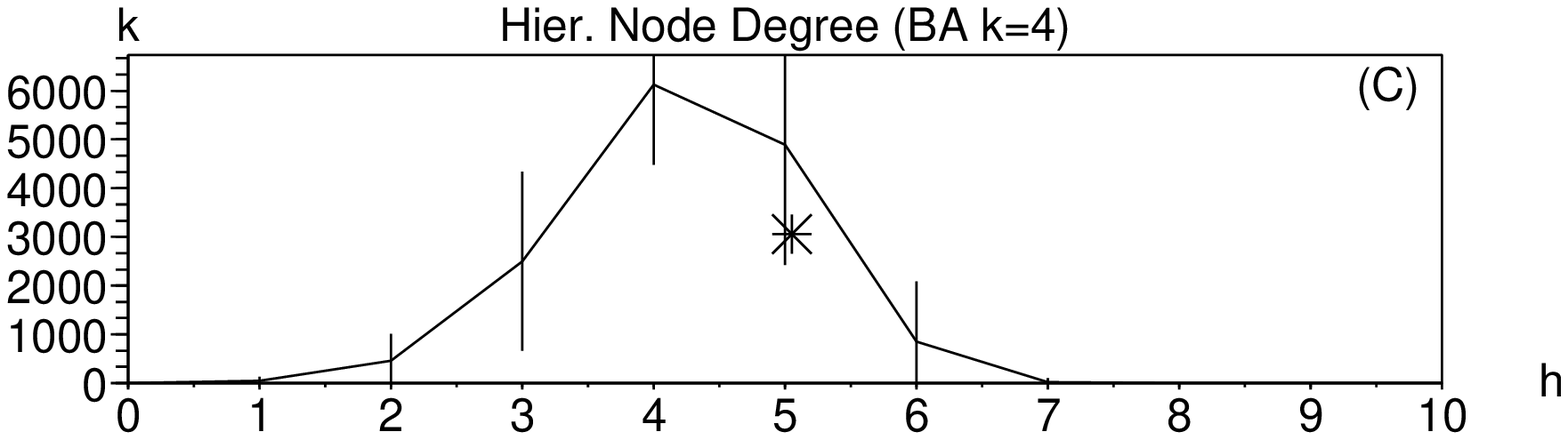}}
   \resizebox{8cm}{2cm}{\includegraphics[]{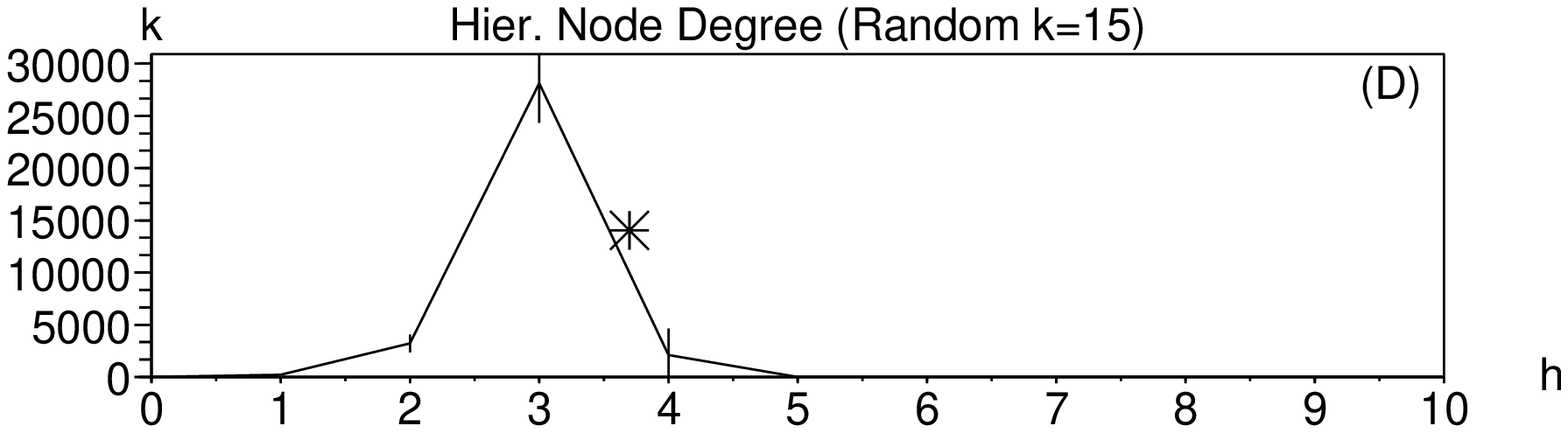}}
   \resizebox{8cm}{2cm}{\includegraphics[]{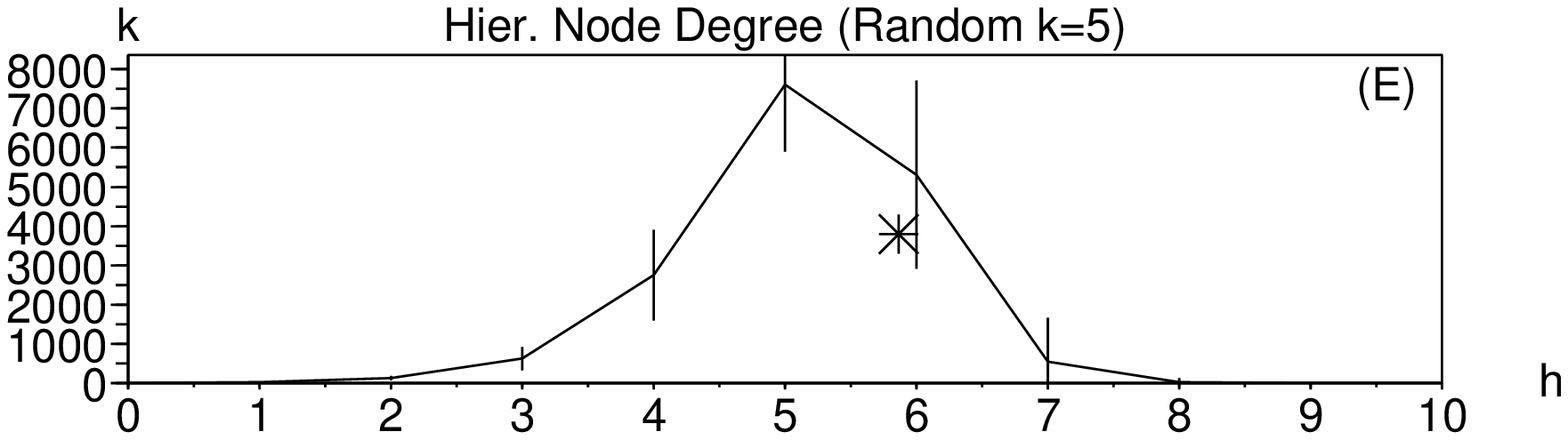}}
   \resizebox{8cm}{2cm}{\includegraphics[]{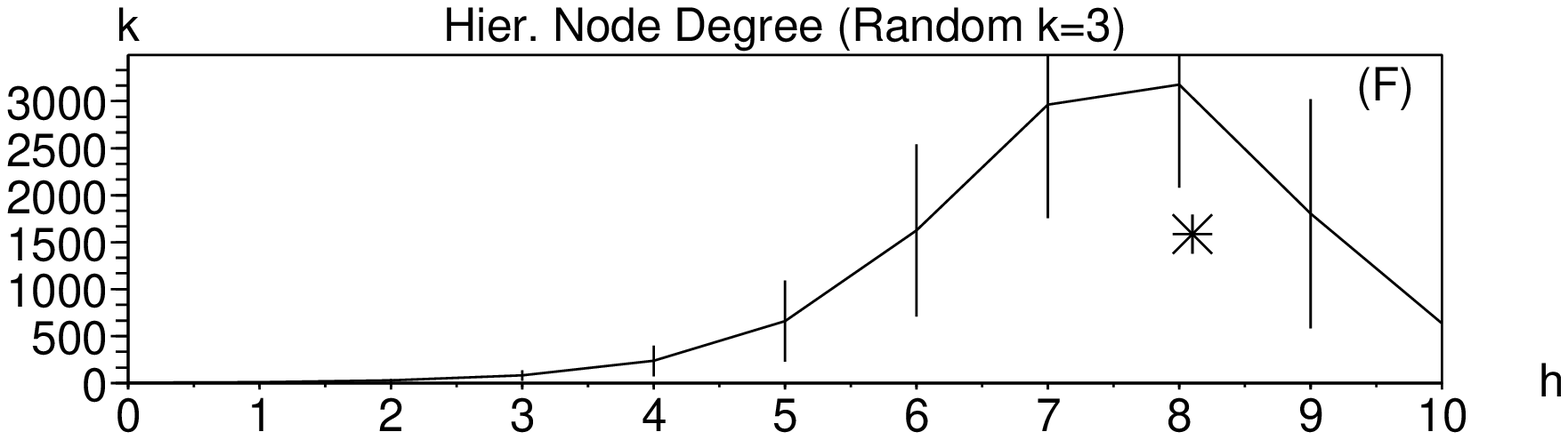}}
   \resizebox{8cm}{2cm}{\includegraphics[]{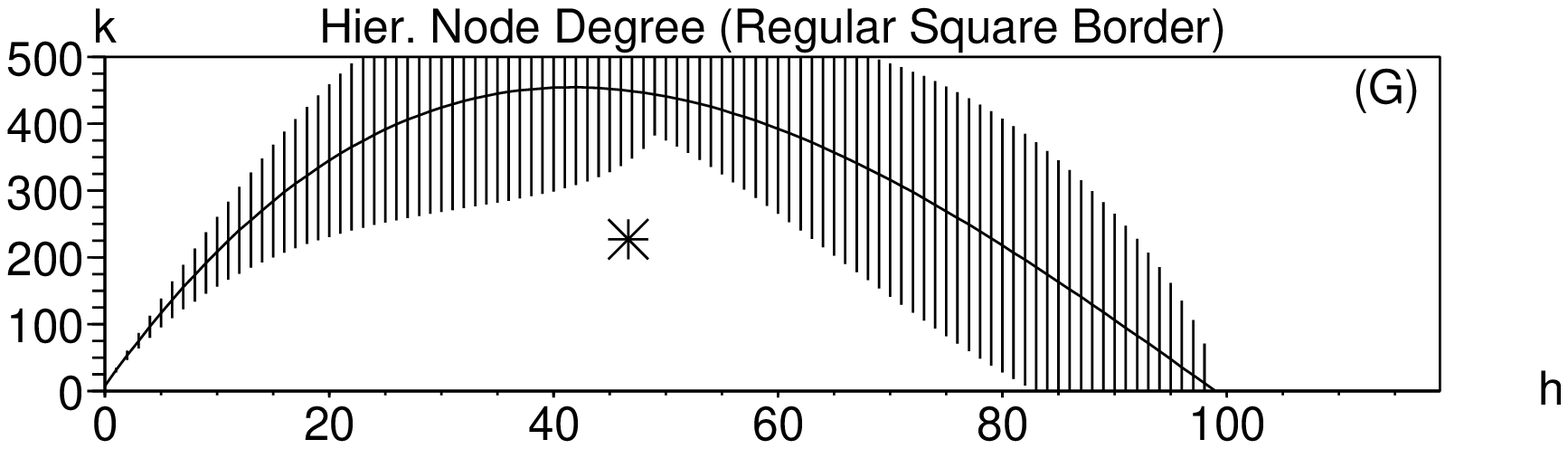}}
   \resizebox{8cm}{2cm}{\includegraphics[]{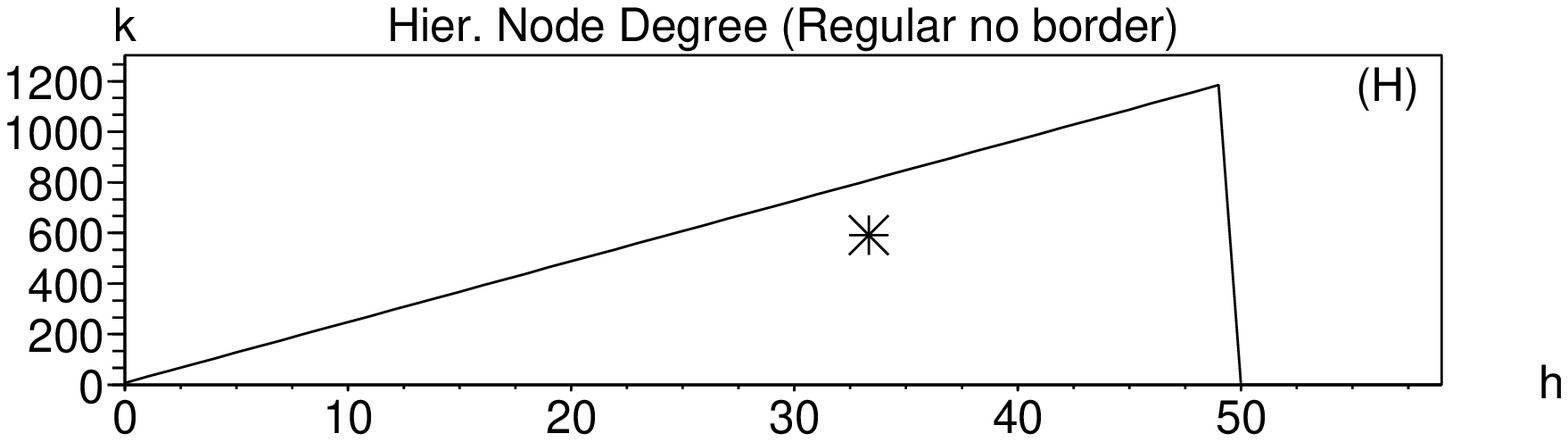}}

   \caption{Hierarchical node degrees obtained for all the considered
   network models. The curves are similar to those obtained for the
   hierarchical number of nodes, except for a expected offset of one
   level.~\label{fig:sim2}}

\end{figure}

The values of hierarchical node degrees, shown in
Figure~\ref{fig:sim2}, are similar to the respective measurements of
hierarchical number of nodes shown in Figure~\ref{fig:sim1}, except
for an expected offset of one hierarchical level to the left.

\begin{figure}
   \resizebox{8cm}{2cm}{\includegraphics[]{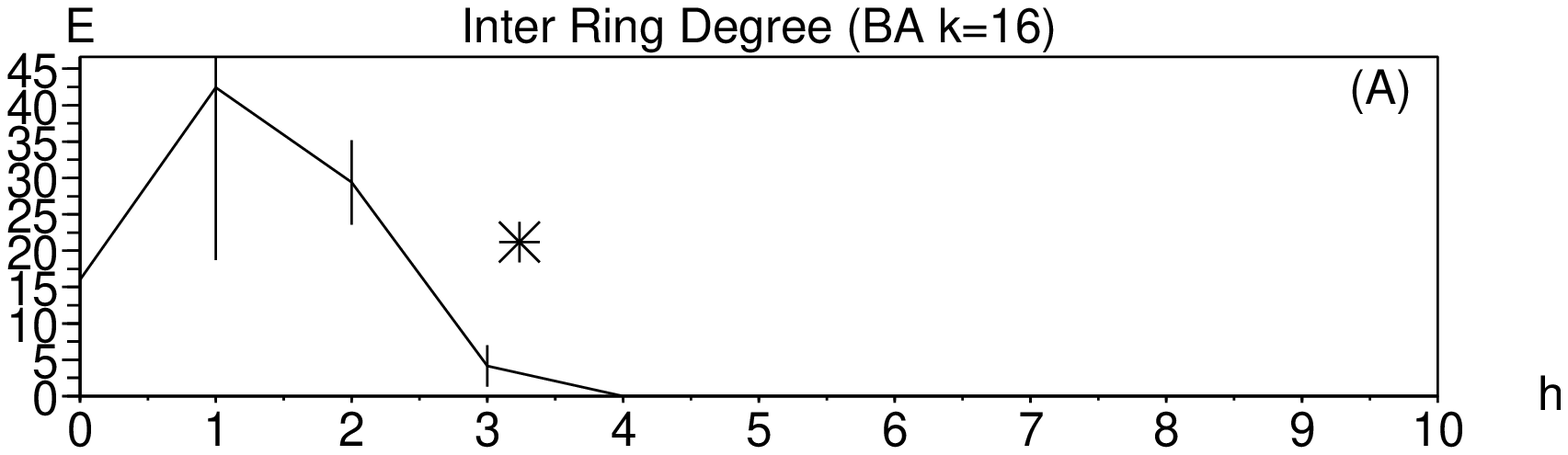}}
   \resizebox{8cm}{2cm}{\includegraphics[]{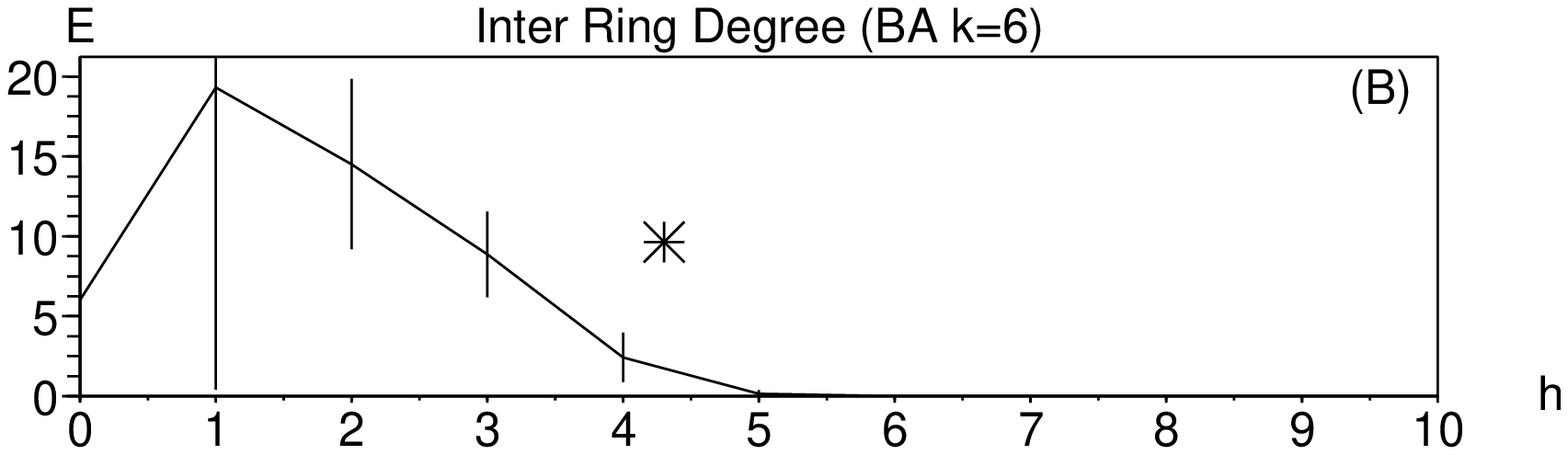}}
   \resizebox{8cm}{2cm}{\includegraphics[]{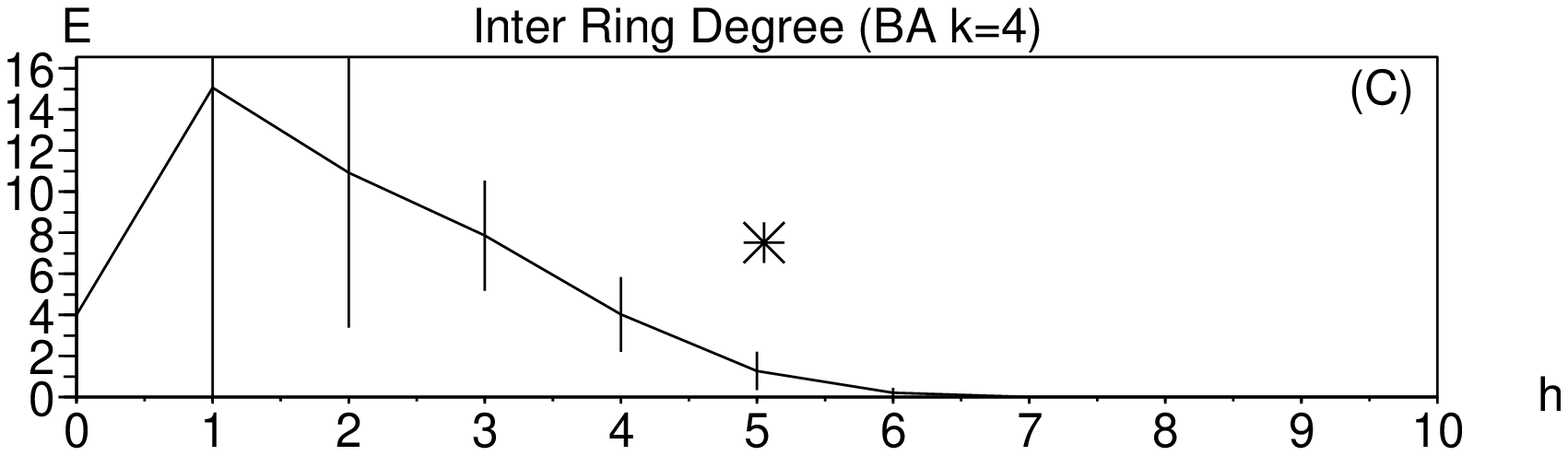}}
   \resizebox{8cm}{2cm}{\includegraphics[]{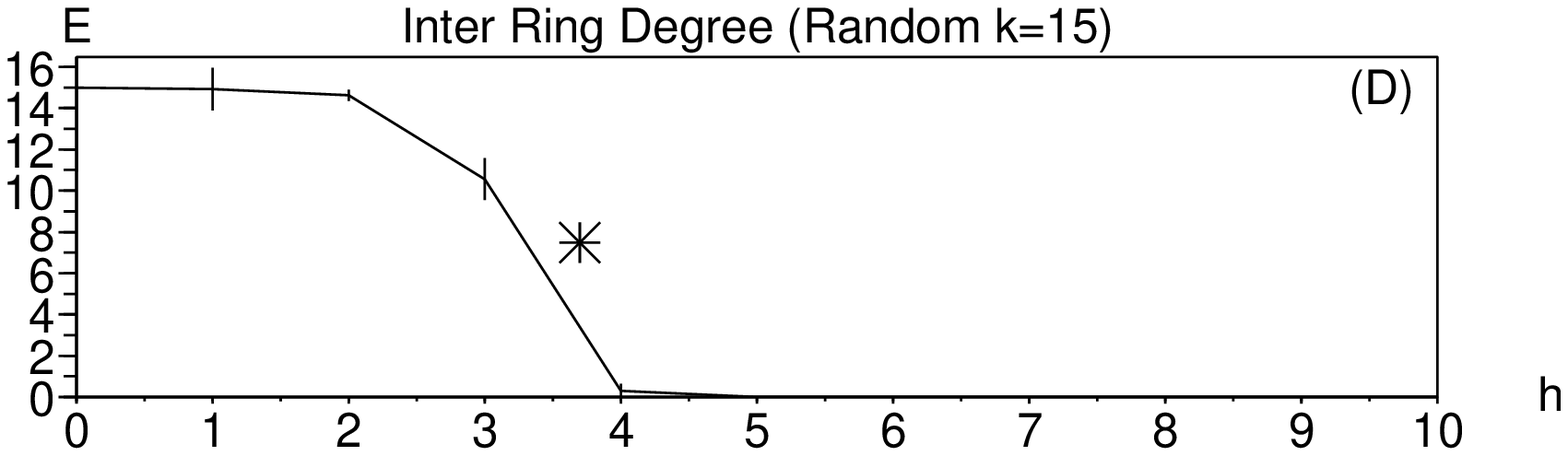}}
   \resizebox{8cm}{2cm}{\includegraphics[]{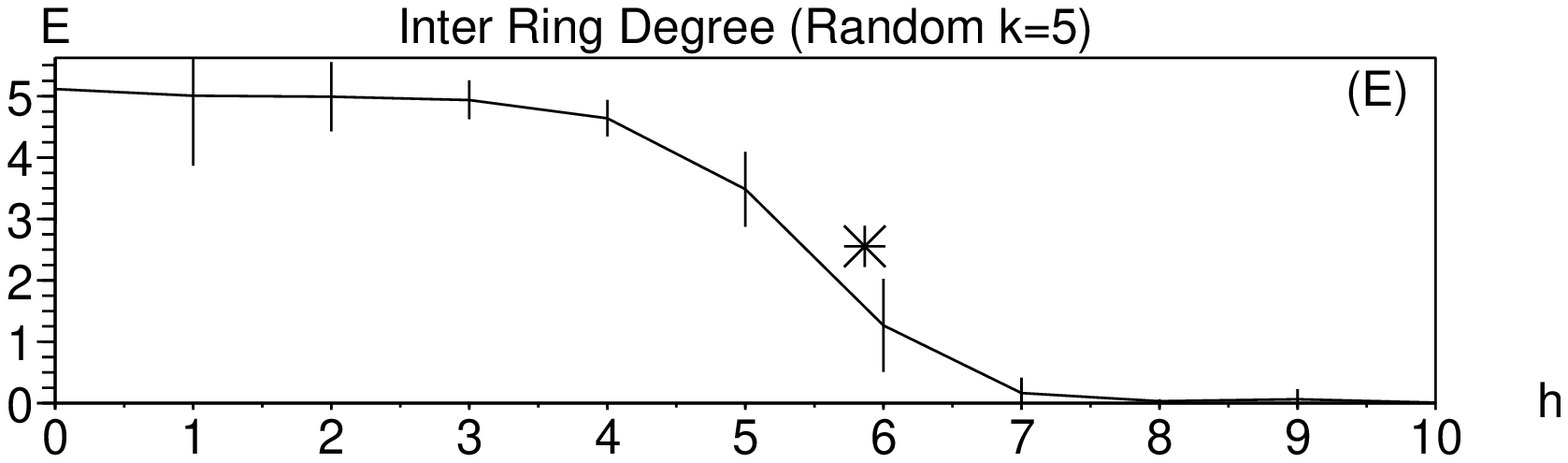}}
   \resizebox{8cm}{2cm}{\includegraphics[]{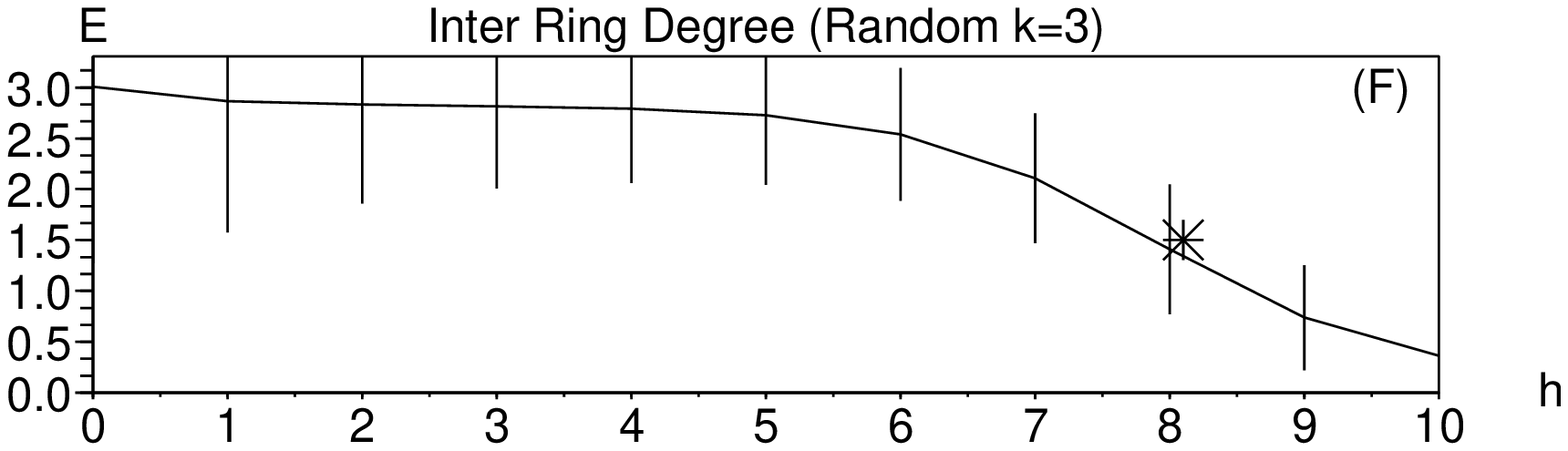}}
   \resizebox{8cm}{2cm}{\includegraphics[]{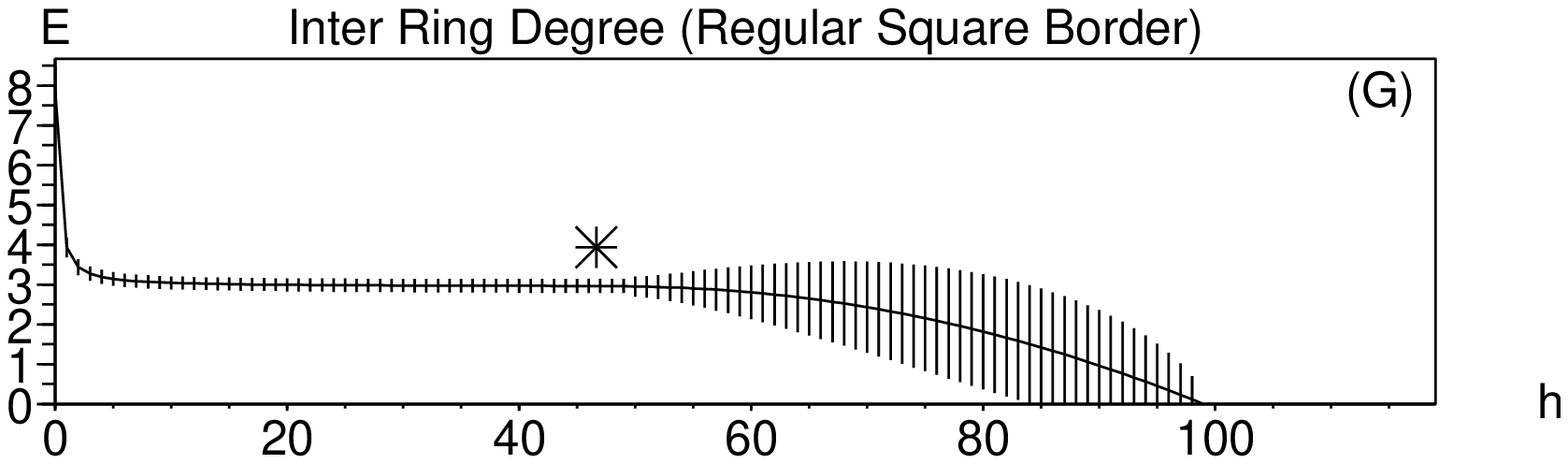}}
   \resizebox{8cm}{2cm}{\includegraphics[]{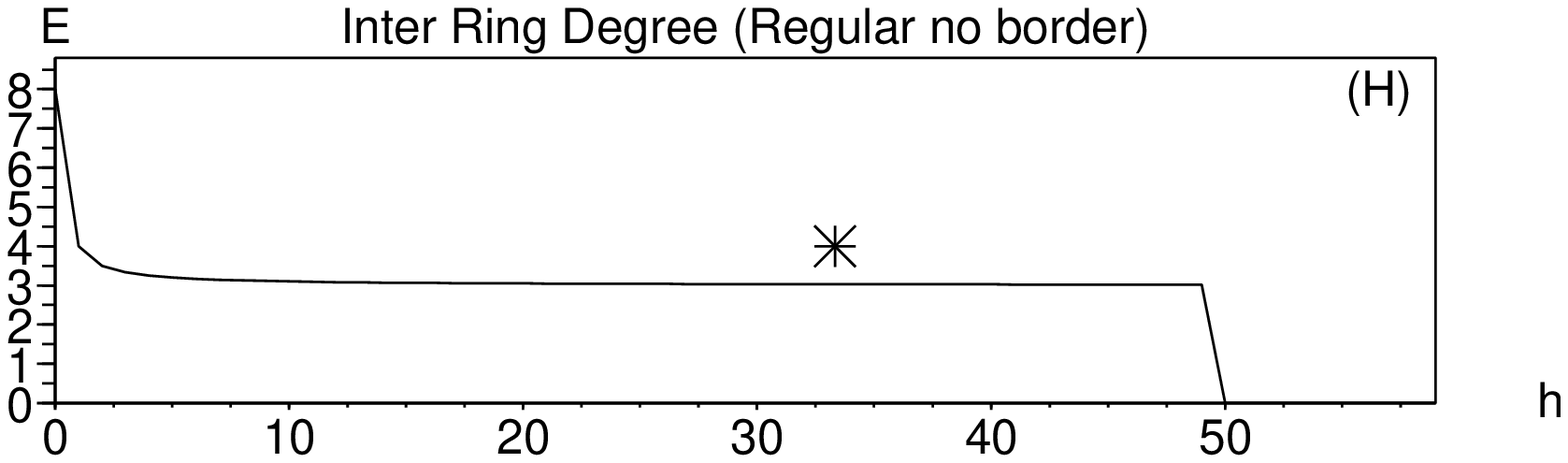}}
     \caption{Inter ring degree values for the considered
     network models.~\label{fig:sim3}}
\end{figure}

All curves obtained for the inter-ring degree, shown in
Figure~\ref{fig:sim3}, are monotonically decreasing after the first
hierarchical level. Again, the curves obtained for the random networks
are less sensitive to variations of the average degree than those
obtained for the Barab\'asi-Albert model. The results for the random
netwoks show wider and smoother curves, while those obtained for
Barab\'asi-Albert tend to be sharper and to concentrate on the left
hand side, implying smaller peaks abscissae which are identical for
the three considered average degrees. Results obtained for the
Barab\'asi-Albert cases also show a peak at the first hierarchical
level and present high variance, this is a consequence of the high
chance of finding a hub on that level. All models, except for the
regular cases, are characterized by presenting the peak of the curve
to the left of the asterisk (i.e. the average shortest path). It is
also interesting to observe that although this measurement is closely
relate to the hierarchical degree, the curves obtained for these two
features (i.e. Figures~\ref{fig:sim2} and~\ref{fig:sim3}) are markedly
different, in the sense that the curves of the hierarchical inter ring
degree obtained for the random model no longer presents the peak
structure as observed in Figure~\ref{fig:sim2}.  The curves obtained
for the regular networks are also interesting, being characterized by
an initial stage of steep decay followed by a plateau which tends to
decrease for higher hierarchical levels in the case of the regular
network with borders.

\begin{figure}
   \resizebox{8cm}{2cm}{\includegraphics[]{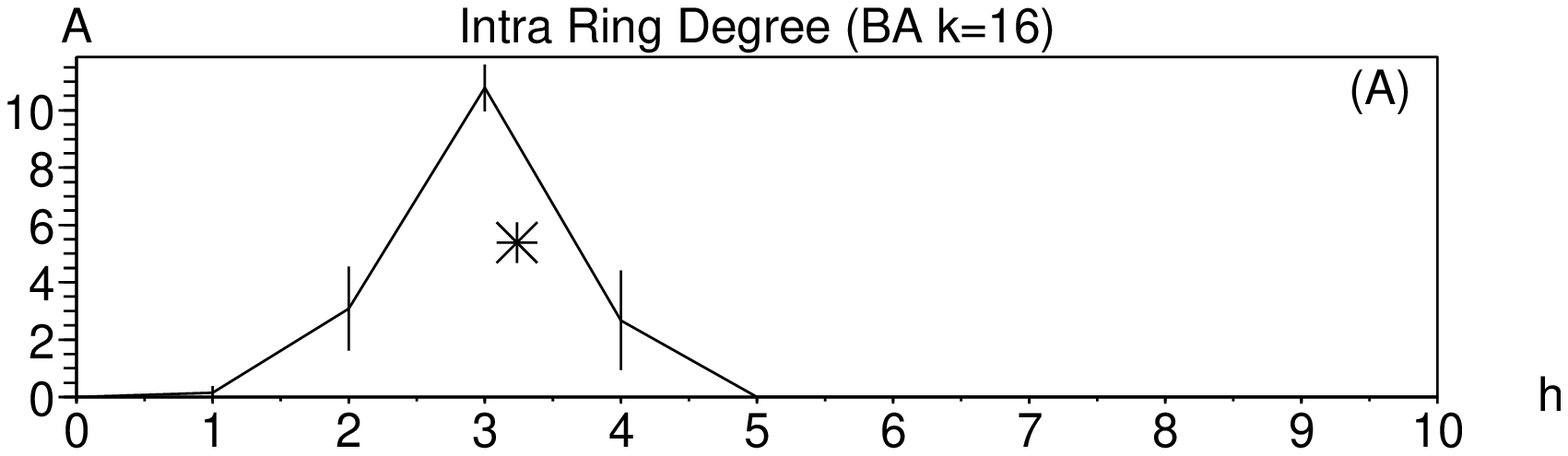}}
   \resizebox{8cm}{2cm}{\includegraphics[]{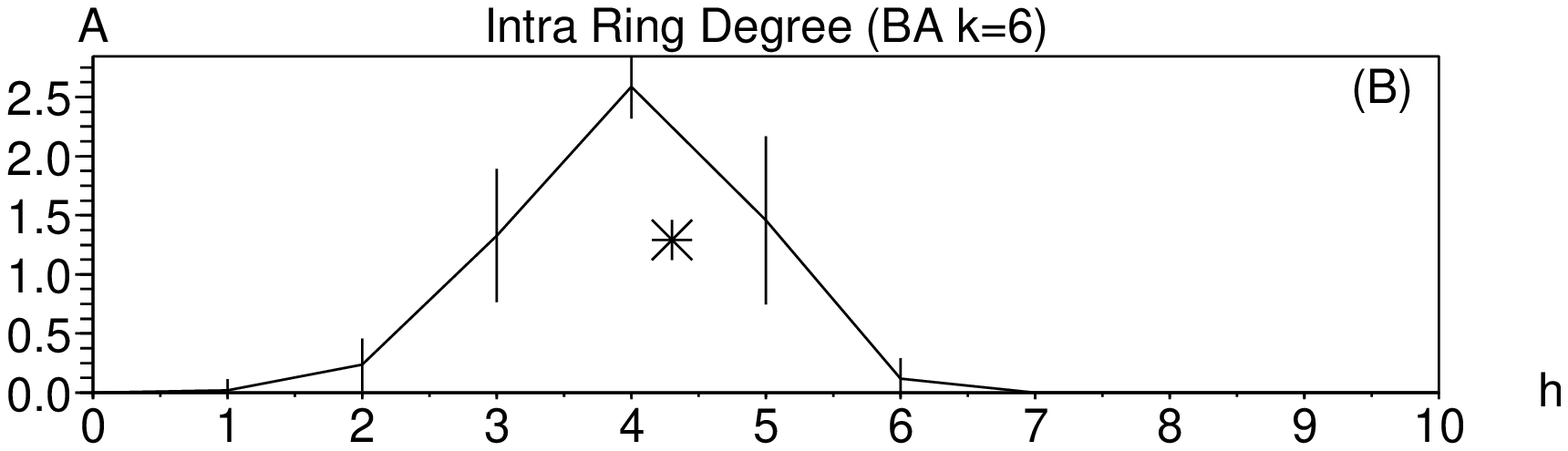}}
   \resizebox{8cm}{2cm}{\includegraphics[]{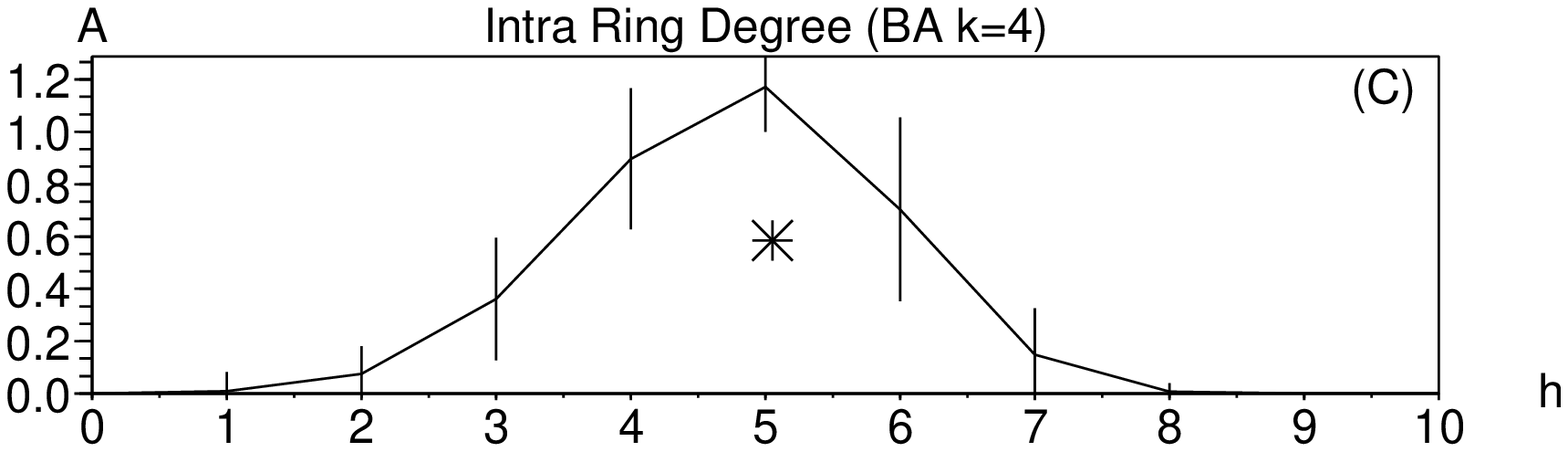}}
   \resizebox{8cm}{2cm}{\includegraphics[]{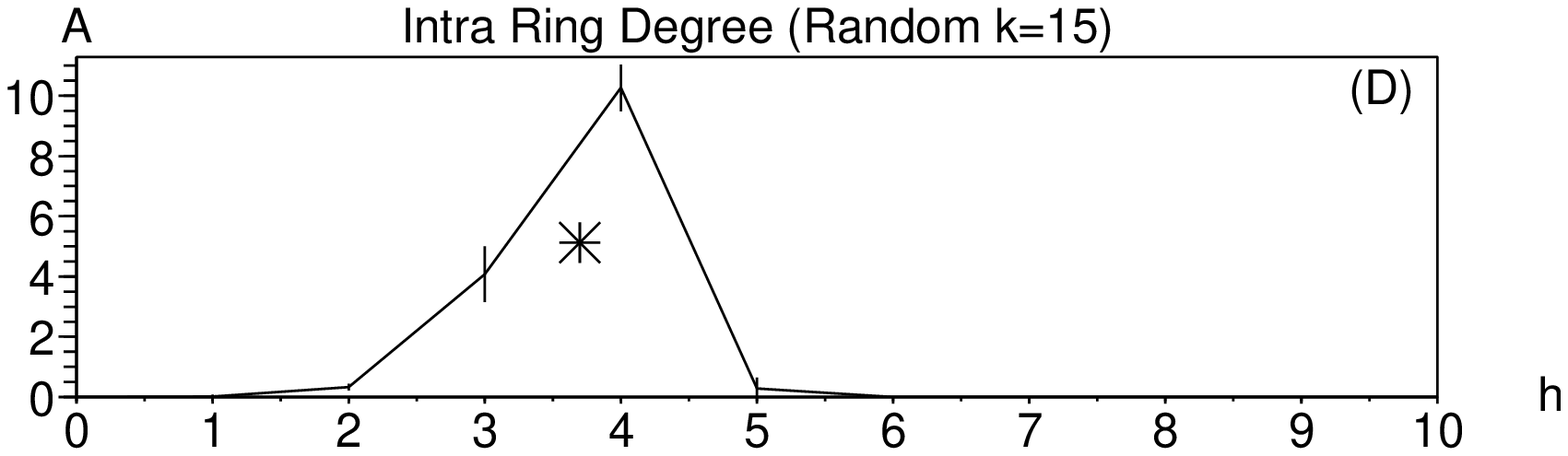}}
   \resizebox{8cm}{2cm}{\includegraphics[]{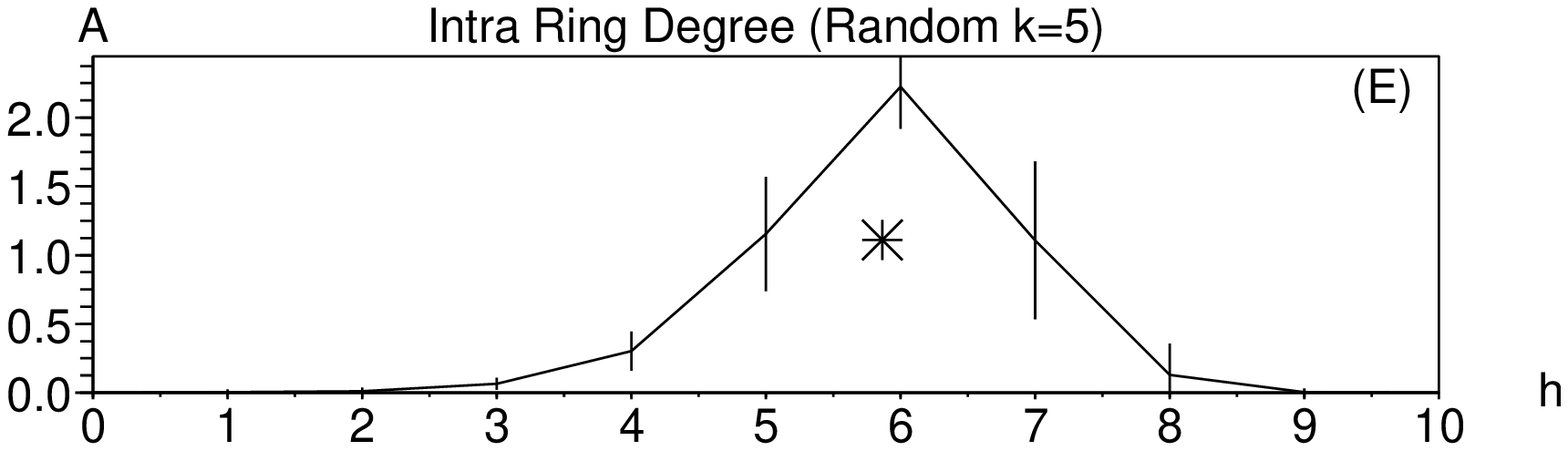}}
   \resizebox{8cm}{2cm}{\includegraphics[]{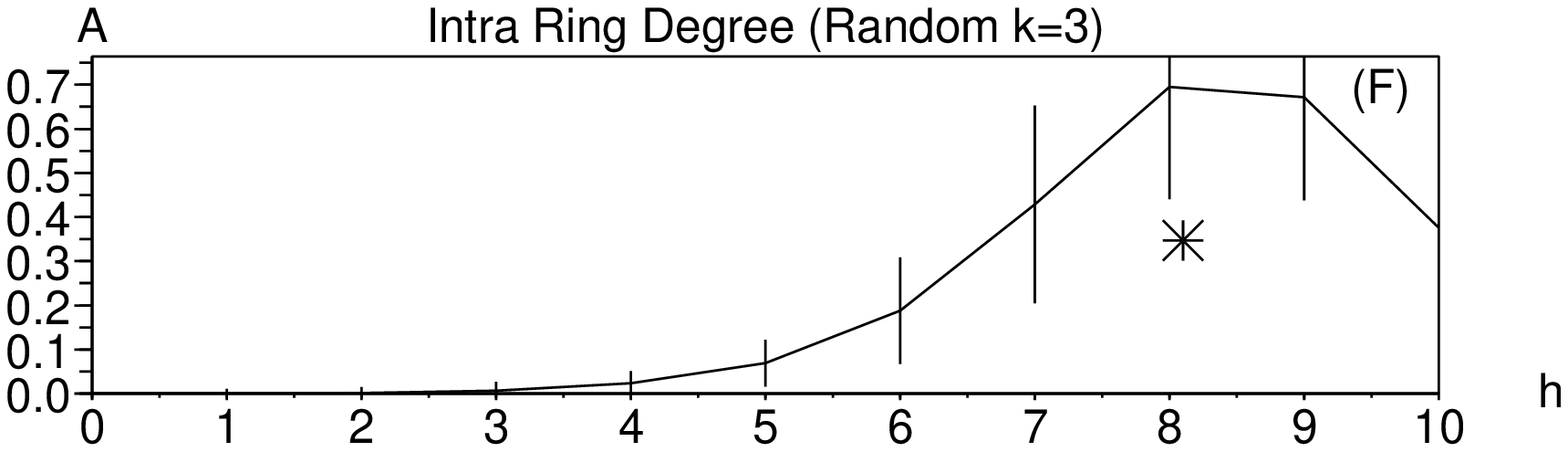}}
   \resizebox{8cm}{2cm}{\includegraphics[]{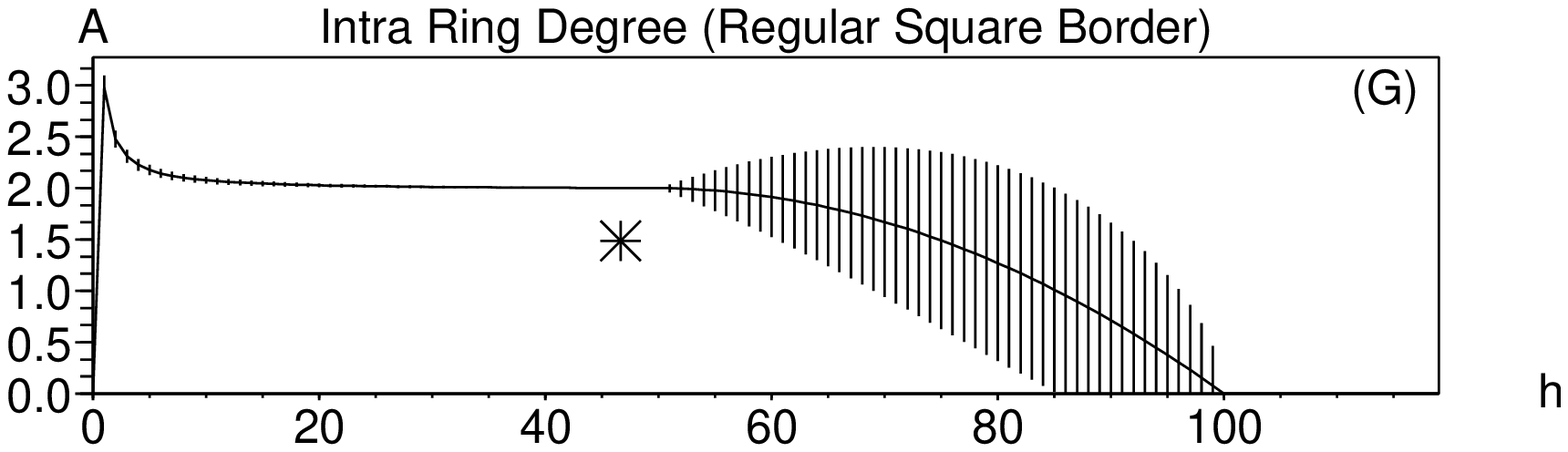}}
   \resizebox{8cm}{2cm}{\includegraphics[]{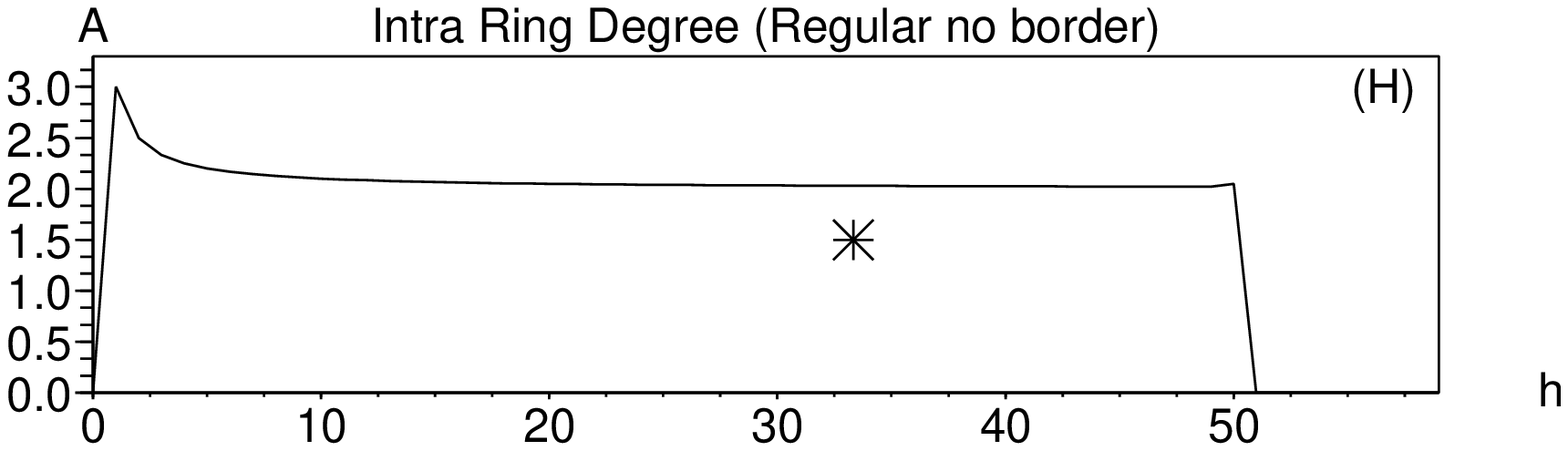}}
   \caption{Intra Ring Degree values for the considered network models.~\label{fig:sim4}}
\end{figure}

The results for intra-ring degree, shown in Figure~\ref{fig:sim4}, are
very similar to the hierarchical number of nodes measurement,
characterized by a peak, except for regular networks, which exhibit a
markedly different evolution resembling the curves obtained for the
inter-ring degree. Note that for regular graphs with no border
effects, the final decreasing part tends to decrease and saturate.
The shape of BA and Random curves are closely similar to those
obtained for the hierarchical number of nodes.

\begin{figure}
   \resizebox{8cm}{2cm}{\includegraphics[]{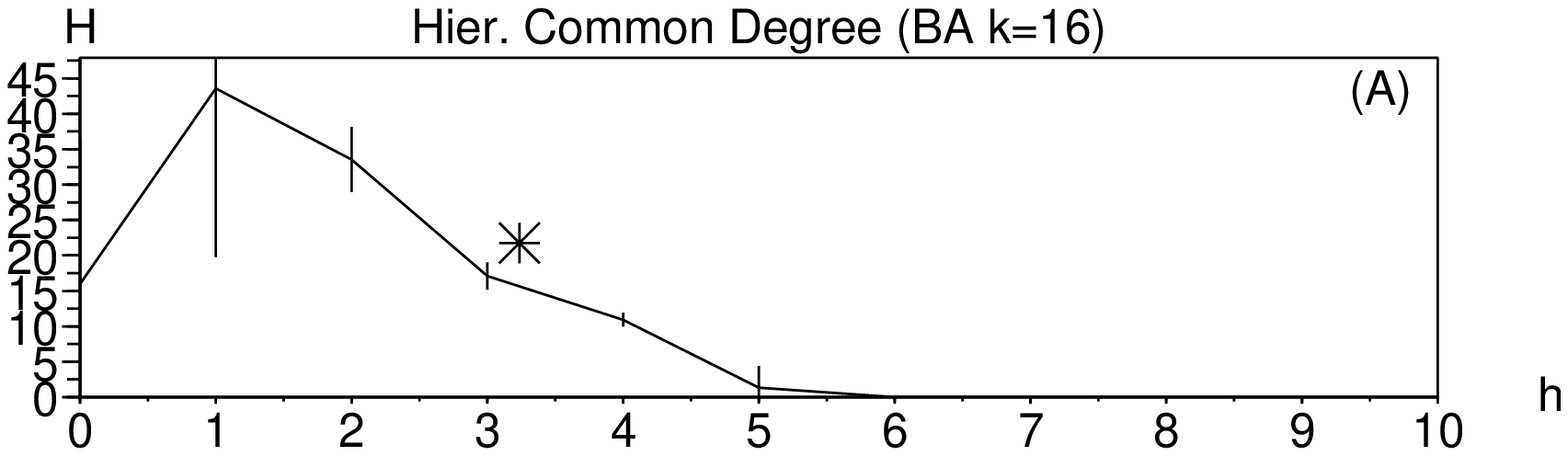}}
   \resizebox{8cm}{2cm}{\includegraphics[]{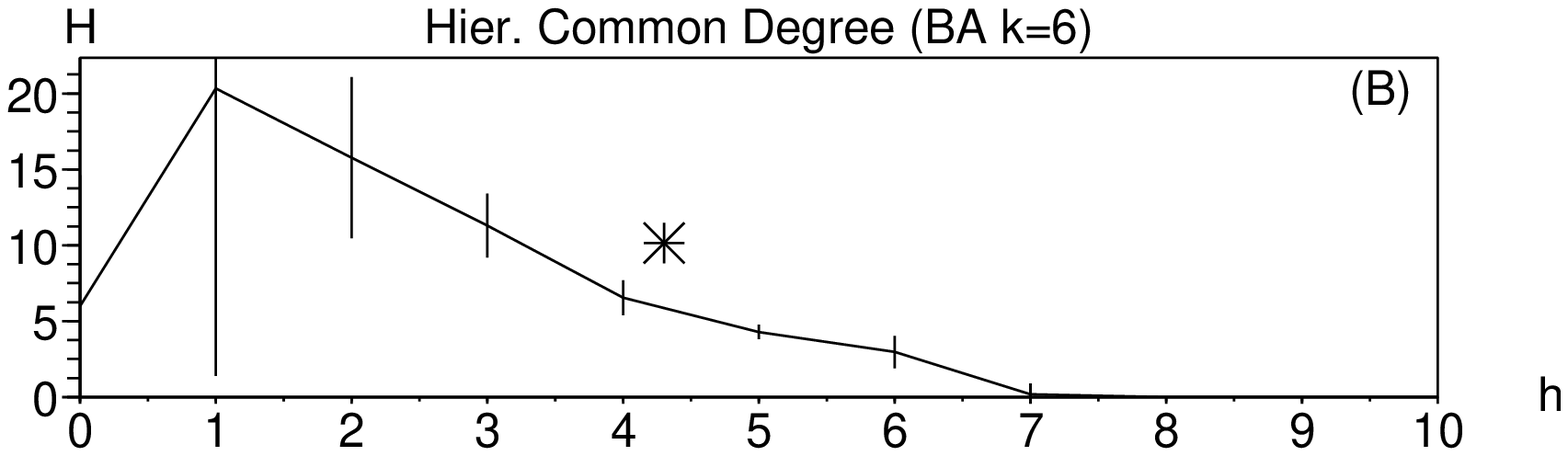}}
   \resizebox{8cm}{2cm}{\includegraphics[]{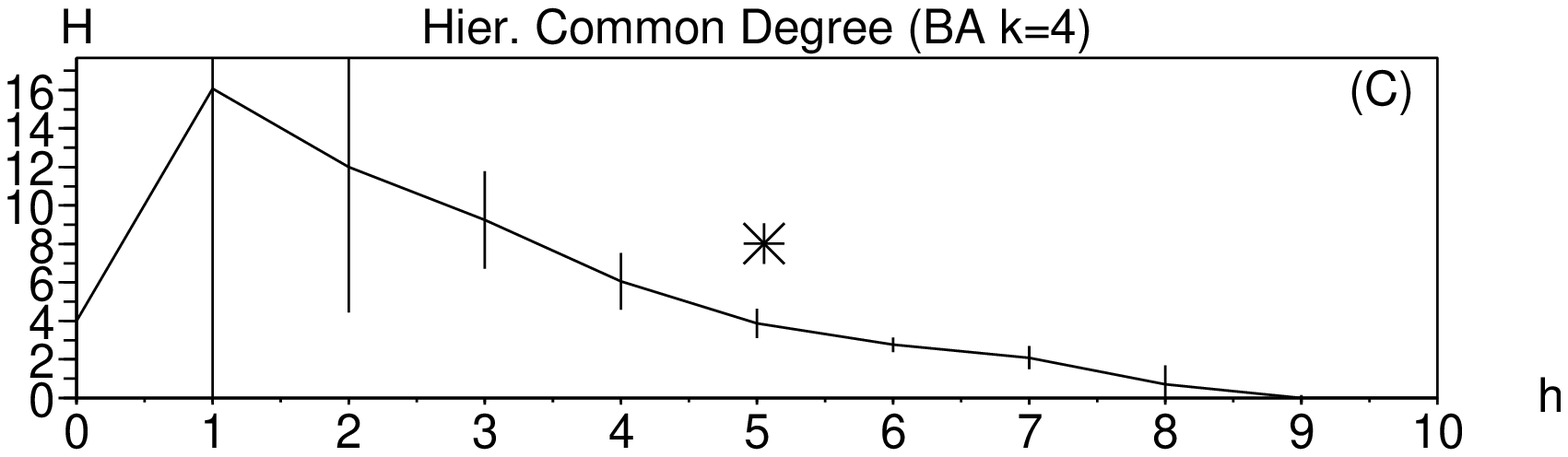}}
   \resizebox{8cm}{2cm}{\includegraphics[]{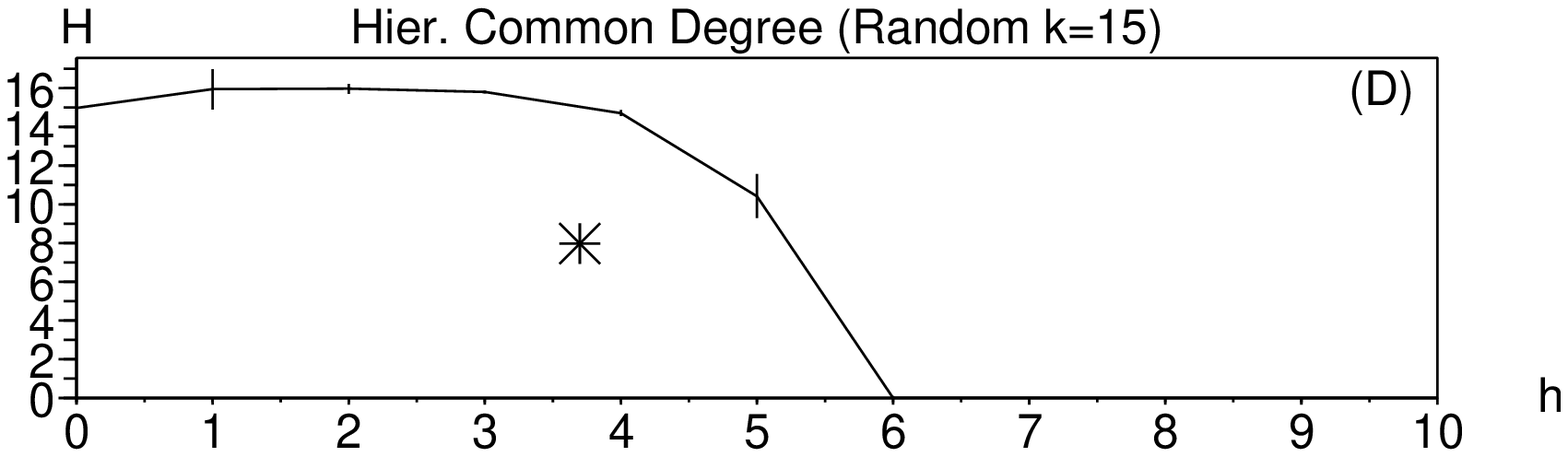}}
   \resizebox{8cm}{2cm}{\includegraphics[]{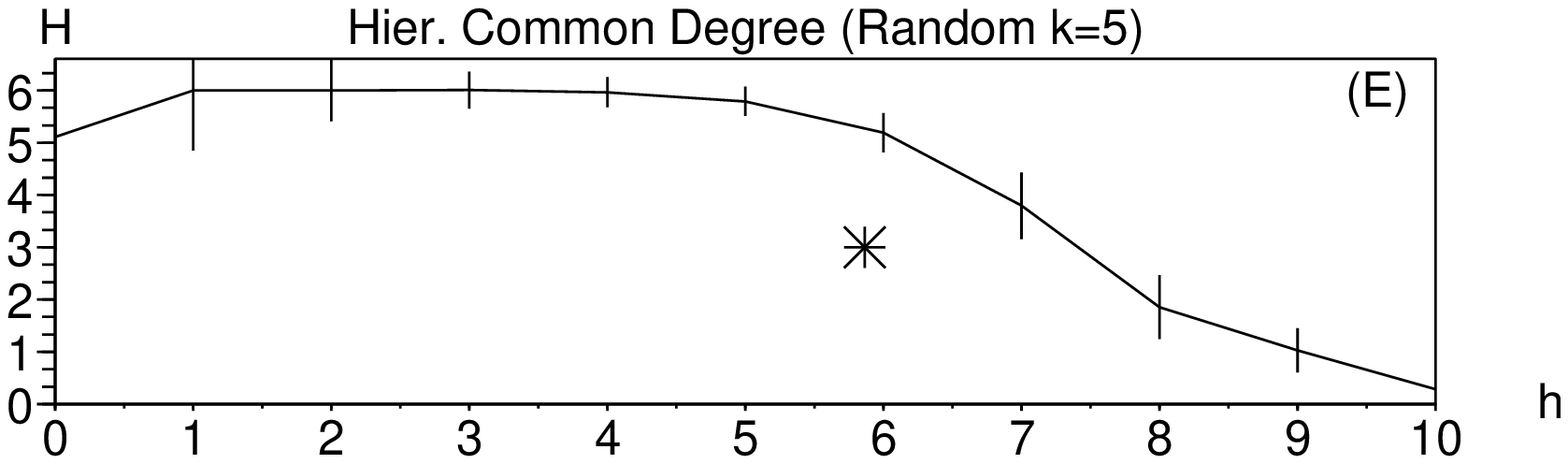}}
   \resizebox{8cm}{2cm}{\includegraphics[]{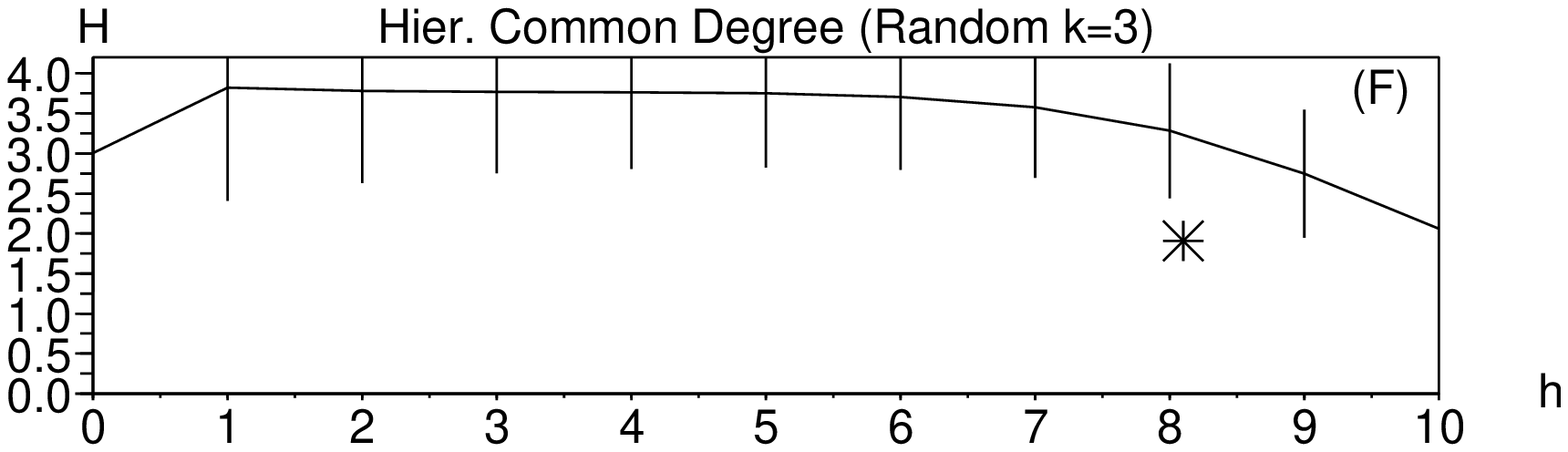}}
   \resizebox{8cm}{2cm}{\includegraphics[]{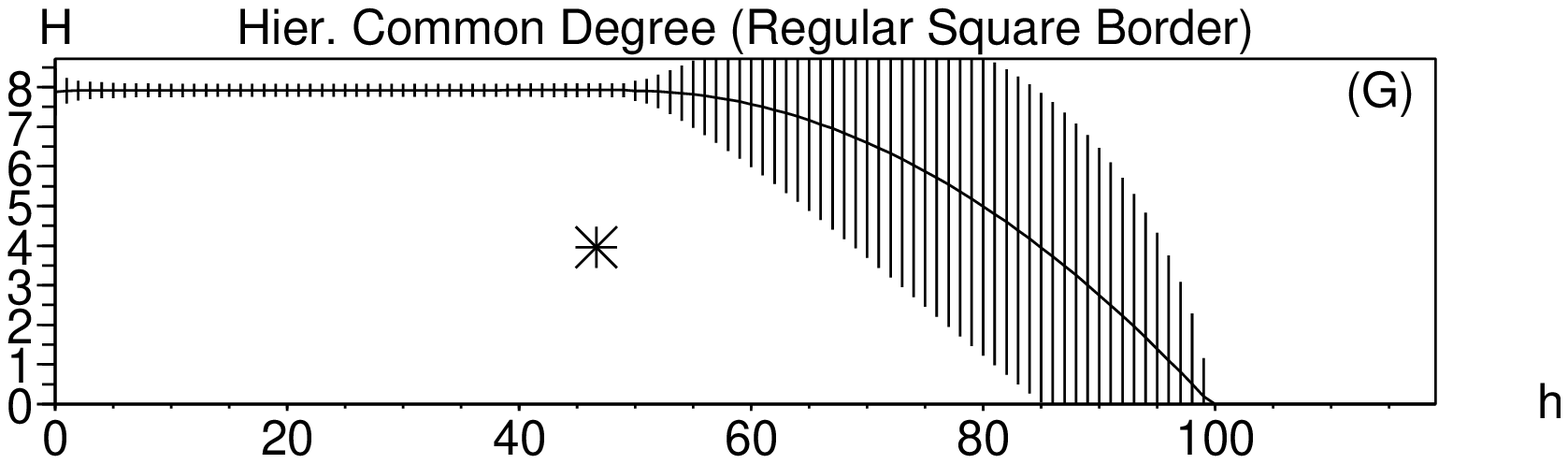}}
   \resizebox{8cm}{2cm}{\includegraphics[]{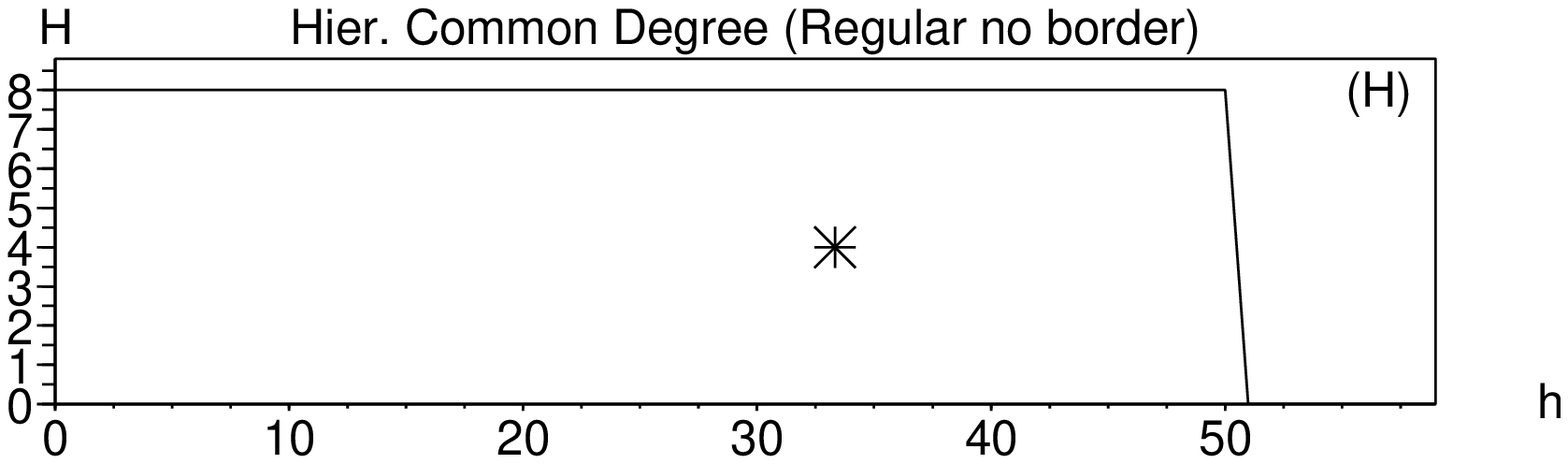}}
   \caption{Hierarchical Common Degree measures with the respective $\pm$ standard deviations obtained for the considered models.~\label{fig:sim5}}
\end{figure}

Figure~\ref{fig:sim5} shows the values of hierarchical common degree
for the considered network models. These distributions are
characterized by a decreasing curve after the first level, excepted
for the regular graphs with no border effects.  Generally, these
curves are similar to those obtained for the inter-ring degrees,
except that the present curves are wider.  Another observation is that
the average hierarchical common degree tends to be higher at the
initial hierarchical levels, which is a consequence of the fact that
the largest hubs present in the BA model tend to be reached sooner,
providing bypasses to the other nodes and therefore reducing the peak
abscissae and number of hierarchical levels.  This is the main reason
why all peaks in the BA networks tend to be displaced to the left hand
side than those in the random networks.  Like with the other
measurements, it can be that the positions of the peaks along the
curves are less affected by variations of the average node degree in
the cases of the BA models.  The curves for random and regular models
resulted similar and characterized by an interval of nearly constant
values at the intermediate part of the curves. This is a direct
consequence of the smaller variance of traditional node degrees in
those two models as compared to the higher variance of the BA cases.

Because the regular models have a fixed number of connections for each
node, the common degree measurement results in a constant curve with
value $k=8$ for the network with border effects.  As some nodes
(i.e. those at the border) do not have exactly the same degree, the
last levels have a smooth decrease but higher standard deviation.

\begin{figure}
   \resizebox{8cm}{2cm}{\includegraphics[]{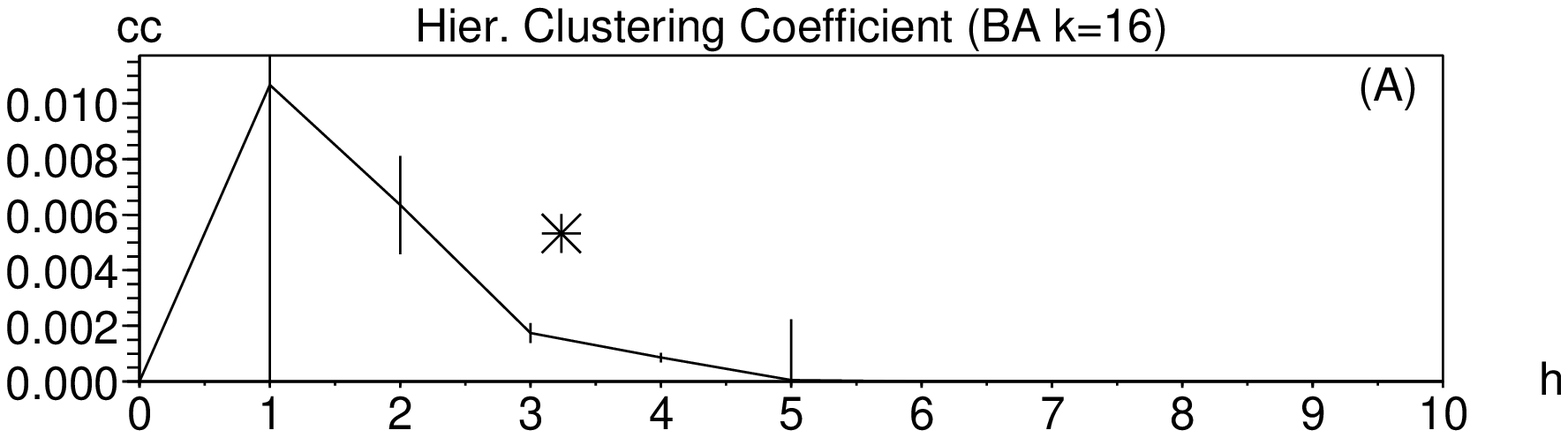}}
   \resizebox{8cm}{2cm}{\includegraphics[]{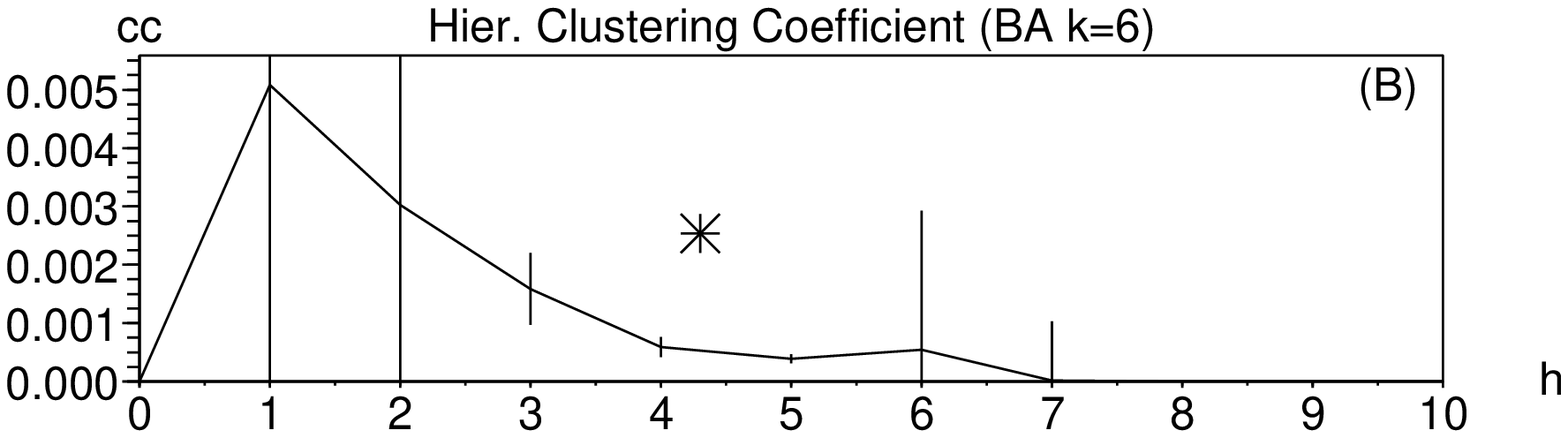}}
   \resizebox{8cm}{2cm}{\includegraphics[]{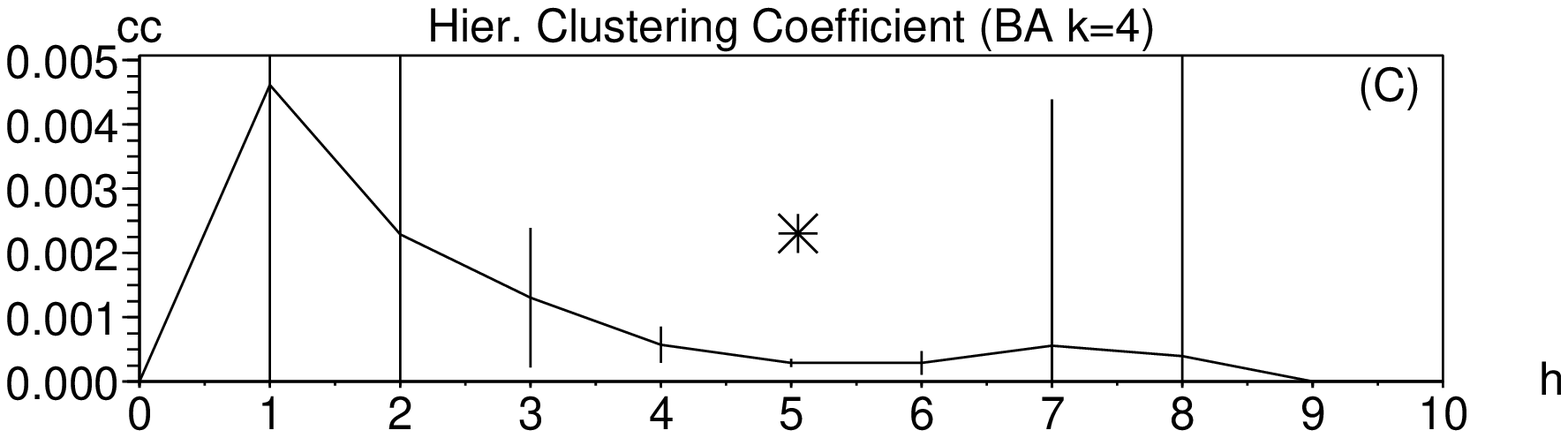}}
   \resizebox{8cm}{2cm}{\includegraphics[]{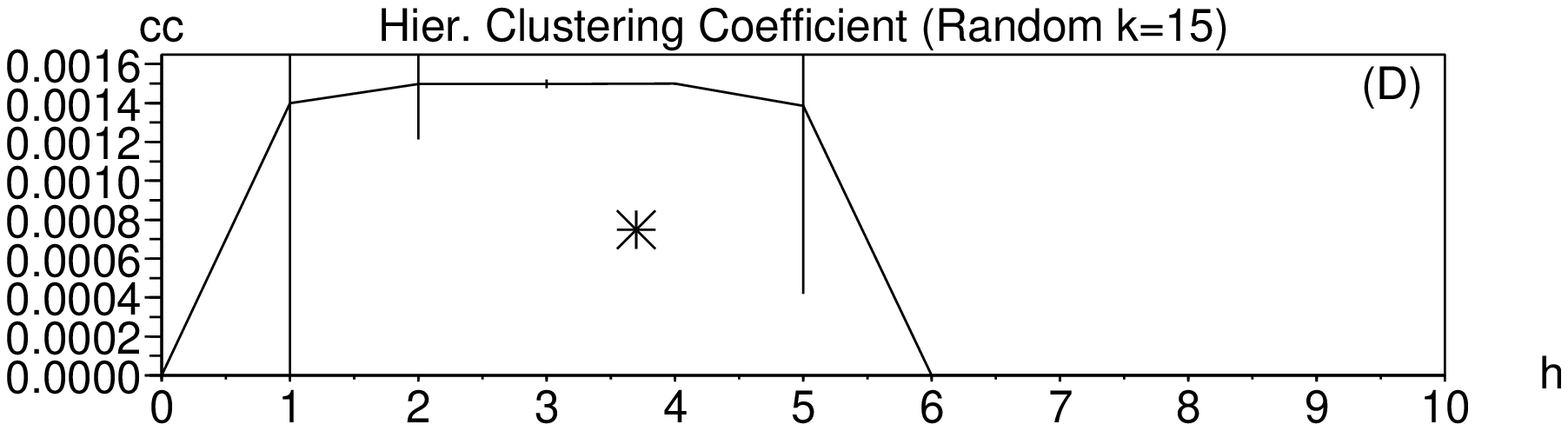}}
   \resizebox{8cm}{2cm}{\includegraphics[]{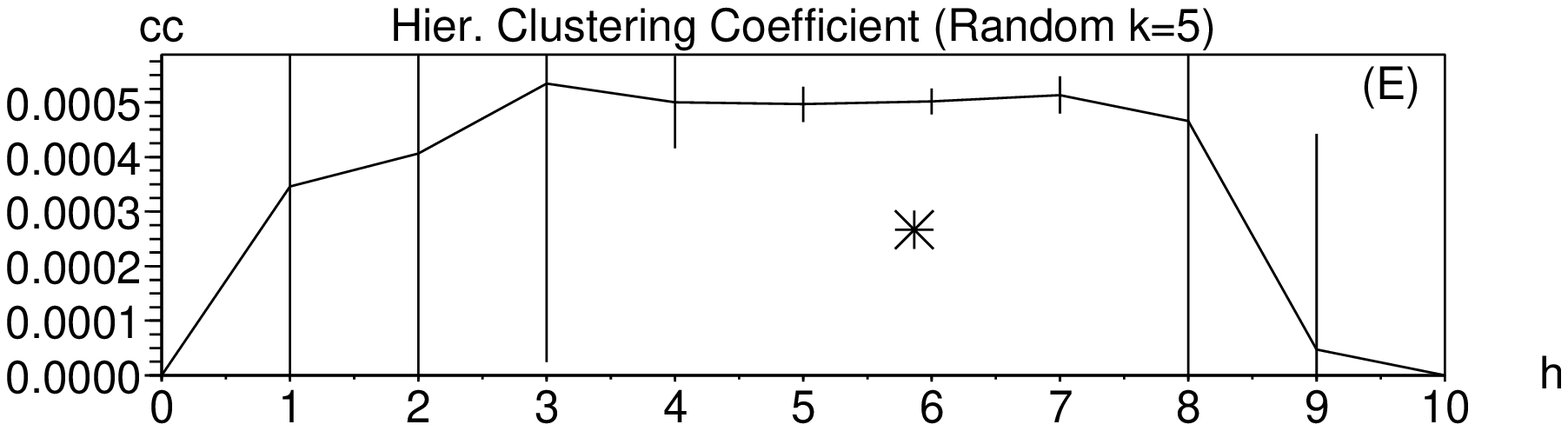}}
   \resizebox{8cm}{2cm}{\includegraphics[]{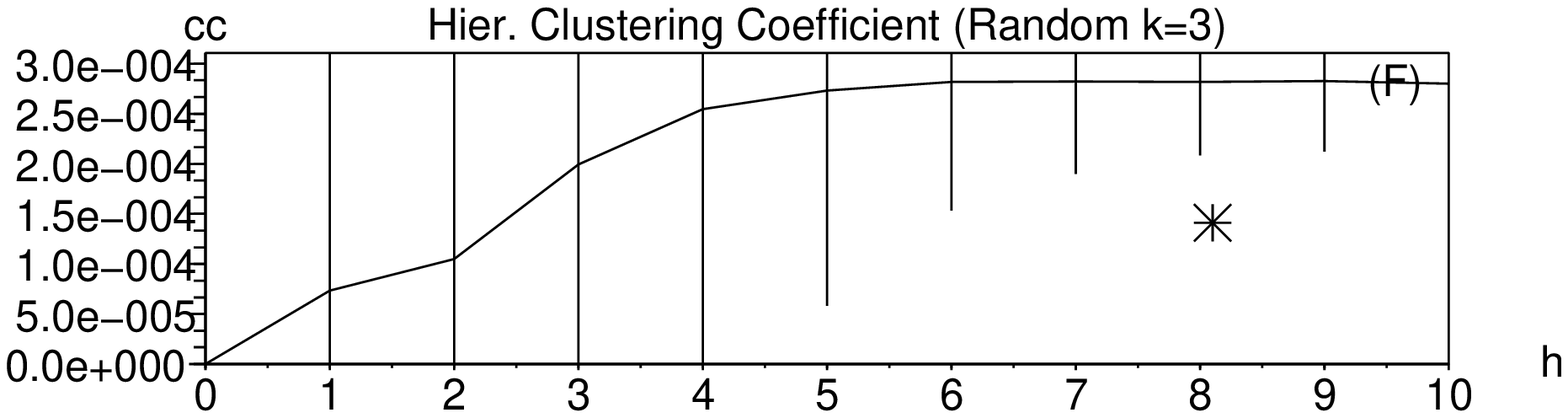}}
   \resizebox{8cm}{2cm}{\includegraphics[]{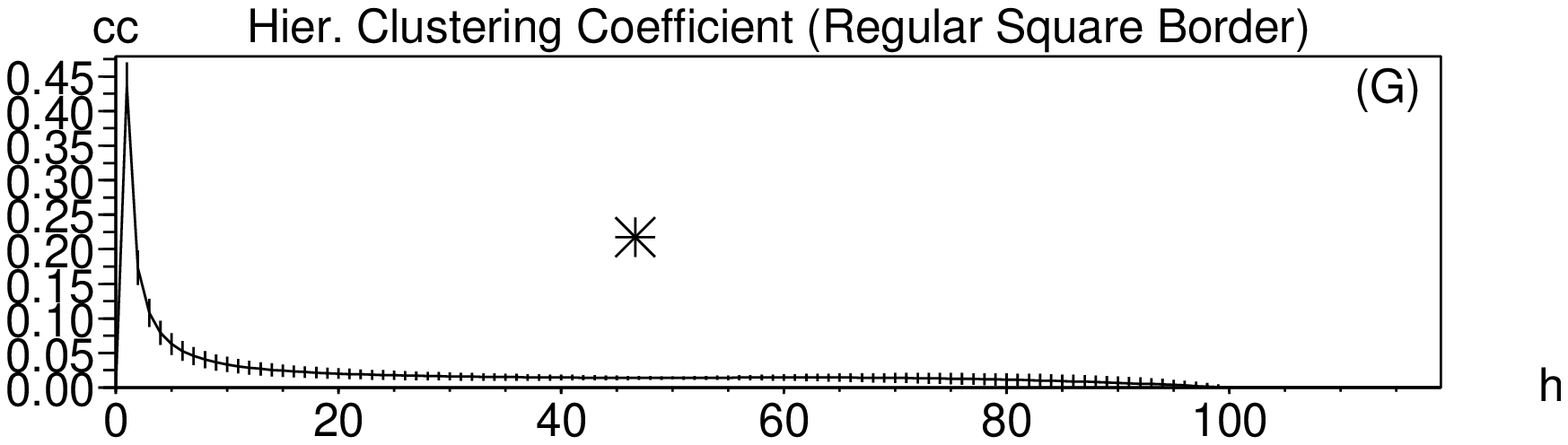}}
   \resizebox{8cm}{2cm}{\includegraphics[]{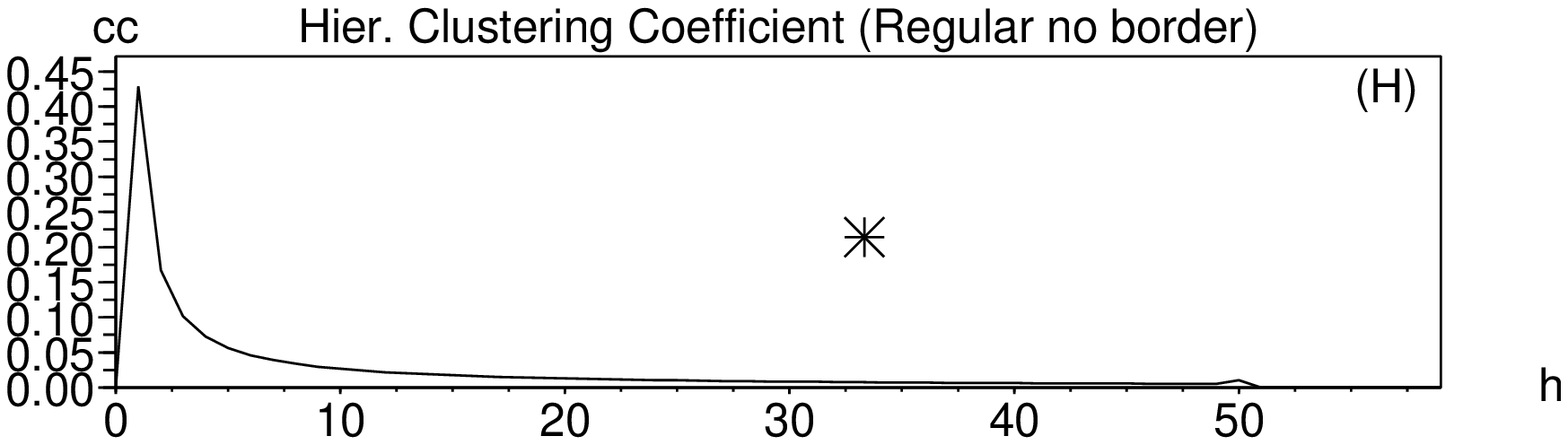}}
   \caption{Hierarchical Clustering Coefficient Degree
   measurements. Note the higher values of standard deviantion
   relatively to the other measurements.~\label{fig:sim6}}
\end{figure}

The curves of hierarchical clustering coefficients resulted the most
distinct among the three considered networks and have the higher
standard deviations, as shown in Figure ~\ref{fig:sim6}. Also
involving an intermediate constant interval, the curves obtained for
the random models correspond to the smallest clustering coefficients
among the models. Therefore, the nodes at each ring of those networks
are characterized by low interconnectivity. The hierarchical
clustering coefficient curves obtained for the BA case, present much
higher values and involve sharper peaks of connectivity, tending to
present another peak along the last levels (see
Figures~\ref{fig:sim6}b-c). The hierarchical clustering coefficient
obtained for the regular networks has a monotonically decreasing
behavior, with values which start higher than those of the two other
considered models.  The monotonic decay observed for this case
(i.e. regular networks) is explained by the fact that both the number
of nodes and edges increase linearly along successive hierarchical
levels for that model (see Equation~\ref{eq:hcc}). Note that the
regular model with and without border are similar.

\begin{figure}
   \resizebox{8cm}{2cm}{\includegraphics[]{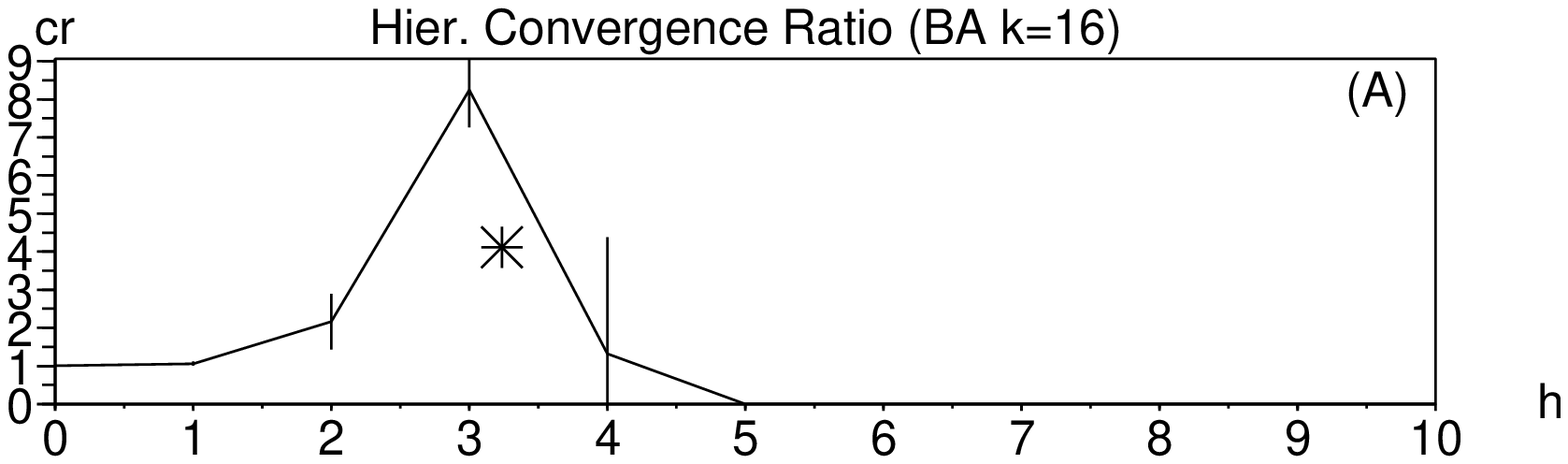}}
   \resizebox{8cm}{2cm}{\includegraphics[]{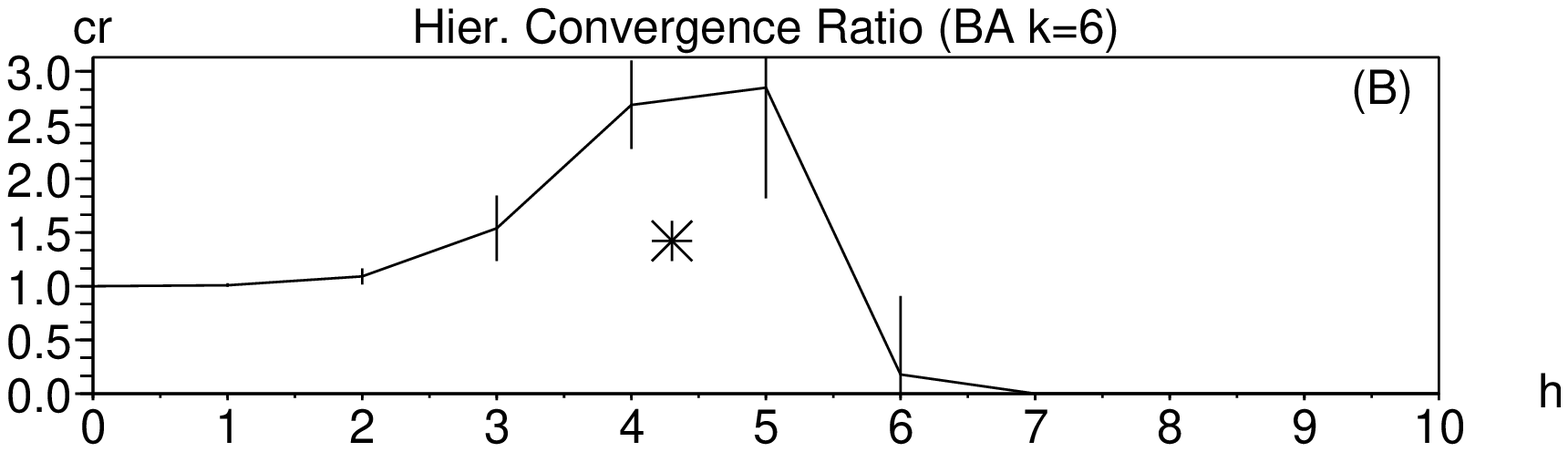}}
   \resizebox{8cm}{2cm}{\includegraphics[]{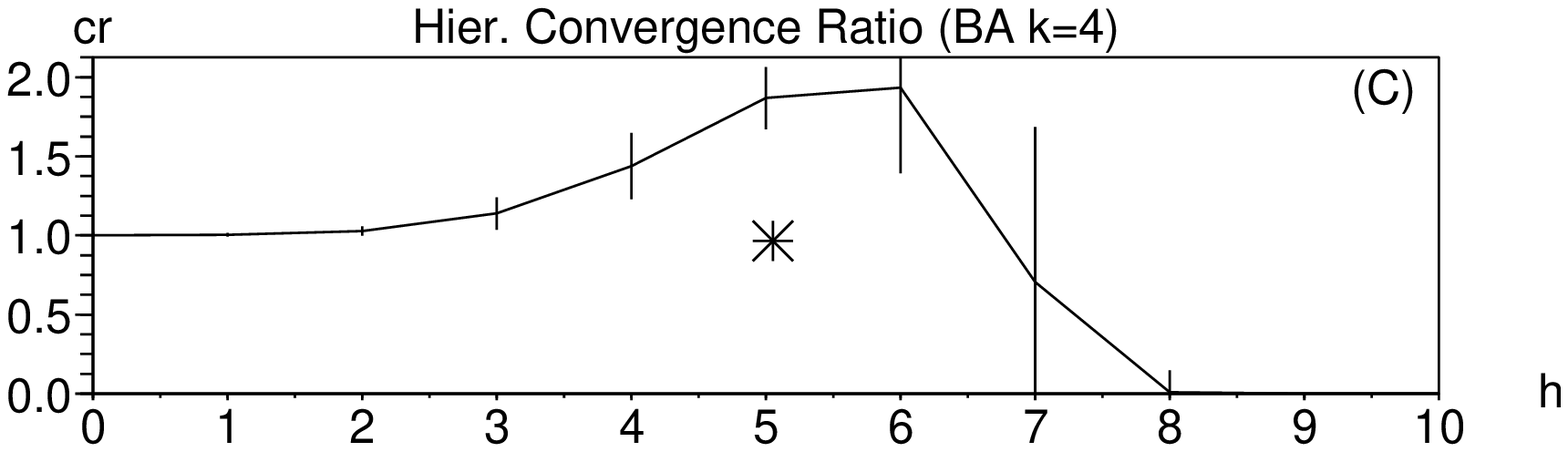}}
   \resizebox{8cm}{2cm}{\includegraphics[]{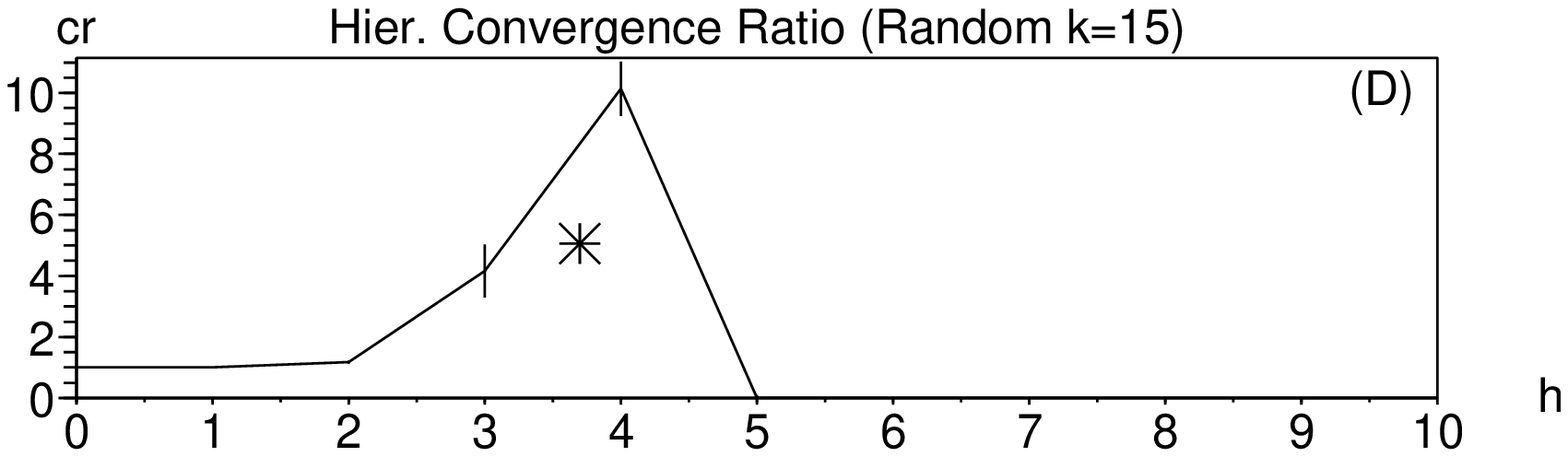}}
   \resizebox{8cm}{2cm}{\includegraphics[]{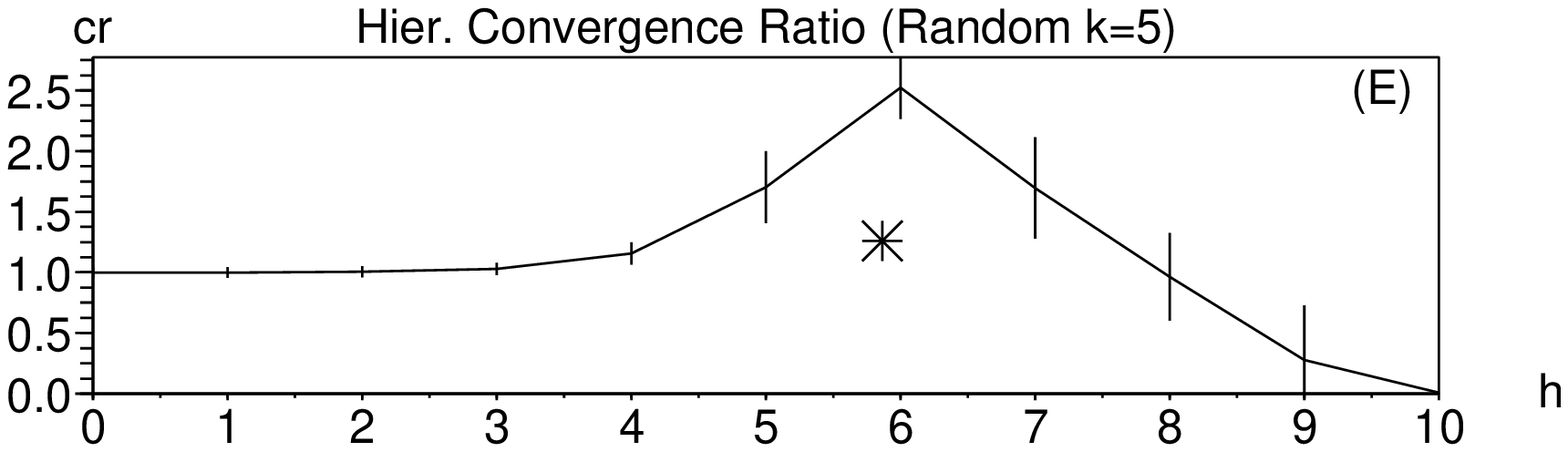}}
   \resizebox{8cm}{2cm}{\includegraphics[]{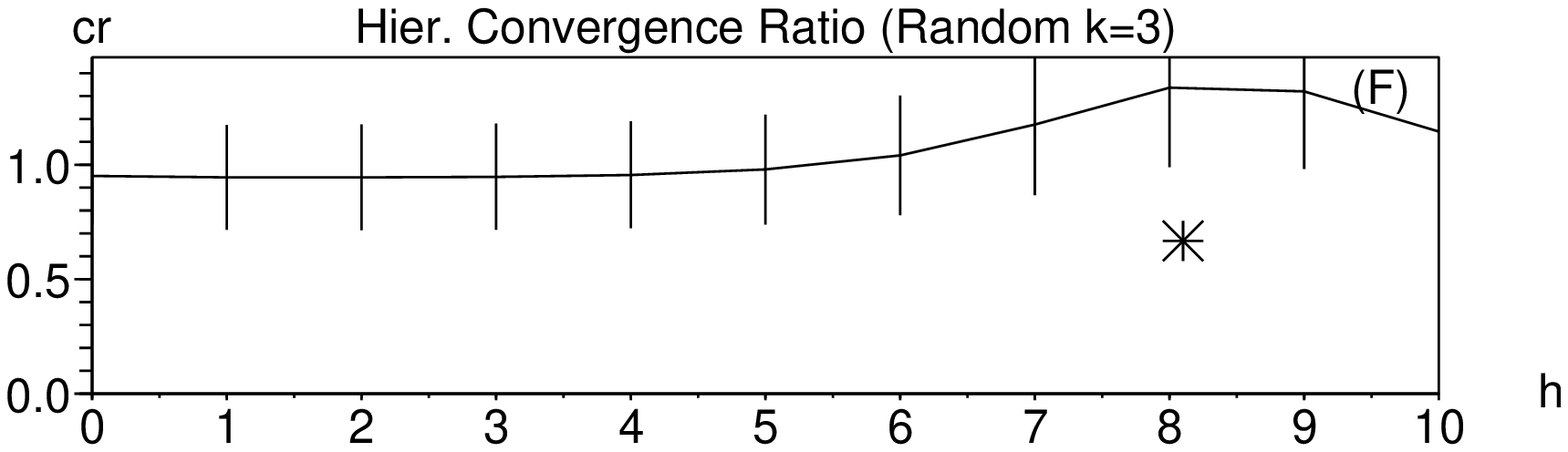}}
   \resizebox{8cm}{2cm}{\includegraphics[]{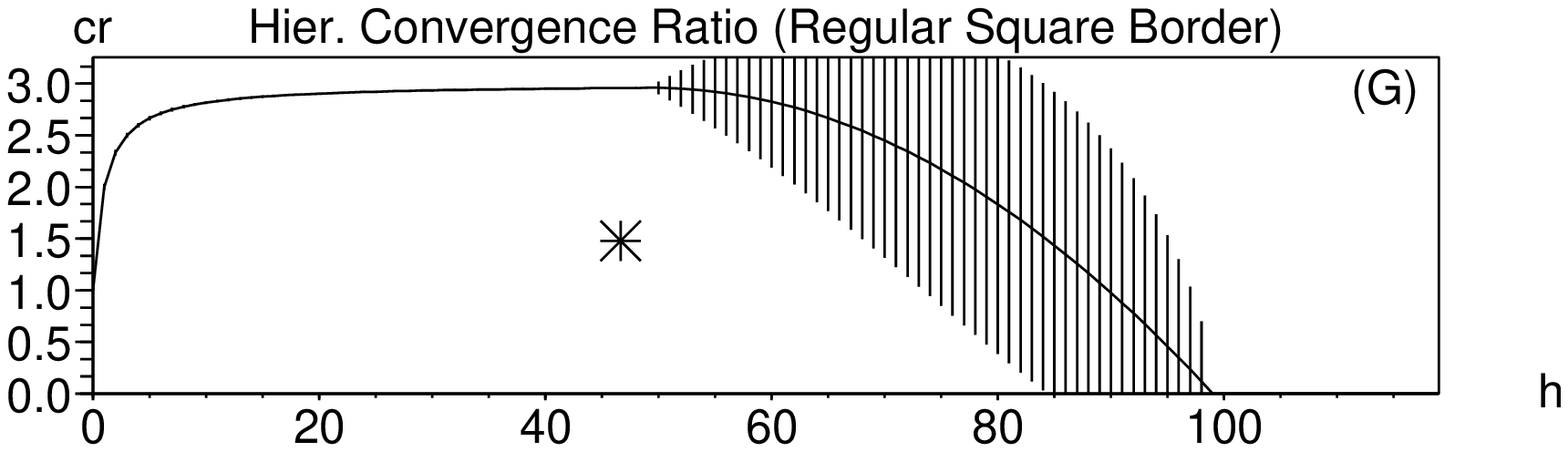}}
   \resizebox{8cm}{2cm}{\includegraphics[]{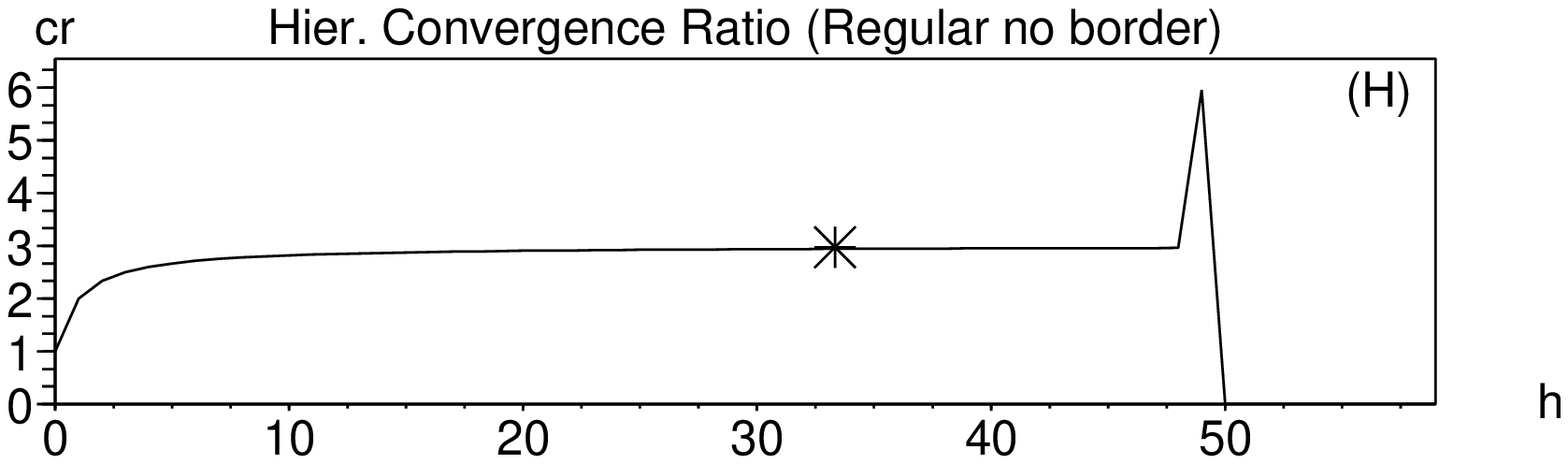}}
   \caption{Convergence Ratio measurements for the considered
   networks.~\label{fig:sim7}}
\end{figure}

The convergence ratios obtained for each of the considered network
models are shown in Figure~\ref{fig:sim7}. These curves are
characterized by similar behavior among themselves with nearly
constant value at the first levels and a peak at the last levels
(except for the regular models), along which the hierarchical
expansion tends to saturate, i.e. after the peak $P$ is reached. Note
also that sharper peaks tend to be obtained for high values of
$k$. The positions of the peaks are near the average shortest paths.

The convergence ratio curves obtained for the regular networks are
also qualitatively similar to those obtained for the other models, but
the bordered graphs lack the peak and have a smooth decay along the
last levels.

Interestingly, among all considered measurements, it was the
hierarchical common degrees and hierarchical clustering coefficients
which provided the most distinctive shapes for each respective network
model. Therefore, such measurements stand out as particularly
promising subsidies for, together with the log-log node degree
density, identifying the category of the network under study.  Such a
possibility is illustrated in the following section.

\section{Application to Real Networks}

The above described hierarchical measurements have also been applied
to characterize three complex networks obtained from real data.
These real networks include: a Edinburgh Associative Thesaurus
network~\cite{edinburgh}, the 1997 US Airports network
(~\cite{Airdatabase}) and a protein-protein interaction
network~\cite{yeast}. The Edinburgh(Word) graph is a empirical
association network created as a set of collected words from human
subjects who are requested to enter words that first come to their
mind after seeing a stimulus word. All the responses are presented
with similar frequency. The detailed procedures of the creation of
Edinburgh graphs can be seen in~\cite{edinburgh}. This network has
23219 nodes and $k\simeq 14$ and is oriented and weighted. A similar
network has been considered in~\cite{whats}, while a preliminary
characterization of such a type of networks by using the
hierarchical node degree has been reported in~\cite{PRL:Costa}. The
protein-protein interaction graph(YEAST), described in
~\cite{yeast}, has 2361 nodes with $k\simeq 3$ where a node
represents a protein and the edge a interaction between the two
respective proteins. The US Airport network is a compilation of
flights between the airports of United States in 1997, where a node
represents an airport and the edge a flight between the two
airports. This network has a total of 332 nodes(airports) and
$k\simeq 6.4$.  All the considered real graphs were compared to
random and BA models with similar node degrees (for the sake of
space economy, not all these graphs are shown in Section
Characterization of Complex Networks Models).

\begin{figure}
   \resizebox{8cm}{2cm}{\includegraphics[]{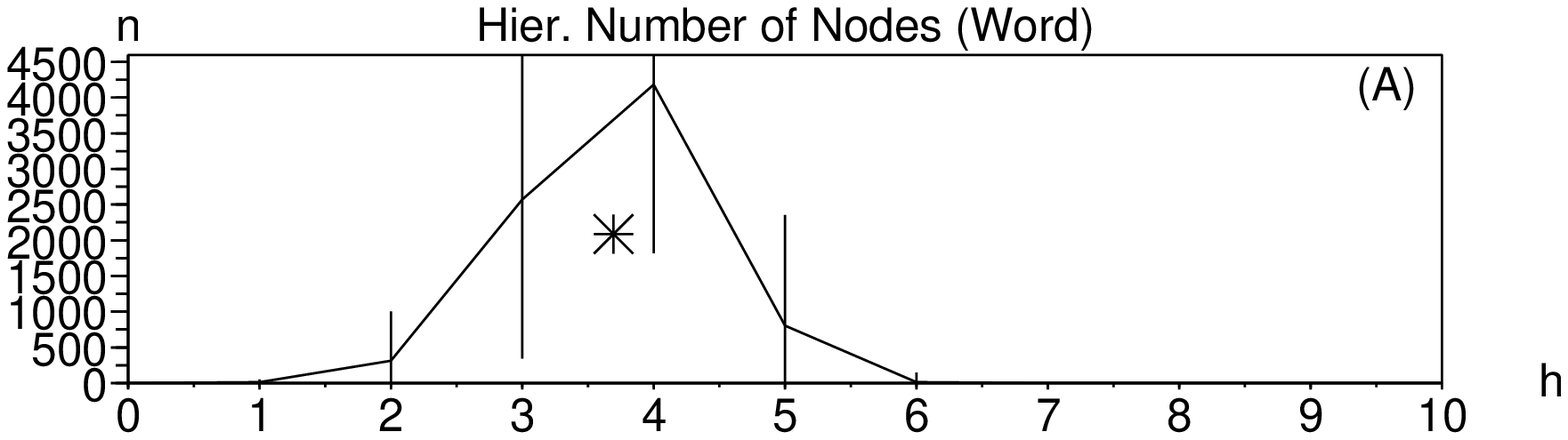}}
   \resizebox{8cm}{2cm}{\includegraphics[]{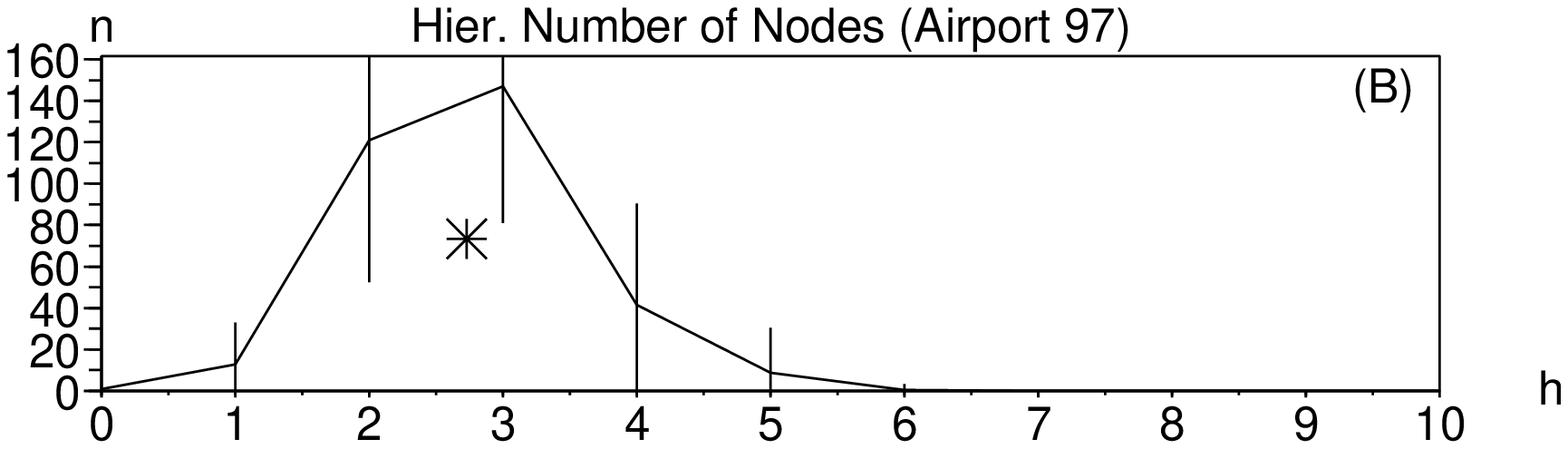}}
   \resizebox{8cm}{2cm}{\includegraphics[]{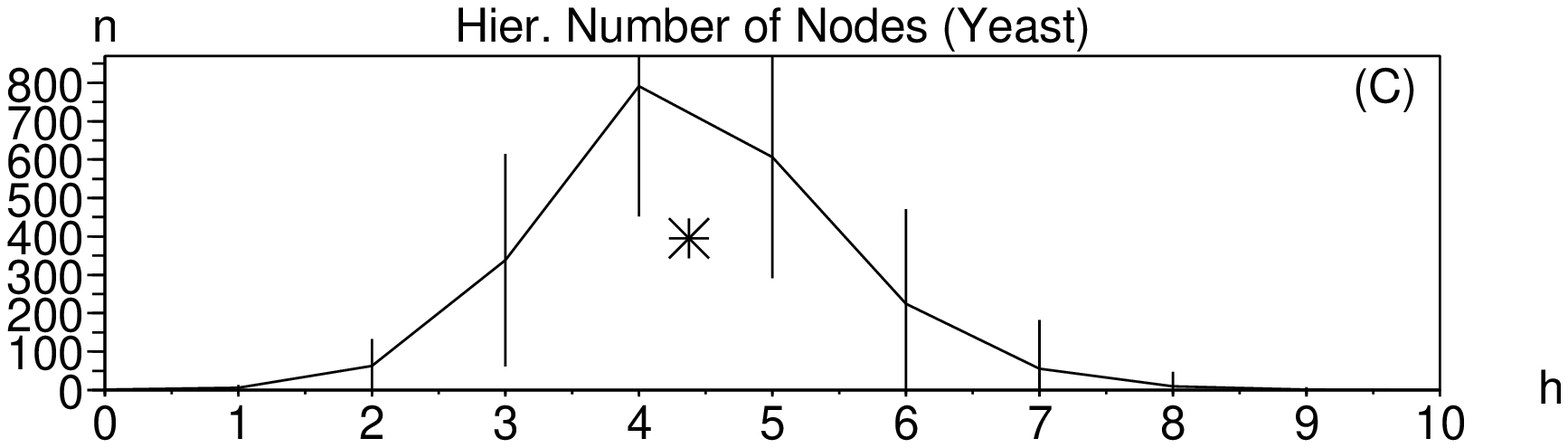}}
   \caption{Hierarchical Number of Nodes obtained for the real
   networks and considered generated networks.~\label{fig:real1}}
\end{figure}

\begin{figure}
   \resizebox{8cm}{2cm}{\includegraphics[]{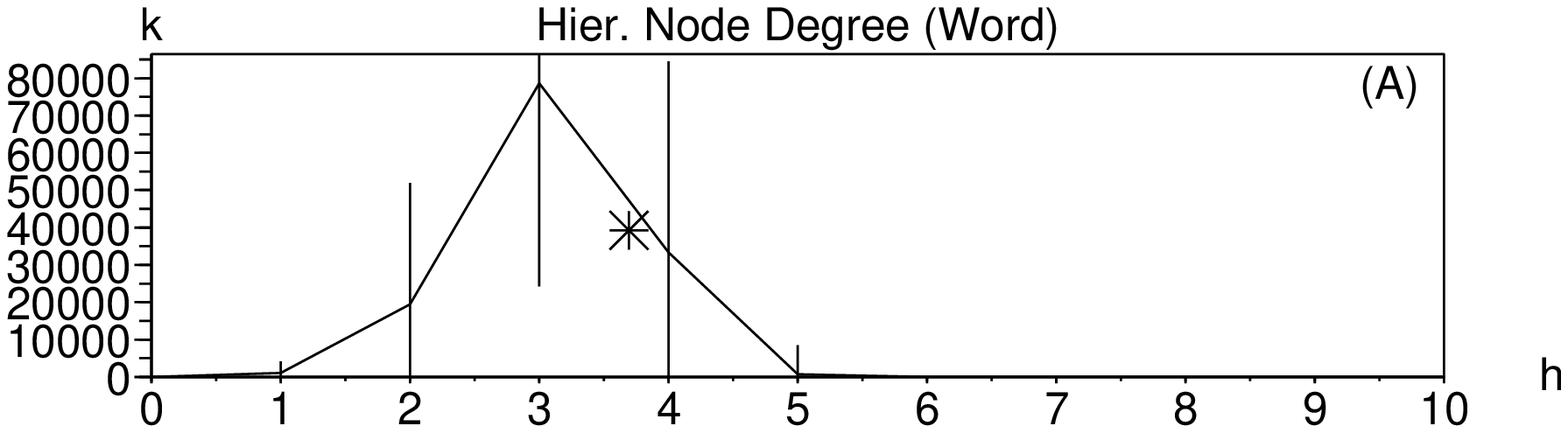}}
   \resizebox{8cm}{2cm}{\includegraphics[]{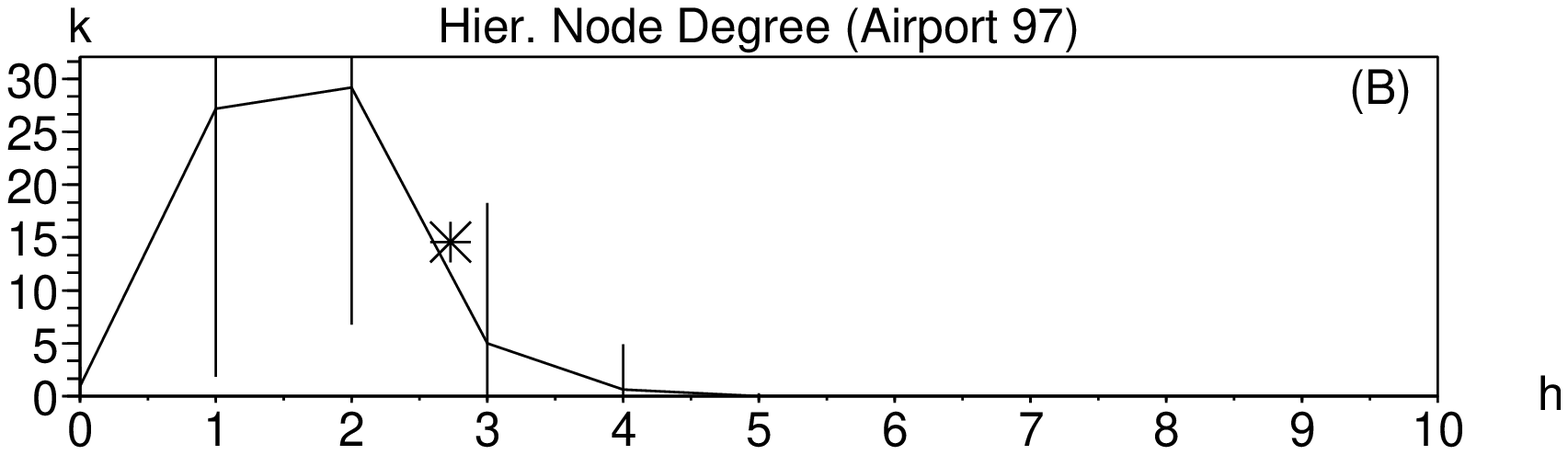}}
   \resizebox{8cm}{2cm}{\includegraphics[]{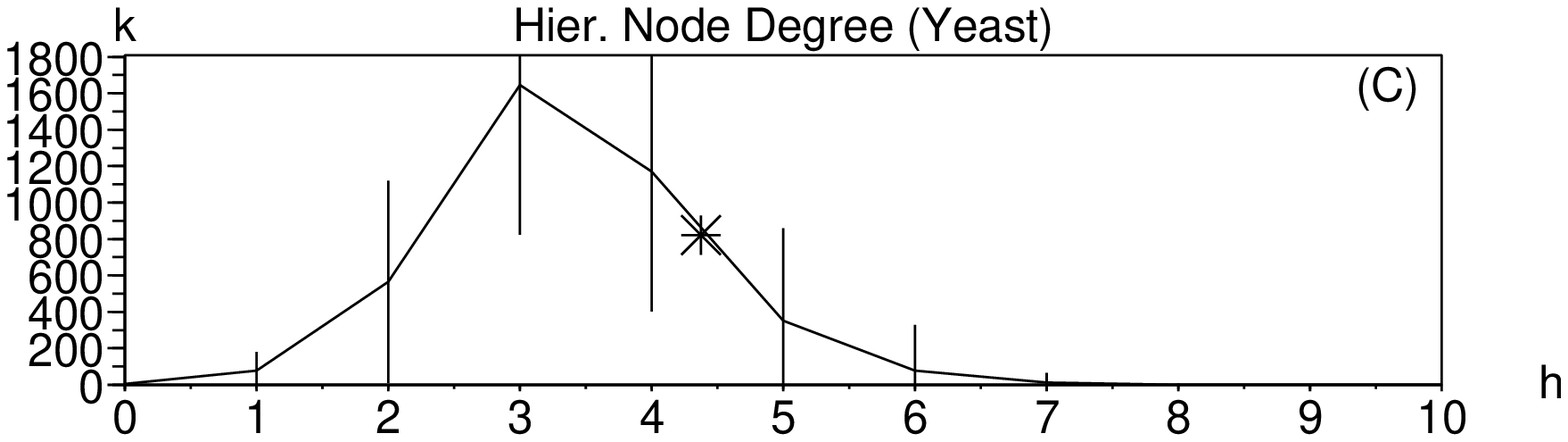}}
   \caption{Hierarchical Node Degree distribution along hierarchical
   levels, same results from Number of Nodes. ~\label{fig:real2}}
\end{figure}

The results for the curves of hierarchical number of nodes and node
degrees are similar as seen in Figure \ref{fig:real1} and
\ref{fig:real2}.  Also, no significant differences were observed
between these results and those obtained for the respective random or
BA simulated networks.

\begin{figure}
   \resizebox{8cm}{2cm}{\includegraphics[]{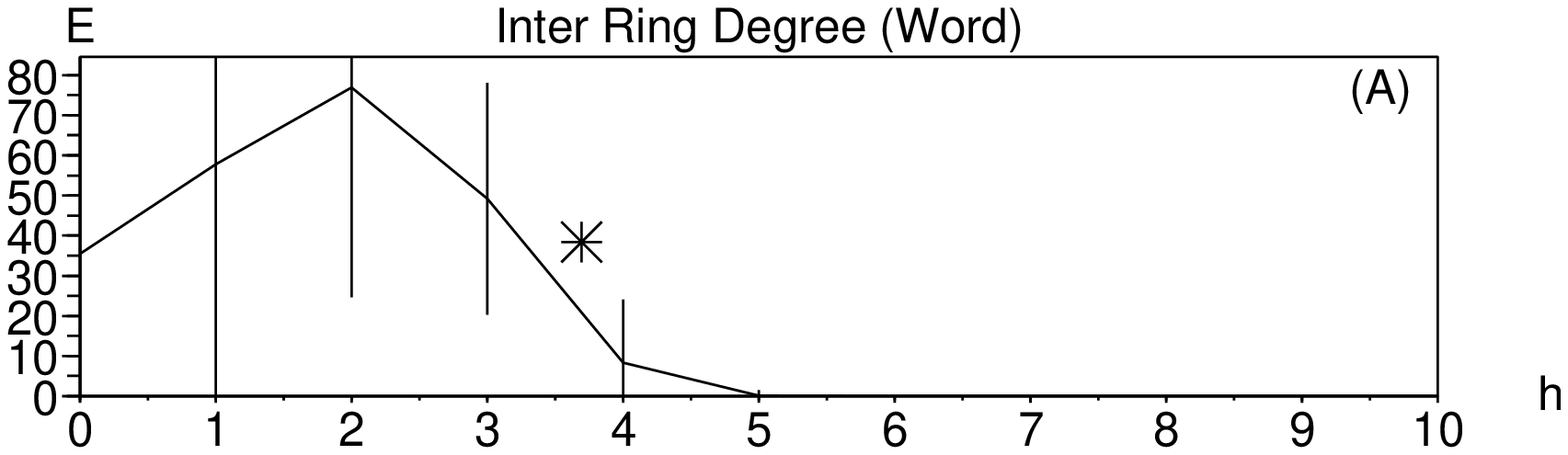}}
   \resizebox{8cm}{2cm}{\includegraphics[]{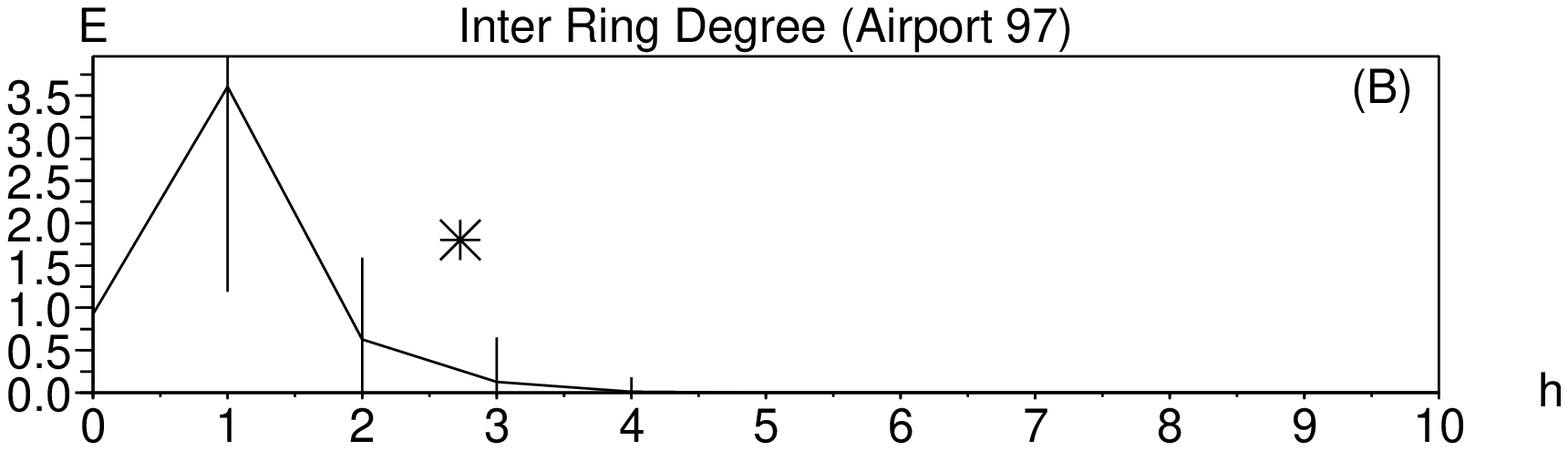}}
   \resizebox{8cm}{2cm}{\includegraphics[]{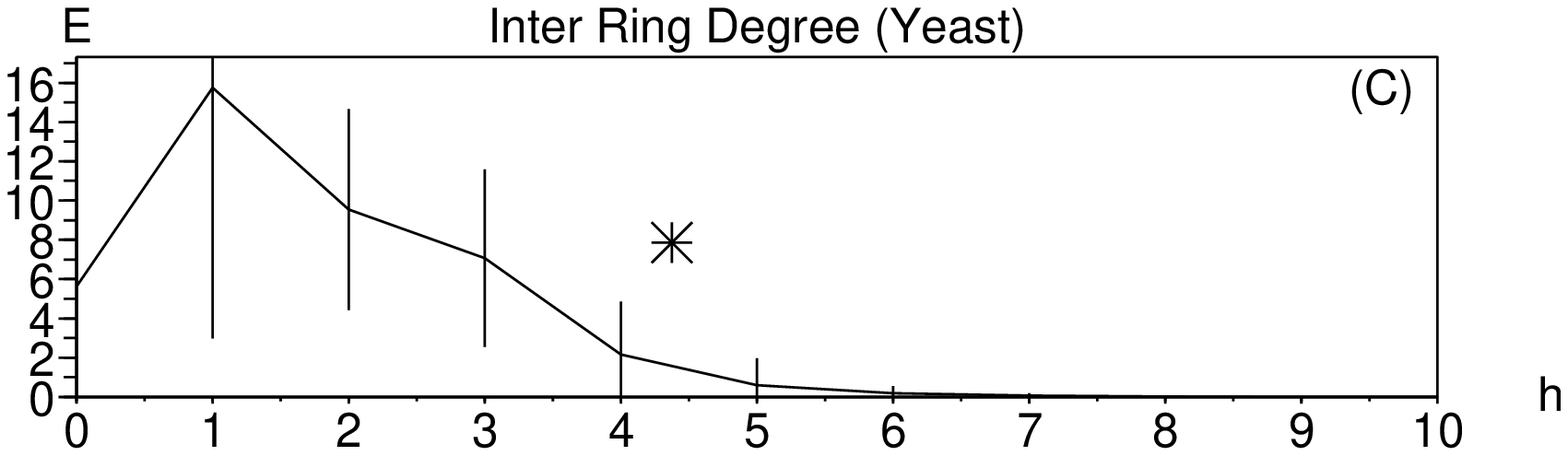}}
   \caption{Inter Ring Degree values for real and generated graphs.~\label{fig:real3}}
\end{figure}

More interesting results have been obtained for the inter-ring
degrees, shown in Figure \ref{fig:real2}. These curves were observed
to be more similar to the respective simulated Barab\'asi-Albert
curves. In fact, all considered real networks are substantially
similar to scale-free networks, being characterized by a high variance
of node degrees and the presence of hubs.

\begin{figure}
   \resizebox{8cm}{2cm}{\includegraphics[]{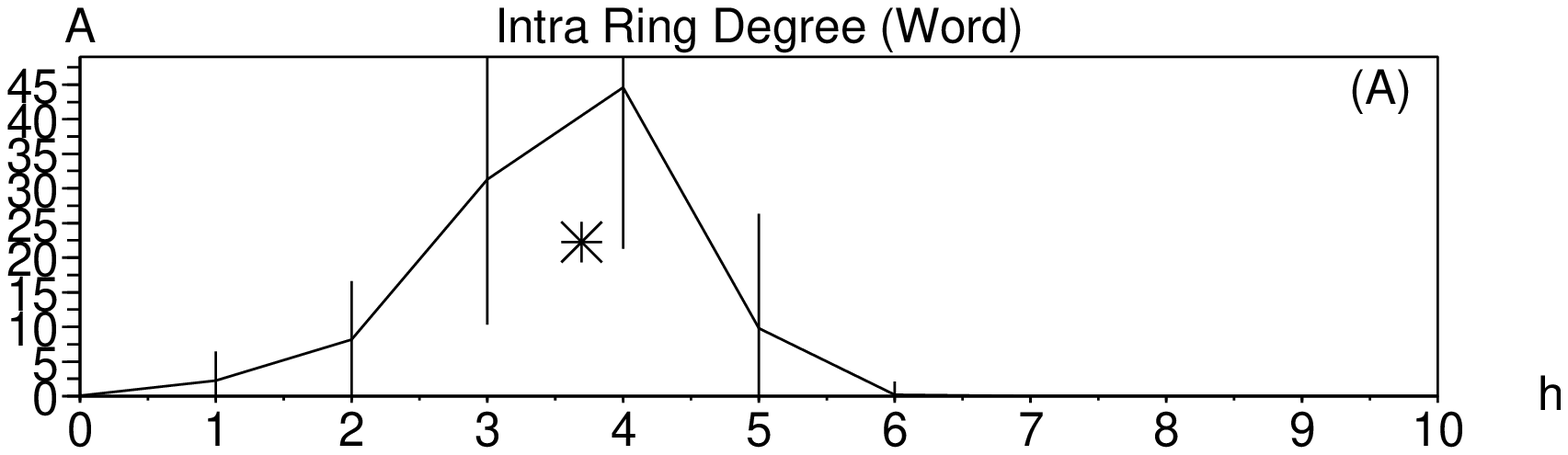}}
   \resizebox{8cm}{2cm}{\includegraphics[]{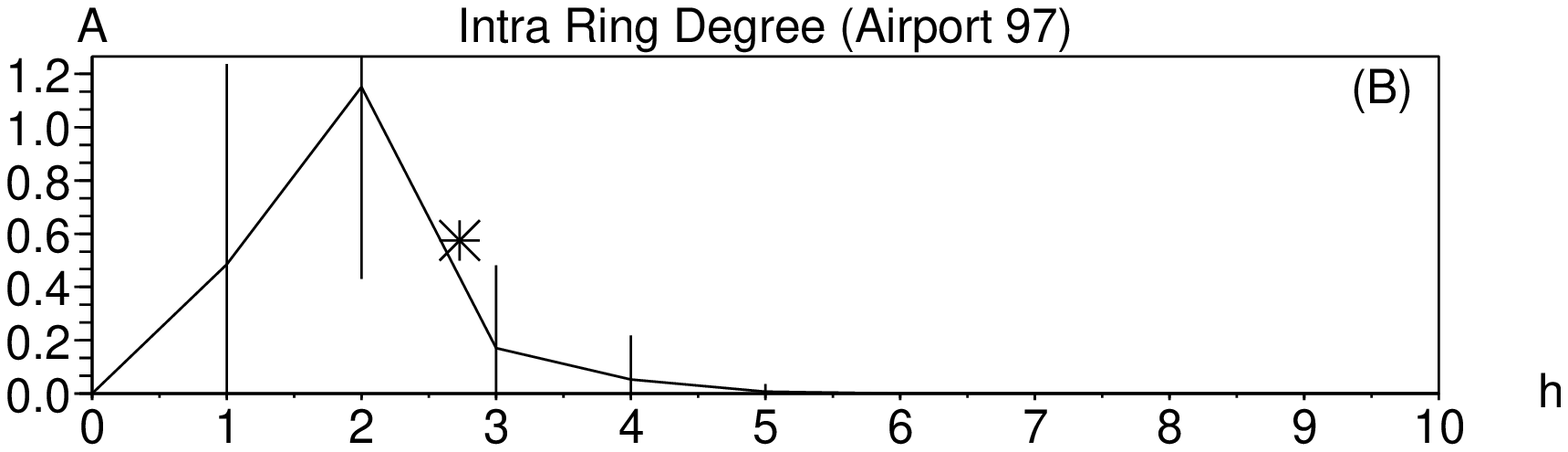}}
   \resizebox{8cm}{2cm}{\includegraphics[]{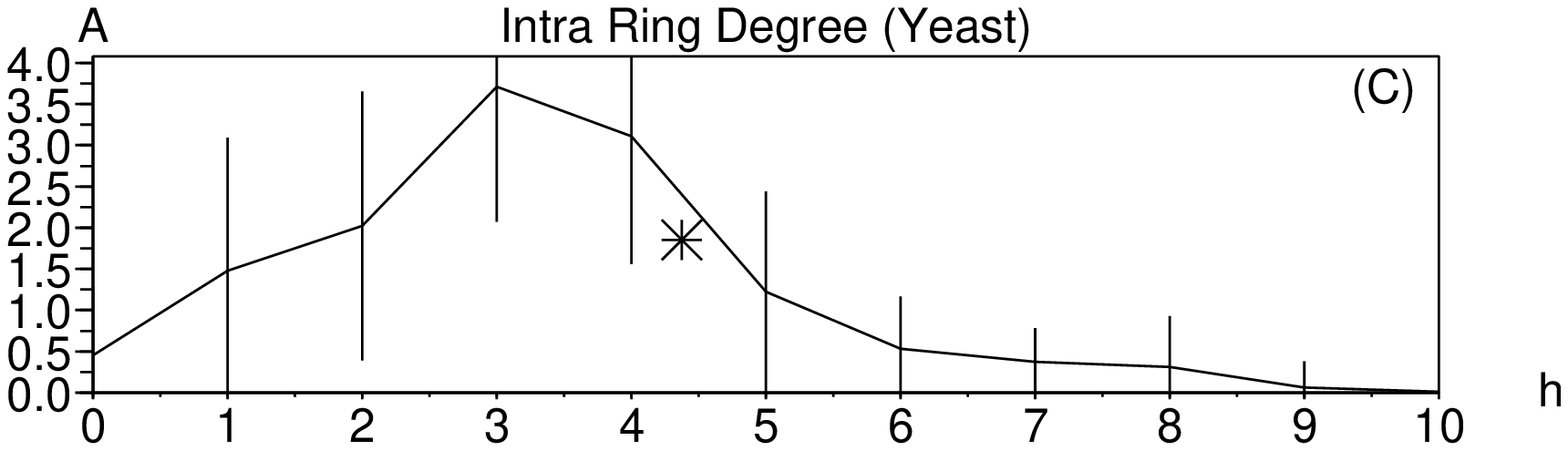}}
   \caption{Intra Ring Degree measurements obtained for the considered
   networks.~\label{fig:real4}}
\end{figure}

The intra ring degrees of the real networks are shown in Figure
\ref{fig:real4}.  Interestingly, the curves obtained for the airport
(b) and yeast(c) present their respective peaks to the left of the
average shortest path (the asterisk position), while in the BA
models the peaks tend to coincide with the asterisks as obtained for
the word network.

\begin{figure}
   \resizebox{8cm}{2cm}{\includegraphics[]{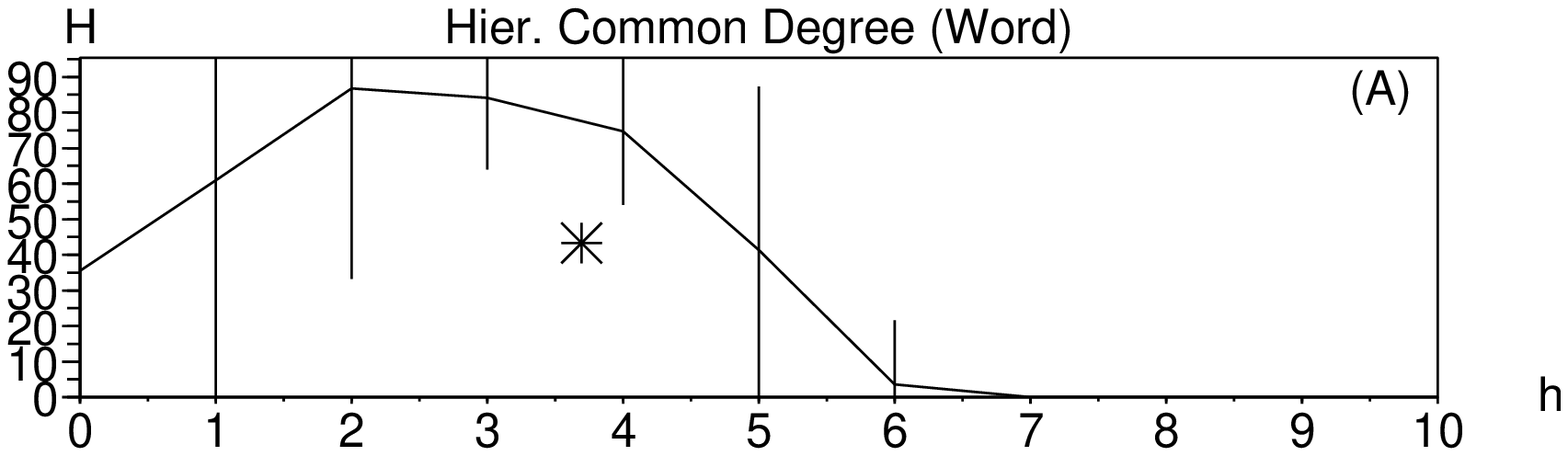}}
   \resizebox{8cm}{2cm}{\includegraphics[]{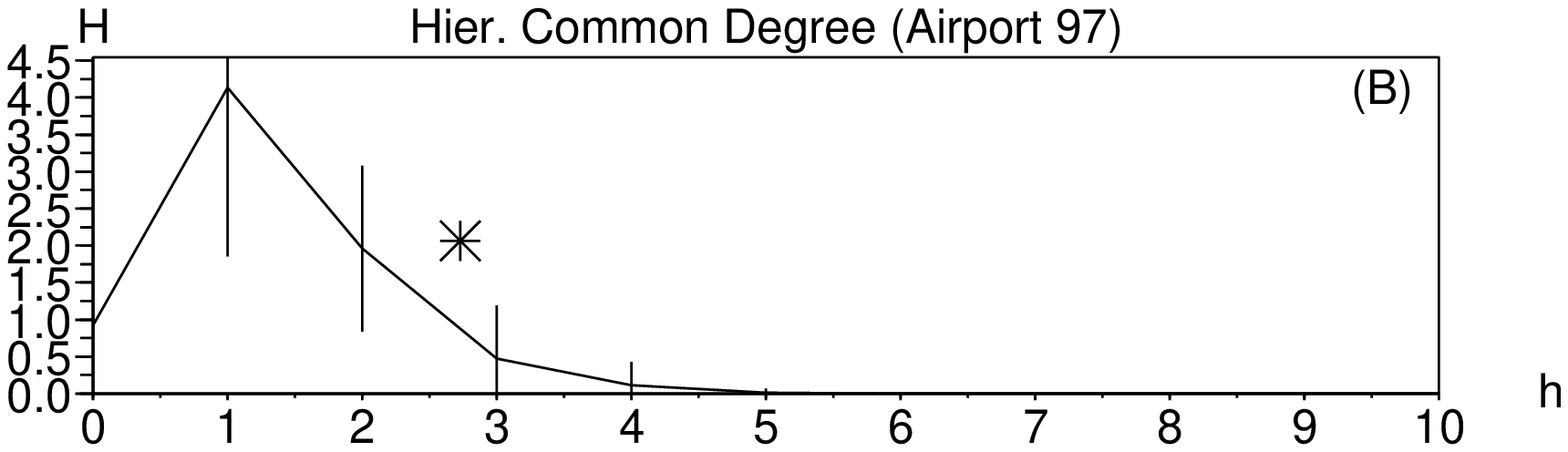}}
   \resizebox{8cm}{2cm}{\includegraphics[]{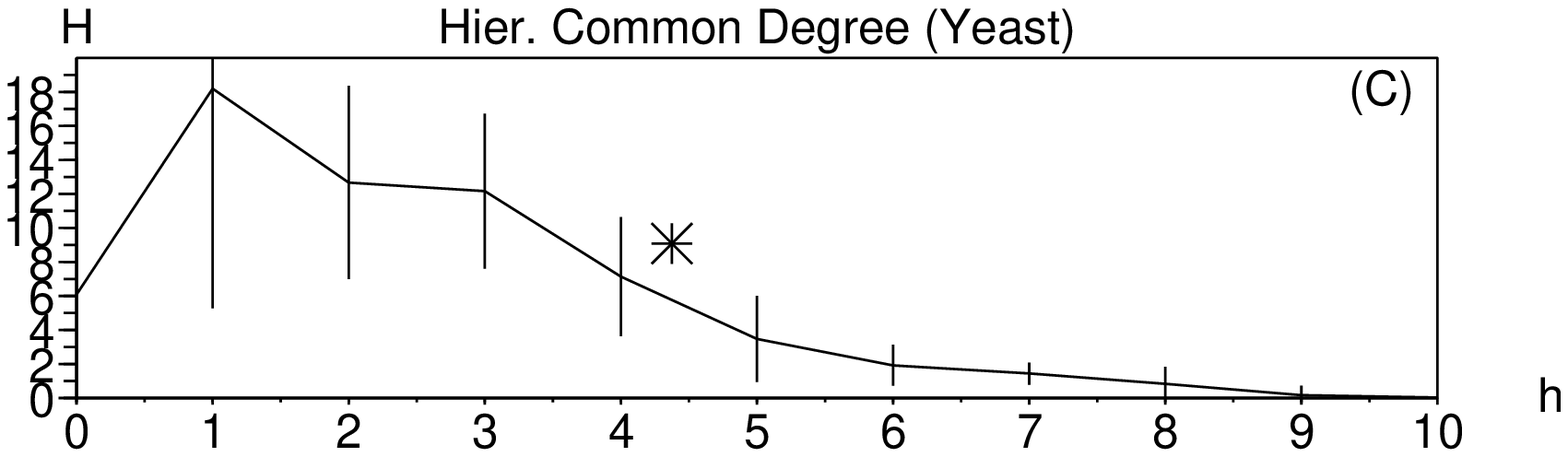}}
   \caption{Hierarchical Common Degree Coefficient of real networks.~\label{fig:real5}}
\end{figure}

Figure \ref{fig:real5} shows the measurements of hierarchical common
degree. The airport(b) and yeast(c) networks curves have a similar
behavior to those obtained for the respective BA curves, with a peak
at the first hierarchical level and a decay.  However the word
network(a) have a mixed behavior, beginning with a increasing curve
like in a BA model, but ending with a convex decay like that
typically observed in random networks.

\begin{figure}
   \resizebox{8cm}{2cm}{\includegraphics[]{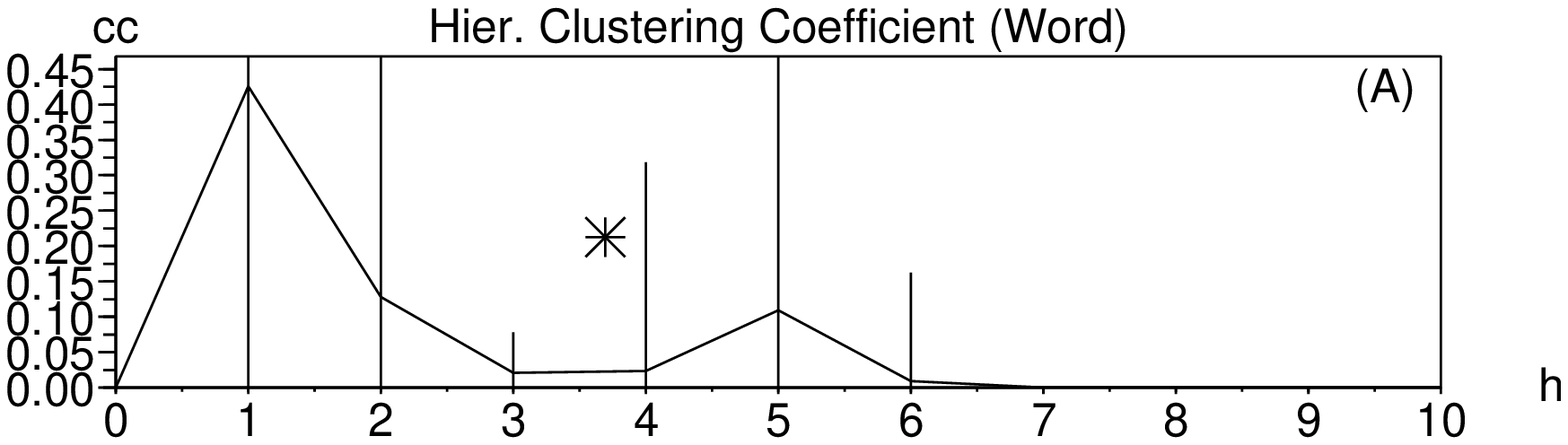}}
   \resizebox{8cm}{2cm}{\includegraphics[]{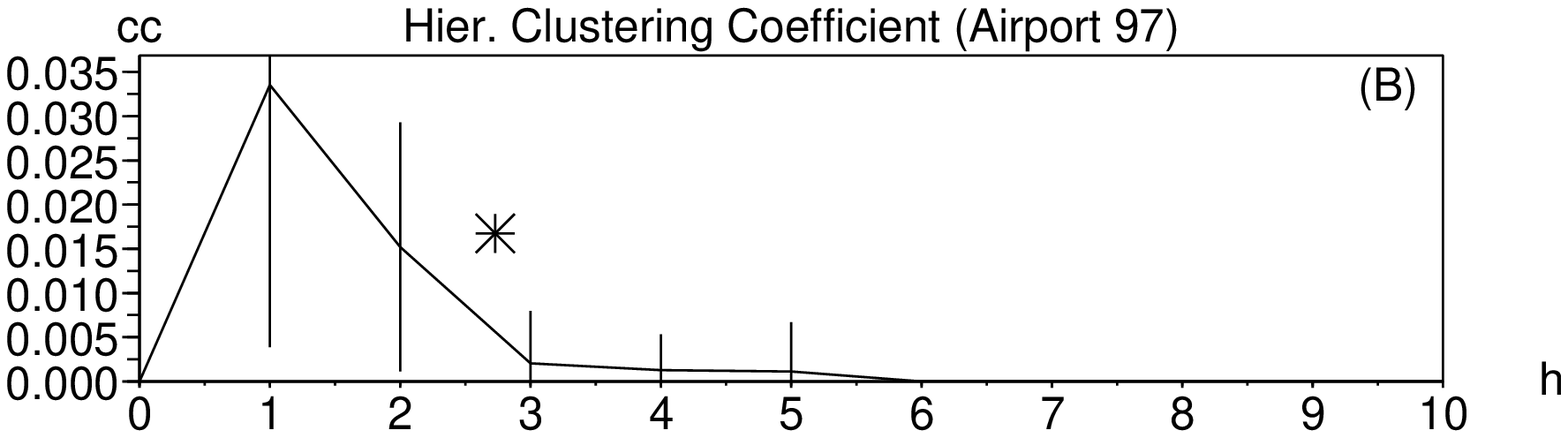}}
   \resizebox{8cm}{2cm}{\includegraphics[]{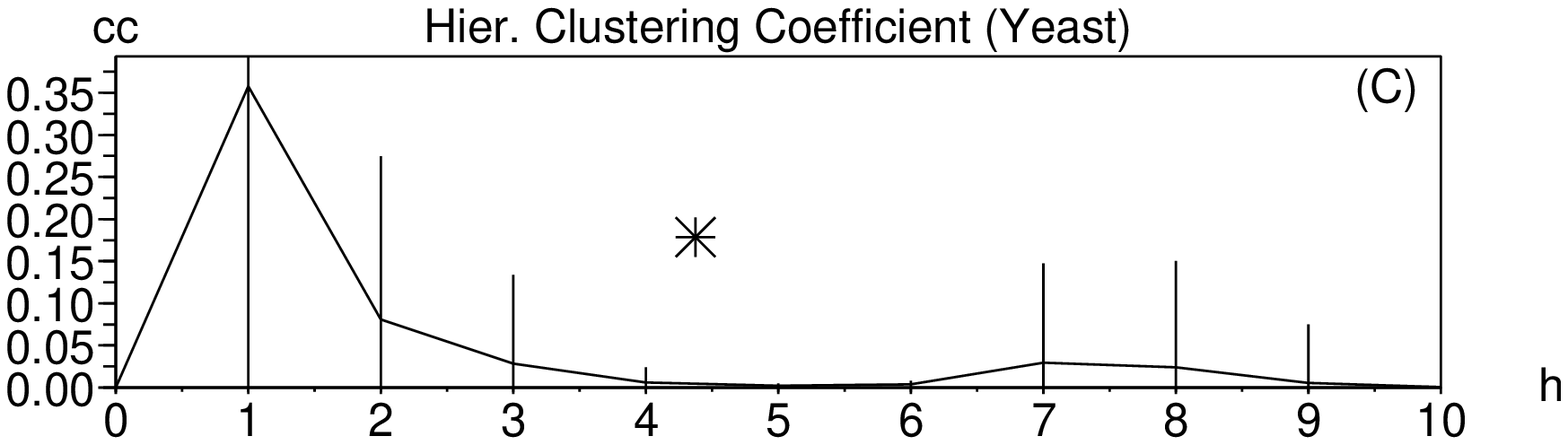}}
   \caption{Hierarchical Clustering Coefficient measures.~\label{fig:real6}}
\end{figure}

The clustering coefficient measurements, shown in Figure
\ref{fig:real6}, substantiate the adherence of the real networks with
respective BA simulated models. Another interesting result which can
be inferred from this figure regards the fact that the hierarchical
clustering coefficients are wider and higher for the word (a) than for
the respective BA simulations.

\begin{figure}
   \resizebox{8cm}{2cm}{\includegraphics[]{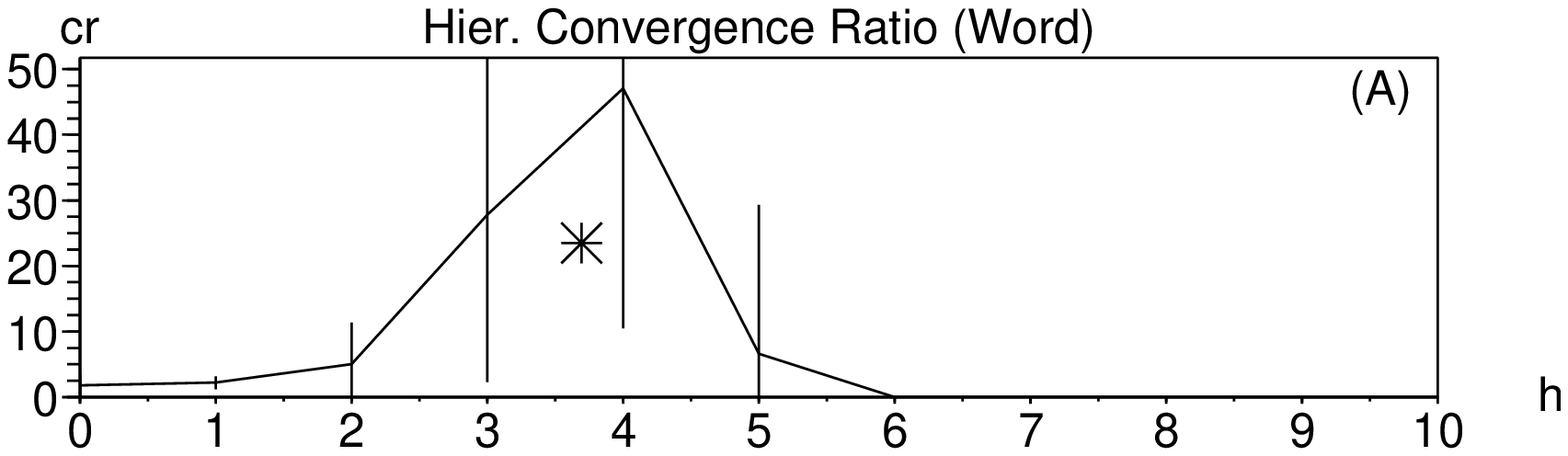}}
   \resizebox{8cm}{2cm}{\includegraphics[]{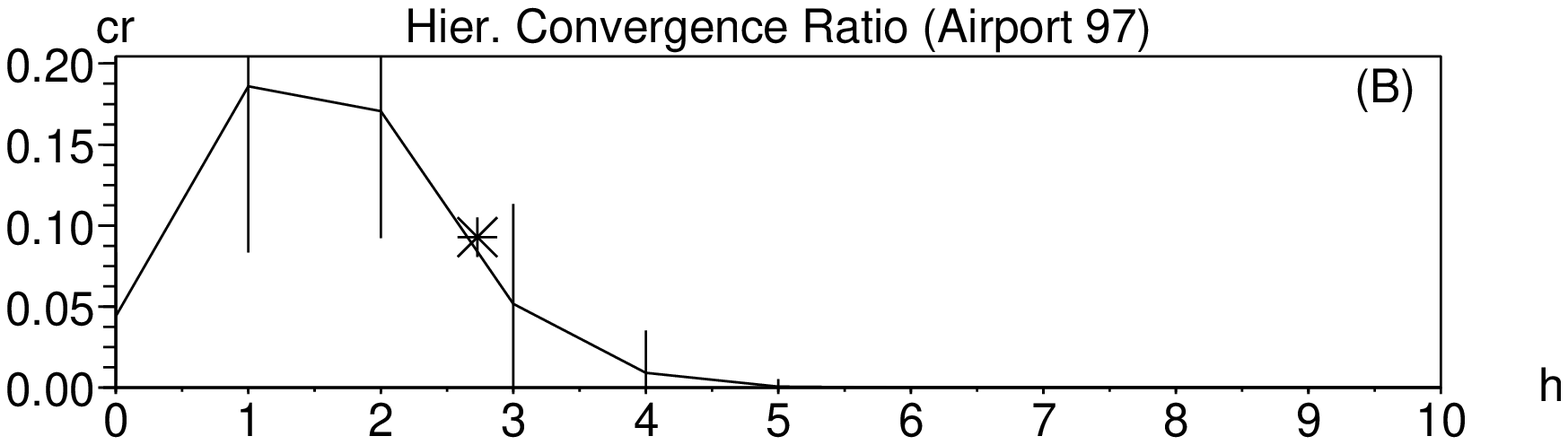}}
   \resizebox{8cm}{2cm}{\includegraphics[]{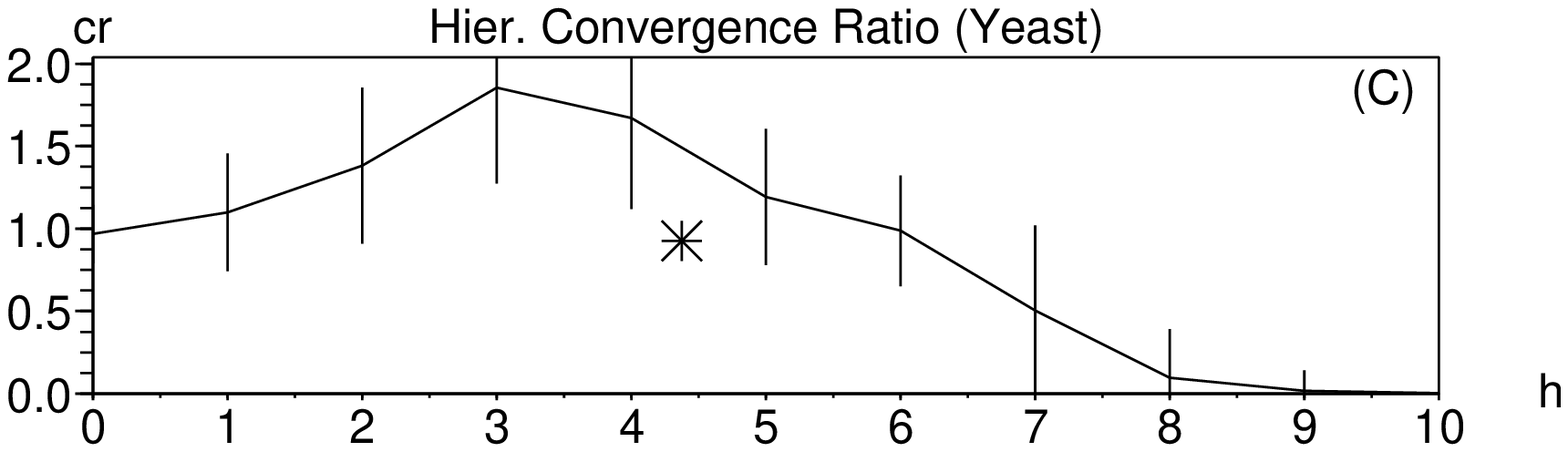}}
   \caption{Convergence Ratio Degree of real networks.~\label{fig:real7}}
\end{figure}

Figure \ref{fig:real7} shows the convergence ratio measurement
values, which yielded the most different curves among the three real
networks and among these and the respective models. The curve for
the word network (a) is more similar to the BA and random model,
being characterized by a low plateau followed by a peak and an
abruptly decrease along the last levels. Different curve profiles
have been obtained for the airport (b) and yeast curves (c).  The
yeast curve presents a wider peak, whose position falls near the
center of the distribution.  The peak of curve obtained for the
airport network resulted displaced to the left hand side, far away
from the average shortest path.  This is a consequence of the fact
that, differently of what is obtained for the yeast, the hubs are
reached after just a few hierarchical levels while starting from
most nodes.  Indeed, we have verified experimentally that the
position and width of the peak of the convergence ratio is
ultimately defined by the distribution of hubs along the hierarchies
after starting from individual nodes.  Therefore, the relatively
narrow peak near the intermediate hierarchical levels obtained for
the word network indicates that the hubs in this structure are
found, in average, after 3 to 5 hierarchical levels. Although also
narrow, the peak of the airport network results in the first levels,
where most hubs are concentrated.  Finally, the wider peak obtained
for the yeast network is a consequence of the fact that the hubs are
distributed more evenly along the hierarchical levels.

\section{Node Categorization through Hierarchical Clustering}

Another possibility for analysis of complex network allowed by the
consideration of hierarchical measurements is the classification of
individual nodes into similar groups. In order to illustrate such a
potential for the characterization of nodes, two complex network are
considered, a Barab\'asi-Albert model (generated with $N=332$ nodes
and $k\simeq 6$ edges) and the airport network with 332 nodes and
$k\simeq 6.4$ edges considered in the last section. This analysis
focuses on the clustering coefficient measurement, which is obtained
for all nodes of such networks. Only the hierarchical levels up to 5
are considered in this example (the use of additional levels tended to
reduce the specificity of the obtained measurements in the case of the
real networks considered in this section).

\begin{figure}
   \resizebox{8cm}{8cm}{\includegraphics[]{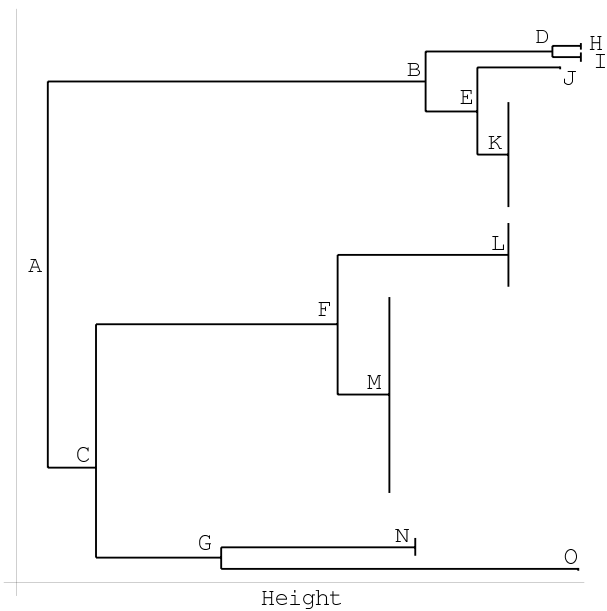}}
   \caption{Dendrogram obtained for the BA model considering the
   hierarchical clustering coefficients of the nodes up to
   hierarchical level 5.  Starting from the righthand side of the
   tree, the nodes are progressively merged into
   clusters in terms of their similarity.~\label{fig:tree1}}
\end{figure}

\begin{figure}
   \resizebox{8cm}{8cm}{\includegraphics[]{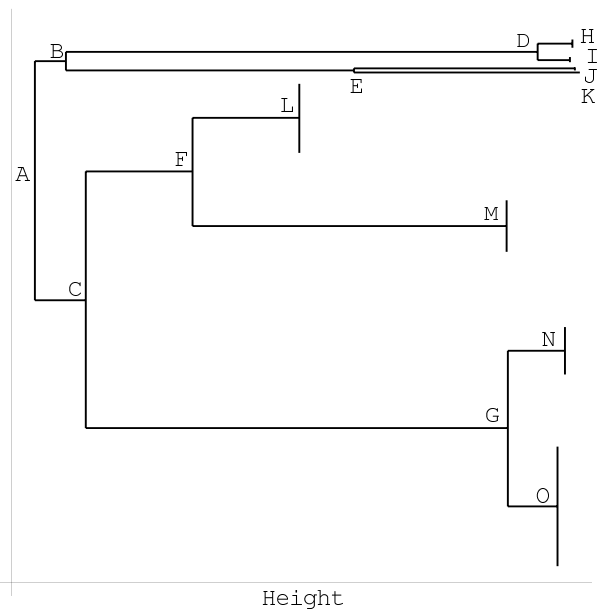}}
   \caption{Dendrogram obtained for the airport network considering the
   hierarchical clustering coefficients of the nodes up to
   hierarchical level 5.~\label{fig:tree2}}
\end{figure}

The hierarchical clustering coefficients are calculated as usual and
supplied to a hierarchical clustering method~\cite{LCosta:book},
namely an agglomerative algorithm, resulting in the trees (also
called \emph{dendrograms}) of measurements shown in figure
\ref{fig:tree1} and figure~\ref{fig:tree2}, respectively to the BA
and airport networks.  For the sake of better visualization, only
the four first hierarchical levels are shown in these figures. The
x-axes in these two three refer to the similarity between nodes.
Starting at the right hand side of the tree, the nodes are merged
with basis on the similarity of their hierarchical clustering
coefficients, yielding the taxonomical categorization of the nodes
into meaningful clusters corresponding to each branching point in
the tree. The y-axes express the size the clusters at the third
hierarchical level.  For instance, the cluster at the top of
Figure~\ref{fig:tree1} contains substantially less nodes than the
third cluster from the bottom of the figure.
bb=0 0 576 164

\begin{figure*}
 \begin{center}
   \includegraphics[scale=.8]{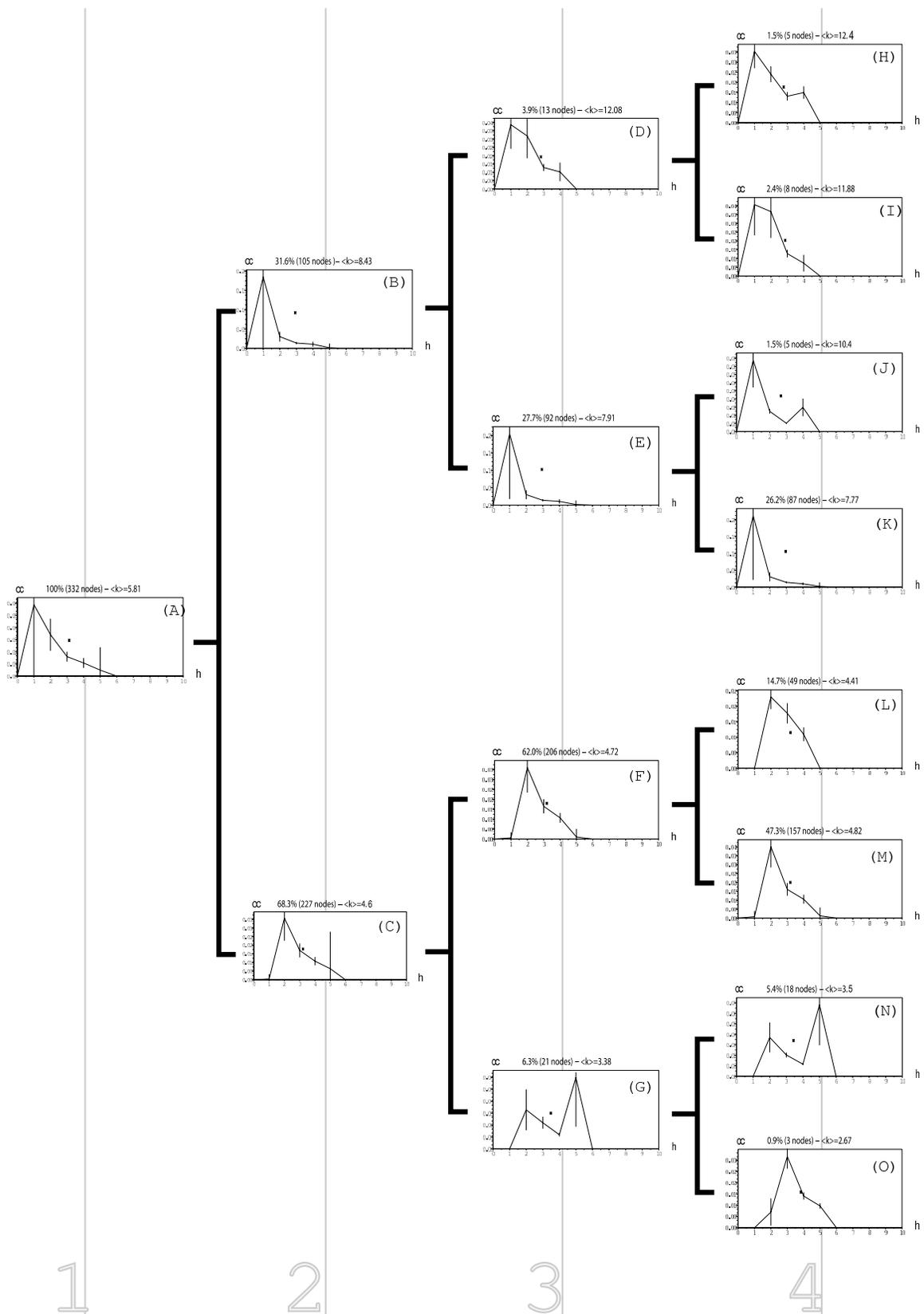} \\

   \caption{Graphs of the average $\pm$ standard deviation of the
   hierarchical clustering coefficient obtained for the BA model.
   Each graph corresponds to the clusters of nodes obtained in the
   four first hierarchical levels of the dendrogram in
   Figure~\ref{fig:tree1}.~\label{fig:tree3}}

  \end{center}
\end{figure*}

\begin{figure*}
 \begin{center}
   \includegraphics[scale=.8]{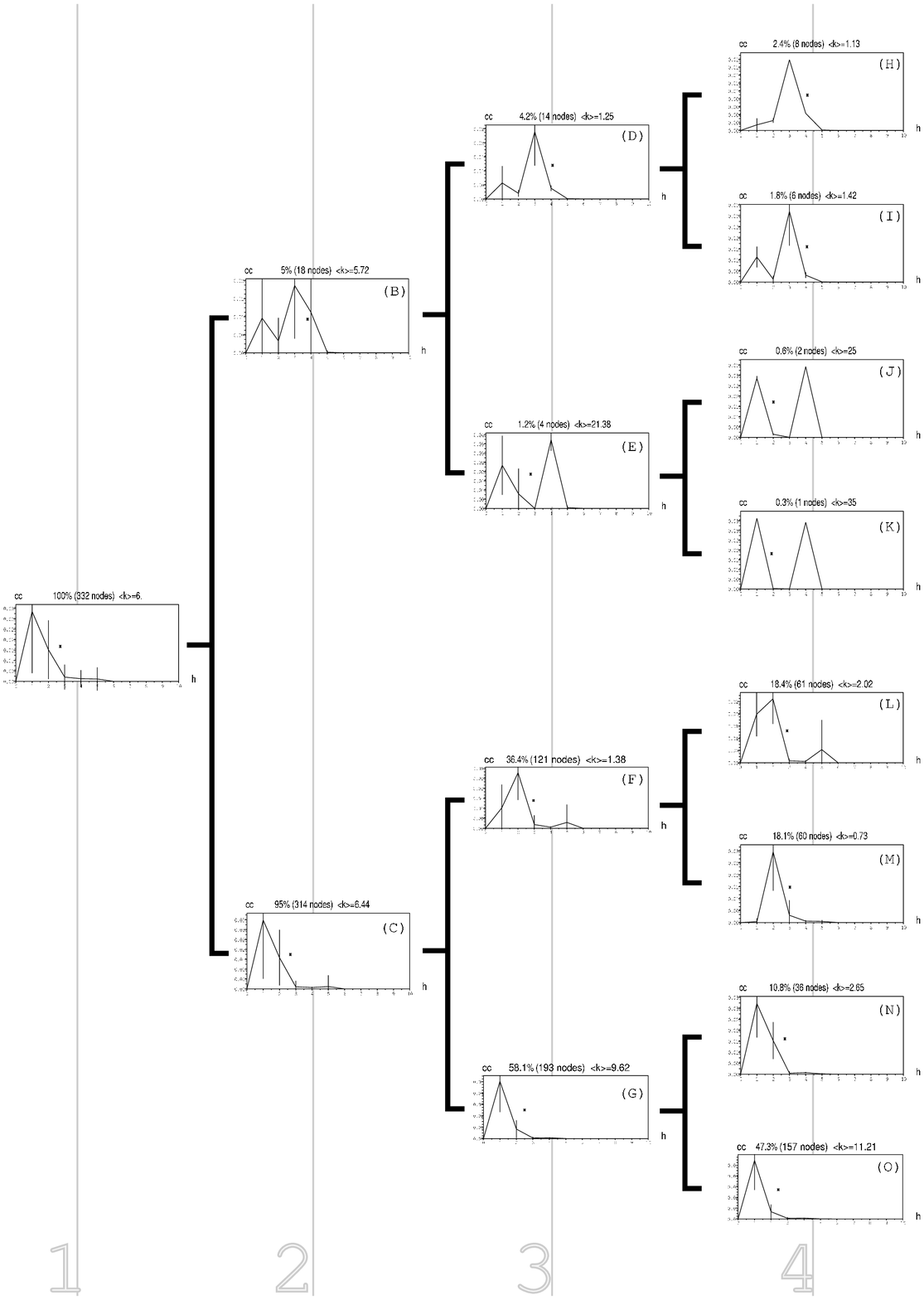} \\
   \caption{Graphs of the average $\pm$ standard deviation of the
   hierarchical clustering coefficient obtained for the airport
   network.~\label{fig:tree4}}
  \end{center}
\end{figure*}

Figures~\ref{fig:tree3} and \ref{fig:tree4} show the graphs of
average $\pm$ standard deviation of the hierarchical clustering
coefficients obtained at each respective level in the dendrograms.
The mean degree and percentage of nodes with respect to the whole
network for each cluster are given above each graph.  Unlike the
dendrograms in Figures \ref{fig:tree1} and \ref{fig:tree2}, the
x-axes of the trees in Figures~\ref{fig:tree3} and \ref{fig:tree4}
do not consider the level of similarity between the groups, which is
done for the sake of better visualization of the graphs obtained for
each cluster of nodes. Starting from the whole network cluster at
the right-hand side of the tree, we can observe the progressive
division of the node hierarchical signatures in terms of subclasses
sharing the basic patterns of hierarchical clustering coefficient
shown in the respective graphs. Such a taxonomical characterization
of the nodes into subclasses provides substantially more
discrimination and characterization than the graphs of average $\pm$
standard deviation obtained considering the whole network such as
those discussed in the previous section. This enhanced potential of
node discrimination and characterization provided by the dendrograms
are particularly useful in the case of networks exhibiting the small
world property, as such cases tend to produce hierarchical
signatures extending over relatively few hierarchical levels.

\section{Concluding Remarks}

This article has addressed, in a didactic and comprehensive fashion,
how a set of hierarchical measurements can be used for the
characterization of important topological properties of complex
networks.  Motivated by the concept of extended neighborhoods and
distances, the identification of hierarchical levels along the
network, with reference to each of its nodes, allows the definition
of a series of useful and informative hierarchical measurements of
the network topology, including hierarchical extensions of the
traditional node degree and clustering coefficient measurements.
The novel concepts of inter and intra-ring degrees, convergence
ratio, edge degree and edge clustering coefficient, as well as their
hierarchical versions, were also introduced here in terms of the
subnetwork generalization described in~\cite{Generalized}.

It has been shown, both analytically and through simulations, that the
hierarchical node degree of a random network has a typical shape
involving a limited number of hierarchical levels while a peak is
observed at its intermediate portion, which is a consequence of the
finite size of the considered networks.  A similar dynamics was
experimentally identified for scale-free and regular network models.
It was also shown, through simulations, that the suggested set of
hierarchical measurements provided a wealthy of information about the
topological structure of the considered models (namely random,
scale-free and regular), allowing the identification of a number of
interesting properties specific to each of those models.  Of
particular interest is the discriminative potential of the
hierarchical common degree and hierarchical clustering coefficient.
The potential of the reported set of hierarchical measurements was
further illustrated with respect to three real networks: word
associations, airport connections and protein-protein interactions.
The comparison of the hierarchical measurements obtained for these
three networks with respective random, regular and BA models with the
same number of nodes and similar node degree indicated that, except
for a few measurements (specific to each model), all the three real
networks were most similar to the BA models.  In the case of the word
associations network, some measurements (i.e.  hierarchical common
degree and inter-ring degree) yielded hierarchical curves which were
similar to random along some parts and similar to BA at other
parts. This network was also verified to present the convergence ratio
most similar to that of a respective BA model.  The concentration of
higher values of convergence ratio at the left hand side of the curves
obtained for the airport network also confirmed the fact that the hubs
in this network are reached much faster than all the other networks
considered in this article.  Contrariwise, the convergence ratio
values obtained for the protein-protein interaction network indicated
that the hubs in this real network are more evenly spaces one
another. As a matter of fact, the convergence ratio resulted in the
most informative of the hierarchical measurements as far as the
analysis of the three real models was concerned.  This is a
consequence of the fact that the presence of a hub at a given
hierarchical level tend to strongly affect the convergence ratio at
that level.

Finally, the current article also proposed and illustrated the
possibility to use the enhanced information provided by the set of
hierarchical measurements in order to organize the nodes of a
network into a taxonomy reflecting the similarities between the
nodes connectivity.  Such a methodology is particularly promising
because the obtained taxonomy can be used to better understand the
main classes of nodes present in a given complex network, i.e. those
classes obtained at the higher levels of the taxonomy.  Indeed,
while the limited number of hierarchical levels present in small
world networks such as random and BA models constrain the potential
of the hierarchical measurements for the discrimination between such
models, the consideration of the main obtained classes of nodes has
been verified to provide further discrimination between the compared
networks.

A series of possible future investigations has been motivated by the
results reported in this article.  First, it would be interesting to
assess in a systematic fashion, and by using multivariate
statistical analysis and hypothesis tests, the potential of each
measurement, as well as their combinations, for discriminating the
possible class of a given network.  Another issue of particular
relevance regards the identification and preservation of hubs
considering not only the immediate neighbors of a node, but of the
neighbors accumulated along growing hierarchical levels.  While such
a possibility has been preliminary considered in~\cite{Generalized},
it would be interesting to consider the preservation of hubs as an
increasing number of hierarchical levels is taken into account. Such
a study is under development with respect to protein-protein
association networks and related results should be futurely
presented. Another study which can complement the results described
in the current work involves the consideration of several types of
small-world networks.  Finally, it would be interesting to apply the
hierarchical measurements for the characterization of several other
real networks such as protein-protein interaction, internet, social
connections, to name but a few.

\vspace{0.3cm}
{\bf Acknowledgment:} Luciano da F. Costa is grateful to FAPESP
(proc. 99/12765-2), CNPq (proc. 308231/03-1) and the Human Frontier
Science Program (RGP39/2002) for financial support.


\end{document}